%% file: main.tex
\documentclass[11pt,preprint]{article}
\pdfoutput=1
\usepackage{jheppub}
\usepackage[dvipsnames]{xcolor}
\usepackage{hyperref}
\usepackage{cleveref}
\usepackage{tikz}
\usepackage[compat=1.1.0]{tikz-feynman}
\usepackage{nccmath}
\usepackage{array}
\usepackage{mathtools}
\usepackage{blkarray}
\usepackage{amssymb}
\usepackage{bbm}
\usepackage{cancel}
\usepackage{slashed}
\usepackage{dsfont}
\usepackage{comment}
\usepackage{bm}
\usepackage{float}
\usepackage{float}
\usepackage{subcaption}

\usetikzlibrary{decorations.markings}
\usetikzlibrary{decorations.pathmorphing}
\usetikzlibrary{intersections}
\usetikzlibrary{calc}

\hypersetup{
    pdfencoding=unicode,
	colorlinks=true,
	urlcolor=Maroon,
	linkcolor=RoyalBlue,
	citecolor=Maroon,
	pdftitle={Soft Factorisation and Exponentiation from Schwinger-Space Geometry},
	pdfauthor={Carolina Figueiredo, Giulio Gambuti, Holmfridur S. Hannesdottir},
	pdfdisplaydoctitle=true,
	pdfstartview=FitH,
	linktocpage=true
}

\tikzset{
    vector/.style={
        decoration={snake, aspect=0.75, mirror, segment length=2mm},
        decorate
    },
    photon/.style={decorate, decoration={snake, amplitude=1pt, segment length=4pt}}
}

\def\d{\mathrm{d}}
\def\be{\begin{equation}}
\def\ee{\end{equation}}

\def\F{\mathcal{F}}
\def\U{\mathcal{U}}
\def\I{\mathcal{I}}
\def\oneloop{1}
\def\e{\mathrm{e}}

\def\D{\mathrm{D}}

\def\GL{\mathrm{GL}}
\def\leq{\leqslant}
\def\geq{\geqslant}

\def\E{\mathrm{E}}
\def\V{\mathrm{V}}
\def\N{\mathcal{N}}

\def\MM{\mathbf{M}}
\def\zz{\mathbf{z}}

\newcommand{\LL}{ {\bf L}}
\newcommand{\Lb}{ {\bf L}}
\newcommand{\Hb}{ {\bf H}}
\newcommand{\Cb}{ {\bf C}}
\newcommand{\Sb}{ {\bf S}}
\newcommand{\Jb}{ {\bf J}}
\newcommand{\Mb}{ {\bf M}}
\newcommand{\Bb}{ {\bf B}}
\newcommand{\Tb}{ {\bf T}}

\newcommand{\pp}{ {\bf p}}

\newcommand{\ab}{\alpha^{-1}}
\newcommand\overmat[2]{%
  \makebox[0pt][l]{$\smash{\overbrace{\phantom{%
    \begin{matrix}#2\end{matrix}}}^{\text{$#1$}}}$}#2}

\def\zz{\mathbf{z}}

\newcommand{\rung}{\boldsymbol{\omega}}
\newcommand{\web}{\boldsymbol{\gamma}}
\newcommand{\jet}{\boldsymbol{\alpha}}

\newcommand{\dd}{\mathrm{d}}
\newcommand{\pl}{\mathrm{pl}}
\newcommand{\npl}{\mathrm{npl}}

\definecolor{darkorange}{HTML}{E28413}
\definecolor{choral}{HTML}{E09891}
\definecolor{darkred}{HTML}{6B0F1A}
\definecolor{darkgreen}{rgb}{0.0, 0.4, 0.0}
\definecolor{darkmagenta}{rgb}{0.55, 0.0, 0.55}
\definecolor{charcoal}{HTML}{343837}

\title{Soft Factorisation and Exponentiation \\ from Schwinger-Space Geometry}

\usepackage{orcidlink}
\author[\!a,\orcidlink{0000-0002-7398-9089}]{Carolina Figueiredo,}\emailAdd{cfigueiredo@princeton.edu}

\author[\!b,\orcidlink{0000-0003-4462-0423}]{Giulio Gambuti,}\emailAdd{giulio.gambuti@physics.ox.ac.uk}

\author[\!c,\orcidlink{0000-0002-5440-2086}]{Holmfridur S. Hannesdottir}\emailAdd{hofie@ias.edu}

\affiliation{$^a$Jadwin Hall, Princeton University, Princeton, NJ, 08540, USA}

\affiliation{$^b$Rudolf Peierls Centre for Theoretical Physics, University of Oxford,
Clarendon Laboratory, Parks Road, Oxford OX1 3PU, UK}

\affiliation{$^c$Institute for Advanced Study, Einstein Drive, Princeton, NJ 08540, USA}

\abstract{
Infrared divergences in Quantum Field Theory govern the low-energy dynamics of many physical theories, and their understanding is a crucial ingredient in predicting the outcomes of collider experiments. We present a novel approach to deriving the structure of these divergences by employing the Schwinger parametrization of Feynman integrals. After using tropical geometry to identify divergent limits, we study the all-orders asymptotic properties of Feynman diagrams via matrix manipulations of graph Laplacians, which allows us to analyse their IR behaviour systematically. We explicitly demonstrate the soft-hard factorization of the integrand for a broad class of diagrams, and reveal that when written in terms of \textit{worldline distances}, topologically distinct diagrams asymptote to the same integrand. 
In particular, for the case of Quantum Electrodynamics, we use this fact to show how ladder-type diagrams combine in Schwinger-parameter space to yield the correct exponentiated soft anomalous dimension. This framework provides a foundation for extending these methods to more complex theories like Quantum Chromodynamics and offers a pathway towards a systematic understanding of infrared divergences in perturbative amplitudes.
}

\preprint{
\begin{flushright}
OUTP-25-03P 
\end{flushright}
}

\makeatletter
\gdef\@fpheader{}
\makeatother

\begin{document} 

\maketitle

\setcounter{page}{2}

\input{Sections/Introduction}

\input{Sections/2loops}

\input{Sections/Schwinger_numerators}

\input{Sections/Factorisation}

\input{Sections/Exponentiation}

\input{Sections/Conclusions}

\acknowledgments
We thank Nima Arkani-Hamed, Fabrizio Caola, Einan Gardi, Aaron Hillman, and Sebastian Mizera for valuable discussions. C.F. is supported  by FCT - Fundacao para a Ciencia e Tecnologia, I.P. (2023.01221.BD and DOI  https://doi.org/10.54499/2023.01221.BD). G.G. was supported by the European Research Council (ERC) under the European Union’s research and innovation programme grant agreement ERC Starting Grant \textsc{hipQCD} (804394) and UKRI Frontier Research Grant, underwriting the ERC Consolidator Grant \textsc{precSM} (UKRI946).
H.S.H. gratefully acknowledges funding provided by the J. Robert Oppenheimer Endowed Fund of the Institute for Advanced Study. This material is based upon work supported by the U.S. Department of Energy, Office of Science, Office of High Energy Physics under Award Number DE-SC0009988.

\newpage
\appendix

\input{Sections/App_Denominators_Diagram}

\input{Sections/App_Loop_momentum_space}

\bibliographystyle{JHEP}
\bibliography{refs}

\end{document}

%% file: Sections/Introduction.tex
\section{Introduction}
Scattering amplitudes lie at the heart of Quantum Field Theory (QFT), connecting theoretical calculations and experimental results. However, when long-range forces are present, such as in Quantum Electrodynamics (QED) and Gravity, the traditional assumptions of scattering theory are violated.  Most notably, the assumption of free asymptotic states fails due to the persistence of long-range forces, giving rise to \emph{infrared (IR) divergences} in perturbative amplitudes. These divergences appear as $1/\epsilon$ poles in dimensional regularization and reflect the inability to isolate interactions at infinite distances. They are of different nature from \emph{ultraviolet (UV) divergences}: While UV divergences can be renormalised away---thanks to the insensitivity of long-distance physics to short-distance details---we cannot use the same argument to eliminate IR divergences, which stem from the low-energy dynamics. To date, IR divergences remain an active area of research, see $e.g.$ ref.~\cite{Agarwal:2021ais} for a review of this rich field.

Two main approaches have been developed to eliminate IR divergences. The first one, the \emph{cross-section method}, traces back to the 1930's, when Bloch and Nordsieck~\cite{Bloch:1937pw} showed that IR divergences in QED are absent in inclusive cross sections. The physical explanation is that low-energy photons are inevitably generated in any scattering process to represent the electromagnetic fields, and when summing over all possible ways of emitting low-energy photons, one gets a finite cross section in QED. However, an analogous theorem for Quantum Chromodynamics (QCD) does not exist as has been established via an explicit counterexample: The cross section for the process $qq \to \mu^+ \mu^- qq + \text{(soft gluons)}$, where the final colours are summed over, is \emph{not} infrared finite~\cite{Doria:1980ak}, as has been discussed in various references since, see $e.g.$ refs.~\cite{DiLieto:1980nkq, Caola:2020xup}. Nevertheless, in any unitary theory, one can resort to the Kinoshita-Lee-Nauenberg (KLN) theorem~\cite{Kinoshita:1962ur, Lee:1964is}, which prescribes that a cross section is infrared finite when summing over all initial and all final states.

Despite the existence of the KLN theorem, determining which exact states are needed for IR finiteness can be complex. Its stronger version implies that summing over initial states \emph{or} final states is sufficient for a finite cross section~\cite{Frye:2018xjj}\footnote{Note the difference to the Bloch-Nordsieck theorem, which prescribes that a sum over the QED process under consideration and any configuration of final state \emph{photons} gives a finite cross section}. However, this version is not practically useful either for defining specific finite observables and follows directly from unitarity: the probability for an initial state to scatter into \emph{anything} is 1, and 1 is IR finite. Instead, \emph{infrared and collinear (IRC) safety} has played a central role in QCD, where observables are constructed to be insensitive to infrared and collinear splittings, $e.g.$ by collecting partons into jets~\cite{Sterman:1977wj, Clavelli:1979md, Farhi:1977sg}. These IRC safe observables, including thrust and jet mass, become calculable in perturbative QCD by resumming large logarithms~\cite{Catani:1991bd,Catani:1992ua} and have thus played an essential role in precision measurements~\cite{Benitez:2024nav,Benitez:2025vsp}, see refs.~\cite{Larkoski:2017jix,Marzani:2019hun} for reviews.

The second strategy to achieve IR finiteness involves tackling the underlying problem of using free asymptotic scattering states, and instead redefine the scattering amplitudes themselves to include the asymptotic long-range interactions. Despite the existence of methods to define finite observables, finite scattering amplitudes are still theoretically appealing, as they are, for example, a prerequisite for using non-perturbative bootstrap methods (see ref.~\cite{Kruczenski:2022lot} for a review). Redefining scattering amplitudes in QFT was originally inspired by a construction of finite amplitudes for charged particles scattering off a Coulomb-potential in non-relativistic quantum mechanics~\cite{dollard1964asymptotic,dollard1971quantum}. It was later generalised to QED by Chung~\cite{Chung:1965zza} and Faddeev-Kulish~\cite{Kulish:1970ut}. They showed how to define QED scattering states incorporating asymptotic dynamics, leading to finite scattering amplitudes. By leveraging modern approaches, one can universally factor out IR divergences from the S-matrix to isolate the effects of the short-distance scattering~\cite{Hannesdottir:2019umk,Hannesdottir:2019opa}.

Naturally, both IRC safety and modifying the definition of scattering amplitudes require a good understanding of how to explicitly compute IR divergences. A remarkable fact is that UV-renormalised scattering amplitudes for hard scattering in QCD \emph{factorise} into a hard, soft and collinear part: 
\begin{equation} \label{eq:master_factorisation}
    \mathcal{M}_n^{\text{ren}} = \prod_{i=1}^n\mathcal{J}_i \cdot \mathcal{S} \cdot \mathcal{H}_n .
\end{equation}
Soft dynamics are captured by the soft function $\mathcal{S}$\footnote{In non-abelian gauge theories $\mathcal{S}$ takes the form of an operator mixing the various colour structures in $\mathcal{H}_n$.} containing divergences from low-energy gluons, collinear dynamics by jet functions $\mathcal{J}_i$, and short-distance interactions by the hard function $\mathcal{H}_n$. The IR divergences in gauge theories were connected $e.g.$ in refs.~\cite{Becher:2009cu, Gardi:2009zv, Becher:2009qa, Gardi:2009qi} to the ultraviolet divergences of Wilson line correlators, which allow for a simplified computation of the soft and jet functions.
The factorisation of amplitudes and observables in QCD has a long and rich history, with foundational work establishing these concepts primarily in momentum space, see $e.g.$~\cite{Collins:1981tt, Collins:1982wa, Collins:1983ju, 
Sen:1982bt, Collins:1988ig, Collins:1989gx, Contopanagos:1996nh,Kidonakis:1998nf,Aybat:2006mz,Feige:2013zla,Feige:2014wja}.
This factorisation can be systematically achieved within Soft-Collinear Effective Theory (SCET)~\cite{Bauer:2000ew,Bauer:2001yt,Beneke:2002ph} (see refs.~\cite{stewart2013lectures, Becher:2014oda} for pedagogical lectures), which provides a field-theoretic framework for separating physics at different scales. 
While the approach we present in this paper is formally distinct, it shares the conceptual goal of isolating these contributions in integration space, offering a complementary perspective on the origins of factorisation.

As explained in the references above, the QCD soft function $\mathcal{S}$ can be expressed as a path ordered\footnote{The path ordering here refers to operators in colour spaces being ordered with respect to the scale $\mu$.} exponential of the so-called soft anomalous dimension $\Gamma_{\text{soft}}$:
\begin{equation}
    \mathcal{S} = \mathbb{P} \exp \int_{\mu}^{\infty} \frac{\d \mu'}{\mu'} \Gamma_{\text{soft}}(\mu') .
\end{equation}
Using either the correspondence between soft IR divergences and UV divergences of Wilson lines or bootstrap methods, the soft anomalous dimension has been computed up to three loops~\cite{Polyakov:1980ca,Arefeva:1980zd,Dotsenko:1979wb,Brandt:1981kf,Korchemsky:1985xj,Korchemsky:1985xu,Korchemsky:1987wg, Mitov:2009sv,Ferroglia:2009ii,Ferroglia:2009ep, Almelid:2015jia,Almelid:2017qju,Liu:2022elt}. At three loops, it exhibits both \emph{dipole} and \emph{quadrupole} terms \cite{Almelid:2015jia}. 
The jet functions $\mathcal{J}_i$ are instead controlled by the collinear anomalous dimensions, which depend on the type of the external state $i$. Both the collinear anomalous dimensions and the cusp anomalous dimension, a component of $\Gamma_{\text{soft}}$, are known at four loops~\cite{Korchemsky:1987wg,Moch:2004pa,Vogt:2004mw,Bruser:2019auj,Henn:2019swt,vonManteuffel:2020vjv,Ravindran:2004mb,Moch:2005id,Moch:2005tm,Agarwal:2021zft}.

An even stronger result holds in QED with massive electrons where infrared divergences \emph{exponentiate} exactly:
\begin{equation} \label{eq:one_loop_exact}
\mathcal{M}_n^{\text{ren}} = \mathcal{S} \cdot \mathcal{H}_n , \quad \text{with } \quad
\mathcal{S} = \exp(\mathcal{S}_{(1)}) \,,
\end{equation}
so that $\mathcal{S}$ is simply an exponential of the one-loop (UV renormalised) divergence $\mathcal{S}_{(1)} \sim 1/\epsilon $. 
Whenever such exponentiation holds, it implies that all IR divergences at $\ell$-loops can be predicted solely from lower-loop information. More precisely, if we know the amplitude up to $k$ loops, we can predict the IR divergences at $\ell$ loops up to and including the coefficient of $1/{\epsilon^{\ell-k}}$. This can be seen by expanding both sides of the equation in powers of the coupling constant, and matching terms on both sides. For example, denoting with subscript $(i)$ the $i$-th loop, and normalizing the tree-level amplitude to 1 for simplicity, \cref{eq:one_loop_exact} gives:
\begin{equation}
    1 + \mathcal{M}_{(1)} + \mathcal{M}_{(2)} + \ldots = 
    \Big(1 +  \mathcal{S}_{(1)} + \frac{1}{2} \mathcal{S}_{(1)}^2 + \ldots \Big)
    \times \left(1 + \mathcal{H}_{(1)} + \mathcal{H}_{(2)} + \ldots \right) \,,
\end{equation}
and when equating terms at the same order one finds
\begin{equation}
\mathcal{M}_{(2)} = \frac{1}{2} \mathcal{S}_{(1)}^2 + \left(\mathcal{M}_{(1)} - \mathcal{S}_{(1)}\right) \mathcal{S}_{(1)}
+ \text{(IR finite)} \, .
\label{eq:expIR}
\end{equation}
That is, the IR divergences of the 2-loop amplitude can be fully determined from the knowledge of the full one-loop amplitude, etc.

The result in \cref{eq:one_loop_exact} was originally worked out using the loop-momentum space representation of Feynman integrals. For the leading ${1}/{\epsilon^{\ell}}$ pole in ladder-type diagrams at $\ell$ loops, a slick argument by Weinberg~\cite{Weinberg:1965nx} involving symmetrizing over different photon attachments to the fermion legs can be used to show how the exponentiation arises. Explaining how the subleading divergences exponentiate, and including various subtleties about $e.g.$ numerator factors and renormalisation, requires the longer explanation provided, for example, in ref.~\cite{Yennie:1961ad} in loop-momentum space, or in the many-body worldline formalism, see refs.~\cite{Feal:2022iyn, Feal:2022ufw}.

In this paper, we explore the structure of IR divergences using an alternative representation of Feynman integrals: the \emph{Schwinger-parameter} formalism. In this approach, the loop momenta are integrated out, see $e.g.$ refs.~\cite{Weinzierl:2022eaz,Hannesdottir:2022bmo} for a review, and we are left with an integral over Schwinger parameters. Here, the worldline perspective, where divergences are associated with the shrinking (UV) or expanding (IR) of Schwinger proper times has proven to be ideally suited for studying divergences using the mathematical methods of \emph{tropical geometry}~\cite{maclagan2015introduction}. It provides a unified framework to analyse both UV and IR divergences by examining the asymptotic behaviour of the integrand.\footnote{As has recently been pointed out~\cite{Gardi:2022khw, Ma:2023hrt, Gardi:2024axt}, the asymptotic structure is not always sufficient to capture the divergences, and one must resort to the \emph{Landau equations}~\cite{Bjorken:1959fd,Landau:1959fi,10.1143/PTP.22.128} to detect all the potential singular regions for finite values of the Schwinger parameters. However, the asymptotic behaviour will be sufficient for the analysis in this paper.} In particular, the polyhedral structure of Symanzik polynomials, built from graph Laplacians, offers a novel framework for understanding the IR behaviour. There has been much recent progress on this topic, for example, in systematically identifying divergent regions~\cite{Jantzen:2012mw,Gardi:2022khw}, implementing subtraction schemes for subleading divergences~\cite{Hillman:2023ezp, Salvatori:2024nva}, and computing the coefficients of UV and IR divergences~\cite{Heinrich:2021dbf, Arkani-Hamed:2022cqe}.
It is therefore natural to ask how this systematic geometric approach can be leveraged to understand IR divergences  to all orders in $\epsilon$ in gauge theories.

As a first step in this direction, the central aim of this work is to show how the factorisation and exponentiation of soft infrared divergences can be clearly worked out within the Schwinger-parameter framework. In this formalism, Feynman integrals for a graph $G$ take the form
\begin{equation}
\label{eq:IG}
    \mathcal{I}_G = 
    i^{E-\ell\D/2}
    \int_0^\infty \prod_{e \in G} \d \alpha_e \, \frac{\N_G}{\U_G^{\D/2}} \exp  \Big[ \frac{i \F_G}{\U_G} \Big],
\end{equation}
where $\U_G$ and $\F_G$ are the first and second Symanzik polynomials of a graph $G$, respectively, and the numerator $\N_G$ is theory dependent, and can be written as a polynomial in the $\alpha_e$, see ref.~\cite{Hannesdottir:2022bmo} or \cref{sec:numerators} of this work for a brief review.
Above $E$ counts the number of propagators in $G$, $\ell$ the number of loops and $\D$ the number of space-time dimensions.
To achieve our goal of demonstrating factorisation and exponentiation, we leverage the fact that the polynomials appearing in Schwinger-parametrized Feynman integrals can be understood in terms of \emph{Laplacian matrices}. This connection, which follows from the matrix-tree theorem (or Kirchhoff's theorem) \cite{1086426}, allows the factorisation and exponentiation theorems to be translated into manipulations of matrices, making the incorporation of numerators, intermediate hard factors, and different limits straightforward. 

A class of diagrams that appears repeatedly in our analysis are \emph{ladder diagrams}, where two (or more) massive lines (jets) in a Feynman diagram are connected via any number of massless lines, see \cref{fig:2loop-diagrams}. Our proofs will involve both planar and non-planar ladder diagrams, but also general \emph{ladder-like} ones, which involve any number of jets. 
We will study the asymptotic form of these diagrams as the Schwinger parameters $\alpha_e$ are scaled by some power of a small parameter $\lambda$
\begin{equation}
    \alpha_e \to \lambda^{-r_e} \alpha_e 
\end{equation}
given by the corresponding tropical \emph{ray} $r$. 
In particular, in this work we will focus on rays associated with IR divergences of the type encountered in QED with massive fermions.
In the rest of this paper we will refer to any such scaling as \emph{soft}, since in momentum space they correspond to limits in which some subset of massless particles have vanishing momenta. In general there can be multiple soft scalings for a single diagram. 

As we will discuss in detail, it turns out that in any soft scaling, the Symanzik polynomials $\F$ and $\U$ satisfy \begin{equation}
\frac{\F}{\U} \to \frac{\F_H}{\U_H} +\frac{\F_S}{\U_S} \,, \qquad \qquad \U \to \U_H \, \U_S \,,
\label{eq:conjecture}
\end{equation}
where the subscripts $H$ and $S$ correspond to hard and soft subgraphs identified by the scaling or ray considered. We will clarify their definition below.

One can straightforwardly see that these equations lead to a factorisation of the integral in~\cref{eq:IG} if $\N_G$ also factorises (note that for a scalar integral, $\N_G=1$).
An interesting observation in Schwinger parameter space which we present in this work is that different Feynman diagrams (such as planar and non-planar ladders) all yield the same \emph{integrand} in the soft limit, while the integration regions differ between different topologies. 
We will see how this property of soft integrands becomes evident only when they are expressed in terms of a set of \textit{worldline variables} which are related to Schwinger parameters by simple linear transformations. 
This is quite different from the loop-momentum space trick of symmetrizing over photon attachments to get an eikonal integrand and a ${1}/{n!}$ factor: instead, here the eikonal integrand is manifest diagram by diagram\footnote{Due to the different integration regions, the leading divergence always comes from the \emph{planar ladder} diagram.} and the sum of all ladder diagrams fills out the entire integration space, leading to factorisation of the integrated amplitude. This structure mirrors the known behaviour of non-Abelian webs in the position-space eikonal formalism~\cite{Frenkel:1984pz,Gardi:2010rn,Gardi:2011yz,Gardi:2013ita}.

This paper is organised as follows:~\cref{sec:two_loop} showcases the main ideas of this paper via concrete examples of one- and two-loop scalar diagrams. It contains an explicit analysis of their integrand in the relevant soft limits as well as of how their domains of integration combine to yield exponentiation of the one-loop divergence. 
In~\cref{sec:numerators} we briefly review a general method to represent Feynman diagram numerators in Schwinger representation. 
In~\cref{sec:factorisation_matrix}, we work out the full proof of integrand factorisation for QED-type diagrams, where numerators and partially scaled diagrams are included and where we work directly at the level of Laplacian matrices. We also include a {\tt Mathematica} notebook showing explicit examples of the matrix definitions and manipulations required. We go on to showing how the integration regions for different diagrams combine in a way that exhibits exponentiation in~\cref{sec:regions}. Appendix~\ref{app:factorisation_diagram} contains details on a diagrammatic proof of factorisation, and appendix~\ref{app:exp-loopmom} connects our analysis in Schwinger parameters to the one in loop-momentum space.

\paragraph{Notation and conventions}

We work in mostly-minus Minkowski signature and in an expansion around $\D=4 - 2\epsilon$ dimensions, where we take $\epsilon \to 0^{-}$ for IR divergences and we will leave the Feynman causal $i\varepsilon$ implicit throughout. Further, we set the dimensional regularisation scale $\mu = 1$ and only reconstruct the dependence on it when necessary from dimensional analysis.
We use $X_{(\ell)}$ to refer to the $\ell$-loop order of the quantity $X$. 
Since in this paper we will mainly work in the Schwinger parametric representation of Feynman integrals, unless otherwise specified we will indicate the ``standard" integration in Schwinger parameter space ($i.e.$ $\mathbb{R}_+^E$) in any number of variables by a single $\int$ symbol.

In the various IR integrations regions we will consider, we will refer to the subset of the edges  close to the mass shell as \emph{soft} part and to the complementary set as the \emph{hard} part. We will use subscripts ``$S$'' and ``$H$'' respectively, to denote various quantities belonging to each part of the diagram.

%% file: Sections/2loops.tex
\section{Warm-up: one and two loops}
\label{sec:two_loop}
\begin{figure}[t]
    \centering
    \input{Figures/2loop}
    \caption{(\textbf{left}) Planar 2-loop ladder; (\textbf{right}) Non-planar 2-loop ladder.}
    \label{fig:2loop-diagrams}
\end{figure}
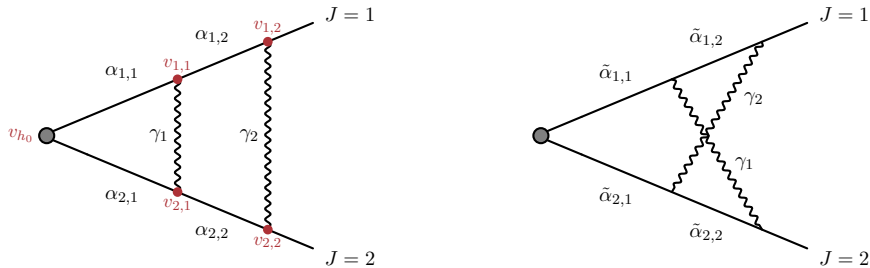
In this section we carry out the analysis of factorisation and exponentiation for the one-loop triangle and two-loop ladder diagrams. The steps taken follow the organisation of the rest of the paper, and illustrate the key aspects of our method. In \cref{sec:factorisation_matrix,sec:regions} we will then discuss how each step is generalised to all orders and to any number of external massive states. 
For simplicity, here we stick to the case where we consider a purely scalar theory (so that there are no numerators), but one in which massive particles exchange a massless one, imitating the QED case. In later sections, we will incorporate a larger class of diagrams, numerators and renormalisation.

The first step is to use tropical geometry to identify the divergent scalings in Schwinger parameter space that lead to IR divergences. We go through this analysis in \cref{sec:div2loop} and explain how the \textit{Newton polytope} of the Symanzik polynomials allow us to detect divergent rays. By doing this at one-loop, we define the \textit{one-loop} soft factor $S_{(1)}$, the same one as appears in the exponential of \cref{eq:one_loop_exact}. 

In \cref{sec:fact2loop}, we then proceed to study the expansion of the integrands for the planar and non-planar ladders at two loops under the divergent scalings. Concretely, using only the recursive definition of the Symanzik polynomials, we can show that, as expected, under the different IR divergent scalings the integrands factorise as the product of hard and soft terms. 

In \cref{sec:2loopMatrix}, we study the same properties as in \cref{sec:fact2loop}, but using a different method: instead of the recursive properties, we use the definition of the Symanzik polynomials via \textit{graph Laplacian} matrices. We show how factorisation of the integrand is a simple consequence of ``factorisation into blocks'' of these matrices. This approach turns out to give a much more uniform description of the hard-soft factorisation, allowing for the study of diagrams with different topologies as well as incorporating numerator factors, as will be explained in subsequent sections.

In \cref{sec:regions2loop}, we show how we can combine the contributions of the different integrands and their respective expansions to define the IR divergent part of the two-loop amplitude. This involves extending the integration domains of the expanded integrands to regions beyond their respective domains of validity. 
In doing so, we will need to carefully add subtraction terms to avoid over-counting divergences. This will allow us to obtain the exponentiated form of the loop-integrated amplitude at two-loops. 
We end in \cref{sec:Renorm2} by commenting on the contributions of diagrams beyond ladders to the IR divergent part of the answer.  

\subsection{Location of IR divergences from tropical analysis}
\label{sec:div2loop}

The first step in our analysis is to understand the definition and origin of the divergent scalings.  This can be done systematically for a general diagram following the steps outlined in ref.~\cite{Arkani-Hamed:2022cqe}, (see also ref.~\cite{Heinrich:2021dbf}). 
In the {\tt Mathematica} file attached with this paper, we provide a code which uses the software {\tt polymake}~\cite{Gawrilow2000,polymake2} to directly output the divergent rays for any given diagram. 

In order to illustrate this procedure explicitly, we review in detail how we can use the tropical methods to derive the divergent scaling of the one-loop diagram, and in this way give an explicit definition of $S_{(1)}$ (which we will need to prove exponentiation at all orders).

\subsubsection*{Divergent rays of the 1-loop diagram}
For simplicity, we start with a one-loop scalar diagram, with
\begin{equation}
\label{eq:Ioneloop}
    \I_{\oneloop} \left[ \begin{gathered} \begin{tikzpicture}[line width=0.8,scale=1,line cap=round]
        \coordinate (o) at (0,0);
        \coordinate (b1) at ($(o) + (30:1)$);
        \coordinate (b2) at ($(o) + (-30:1)$);
        \coordinate (c) at ($(o) + (30:2)$);
        \coordinate (d) at ($(o) + (-30:2)$);
        \path (b1) -- (c) coordinate[pos=0.4] (p1t); 
        \path (b2) -- (d) coordinate[pos=0.4] (p1b);
        \draw[] (o) -- (c);
        \draw[] (o) -- (d);
        
        \draw[photon] (p1t) -- (p1b) node[midway,scale=0.7,xshift=-10] {$\gamma_1$};
        
        \node[scale=0.7,xshift=-30,yshift=2] at (p1t) {$\alpha_{1,1}$};
        \node[scale=0.7,xshift=-30,yshift=1] at (p1b) {$\alpha_{2,1}$};
     
        \filldraw[color=black,fill=gray] (o) circle[radius=0.1];
        \node[scale=0.7,xshift=20,yshift=5] at (c) {$J=1$};
        \node[scale=0.7,xshift=20,yshift=-5] at (d) {$J=2$};
    \end{tikzpicture}\end{gathered}\right]  = i^{3-\D/2} \int_{\alpha_e > 0}  \frac{\d^3 \alpha_e}{\U_{(1)}^{\D/2}} \exp\left[ i \, \frac{\F_{(1)}}{\U_{(1)}}\right] \,.
\end{equation}
Mirroring QED, we take the straight lines to be massive with mass $m$ and the wavy line to be massless, and we will refer to the wavy lines as photons. 
The one-loop Symanzik polynomials are given by
\begin{align}
    \U_{(1)} & = \alpha_{1,1}+\alpha_{2,1}+\gamma_1 \,, \label{eq:Upol}\\
    \F_{(1)} & = s \alpha_{1,1} \alpha_{2,1} + m^2 (\alpha_{1,1} \gamma_1 + \alpha_{2,1} \gamma_1)- m^2 (\alpha_{1,1}+\alpha_{2,1}) (\alpha_{1,1}+\alpha_{2,1}+\gamma_1) \,,
\end{align}
where we have defined $s=(p_1+p_2)^2$, with $p_J$ the momentum associated with the external leg $J$.
Note that, without any limit or expansions, the second Symanzik polynomial simplifies to
\begin{equation}
    \F_{(1)} = s \alpha_{1,1} \alpha_{2,1} - m^2 (\alpha_{1,1}+\alpha_{2,1})^2 \,,
    \label{eq:Fpol}
\end{equation}
which is the same $\F$ polynomial as the one for a scalar bubble diagram with incoming momentum squared $s$ and equal internal masses $m$. However, crucially, the $\U$ polynomial differs from that of a bubble diagram since it contains the additional Schwinger parameter $\gamma_1$.

Since the argument of the exponential in~\cref{eq:Ioneloop} is homogeneous of degree one in the $\alpha_e$ we can write it as a projective integral in Feynman representation as 
\begin{equation} \label{eq:one-loop-Symanzik}
    \I_{(1)}  = \Gamma \left(3-\D/2 \right) \int_{\alpha_e > 0} \frac{\d^3 \alpha_e}{\GL (1)} \frac{1}{\F_{(1)}^{3-\D/2} \U_{(1)}^{\D-3}} \,.
\end{equation}
The overall $\GL(1)$ covariance can be used to set, for example, $\alpha_{1,1}+\alpha_{2,1}+\gamma_1=1$, or any of the individual Schwinger parameters, $\alpha_{1,1}$, $\alpha_{2,1}$ and $\gamma_1$, to one.

This integral has multiple possible divergences, which can be detected by solving the \emph{Landau equations}~\cite{Landau:1959fi,Bjorken:1959fd,10.1143/PTP.22.128}. Here, we distinguish between UV/IR divergences and the \emph{kinematic singularities}, which occur only for specific values of kinematic data. For example, $\I_{(1)}$ has a kinematic singularity if
\begin{equation}
    s = 4 m^2, \qquad \alpha_{1,1} = \alpha_{2,1} \,.
\end{equation}
In contrast with UV/IR divergences, this singularity occurs only at one point in the kinematic space, $i.e.$ at $s=4m^2$. Depending on the space-time dimension $\D$, it can lead to a logarithmic or square-root type branch point of the function $\I_1$ in the complex $s$ plane. Locating these singularities is crucial to studying the analytic structure of Feynman integrals, but will not be a concern in this paper.\footnote{See refs.~\cite{Fevola:2023kaw,Fevola:2023fzn,Helmer:2024wax,Caron-Huot:2024brh,Correia:2025yao} for recent work on solving the Landau equations systematically.}

In this paper, we will instead be concerned with singularities that prevail at \emph{any} values of $s$ and $m^2$, which lead to UV/IR divergences of the Feynman integral and must be regulated, $e.g.$ in dimensional regularization for the integral to be well-defined. A class of these singularities come from configurations where $\alpha_e$ are at the boundaries of the integration contour, that is, $\alpha_e \to \infty$ or $\alpha_e \to 0$, for $\alpha_e \in \{\alpha_{1,1},\alpha_{2,1},\gamma_1\}$. Intuitively, the integral in~\cref{eq:Ioneloop} is the integral over the worldline action for each internal particle, where $\alpha_e$ is the proper time along the worldline for each internal particle $e$. Very crudely, the configurations where $\alpha_e \to 0$ then correspond to the worldlines shrinking, signalling contact interactions and UV divergences, while those obtained for $\alpha_e \to \infty$ correspond to IR divergences (note that in general this identification is not valid in the projective representation of \cref{eq:one-loop-Symanzik}).
We point out that there exist examples of values of $\alpha_e$ in the bulk of integration contour can also lead to UV/IR divergences~\cite{Jantzen:2012mw,Ananthanarayan:2018tog,Gardi:2024axt} of Feynman integrals, although those types of divergences do not occur in the diagrams we consider in this paper, so we will not worry about them.

Going back to our one-loop example, we start by fixing $\gamma_1=1$, and perform the change of variables, $\alpha_{1,1} \to \e^{\tau_1}, \alpha_{2,1} \to \e^{\tau_2}$. To derive the divergent directions in $\tau_1,\tau_2$ space, we consider the Newton polytopes\footnote{The Newton polytope $\text{Newt}(P)$ of a polynomial P is the convex hull of the points given by the exponents of its monomials. Its geometric properties, particularly its outer normal vectors, encode the dominant scaling behaviours of the polynomial.} of the Symanzik polynomials:
\begin{equation}
    \mathbf{U}_{(1)} = \text{Newt}(\U_{(1)}), \qquad \mathbf{F}_{(1)} = \text{Newt}(\F_{(1)}) \,.
\end{equation}
In particular, to study to potential directions for which the integral from~\cref{eq:one-loop-Symanzik} can diverge, it suffices to consider the Minkowski sum $\mathbf{S}_{(1)} = \mathbf{U}_{(1)} \oplus \mathbf{F}_{(1)}$. The resulting Newton polytope $\mathbf{S}_{(1)}$ is the convex hull of the points
\begin{equation}
\{2, 1\}, \{3, 0\}, \{1, 2\}, \{1, 2\}, \{2, 1\}, \{0, 3\}, \{1, 1\}, \{2, 0\}, \{0, 2\}
\end{equation}
and looks as follows in the $\tau_1$, $\tau_2$ space:
\begin{equation}
\begin{gathered}
\begin{tikzpicture}
\coordinate (A) at (0,2);
\coordinate (B) at (0,3);
\coordinate (C) at (1,2);
\coordinate (D) at (2,1);
\coordinate (E) at (3,0);
\coordinate (F) at (2,0);
\coordinate (G) at (1,1);

\fill[Maroon!20] (A) -- (B) -- (C) -- (D) -- (E) -- (F) -- (G) -- cycle;
\draw[thick] (A) -- (B) -- (C) -- (D) -- (E) -- (F) -- (G) -- cycle;

\coordinate (MAB) at ($(A)!0.5!(B)$);
\draw[->,Maroon,thick] (MAB) -- ++(${-0.5*sqrt(2)}*(1,0)$); 

\coordinate (MBC) at ($(B)!0.5!(E)$);
\draw[->,Maroon,thick] (MBC) -- ++(0.5,0.5);

\coordinate (MEF) at ($(E)!0.5!(F)$);
\draw[->,Maroon,thick] (MEF) -- ++(${-0.5*sqrt(2)}*(0,1)$);

\coordinate (MFA) at ($(F)!0.5!(A)$);
\draw[->,Maroon,thick] (MFA) -- ++(-0.5,-0.5);

\draw[->,thick] (-0.5,0) -- (4,0) node[right] {$\tau_1$};
\draw[->,thick] (0,-0.5) -- (0,4) node[above] {$\tau_2$}; 

\node[below left] at (A) {$(0,2)$};
\node[above left] at (B) {$(0,3)$};
\node[below right] at (E) {$(3,0)$};
\node[below left] at (F) {$(2,0)$};

\node[Maroon,left,xshift=-5] at ($(MAB) + (-0.5, 0)$) {$r_1$};
\node[Maroon,above,xshift=20,yshift=20] at ($(MBC) + (-0.2, -0.2)$) {$r_2$};
\node[Maroon,below,yshift=-10] at ($(MEF) + (0, -0.4)$) {$r_3$};
\node[Maroon,left,xshift=-15,yshift=-12] at ($(MFA) + (0.3, -0.3)$) {$r_4$};
\end{tikzpicture}
\end{gathered}
\end{equation}
The outward-pointing normals to the facets of $\mathbf{S}_1$, $i.e.$ the duals \emph{rays}, are given by
\begin{equation}
    r_1 = (-1, 0), \quad 
    r_2 = (1,1), \quad
    r_3 = (0,-1), \quad
    r_4 = (-1,-1)\,,
\end{equation}
representing the possibly divergent directions. To diagnose whether such a ray produces an IR divergence, we have to check (i)  whether it comes from a low-energy configurations, and (ii) whether the direction leads to a divergent power counting of the integrand in $\D=4$ spacetime dimensions.

The former condition (i) can be obtained by considering the corresponding value of the loop momenta, which is given by~\cite{Hannesdottir:2022bmo}
\begin{equation}
\label{eq:mom-interpretation}
    q_e^\mu = \frac{1}{\U} \sum_T p_{T,e}^\mu \prod_{e' \not\in T} \alpha_{e'}\,,
\end{equation}
where $q_e^\mu$ is the momentum flowing through edge $e$, $T$ are the spanning trees of the Feynman diagram, and $p_{T,e}$ is the external momentum that flows through the edge $e$ in the tree $T$. 
In our simple example we find
\begin{equation}
        q_1^{\mu} = \frac{\gamma_1 p_1^{\mu} + \alpha_{2,1}(p_1+p_2)^{\mu} }{\alpha_{1,1}+ \alpha_{2,1} + \gamma_1}, \, \quad 
        q_2^{\mu} = \frac{\gamma_1 p_2^{\mu} + \alpha_{1,1}(p_1+p_2)^{\mu} }{\alpha_{1,1}+ \alpha_{2,1} + \gamma_1}, \, \quad 
        q_3^{\mu} = \frac{ \alpha_{2,1} p_2^{\mu}  -\alpha_{1,1} p_1^{\mu} }{\alpha_{1,1}+ \alpha_{2,1} + \gamma_1} \, .
\end{equation}
So in our one-loop case where $\gamma_1 = 1$, we see that 
\begin{equation}
    (q_1,q_2,q_3) \rightarrow 
    \begin{cases}
        (\frac{p_1 + \alpha_{2,1}(p_1+p_2) }{\alpha_{2,1} + 1},\frac{ p_2 }{\alpha_{2,1} + 1},\frac{ \alpha_{2,1} p_2 }{\alpha_{2,1} + 1}),  &\quad \text{along} \; r_1 \\
        (\frac{\alpha_{2,1}(p_1+p_2) }{\alpha_{1,1} +\alpha_{2,1}},\frac{\alpha_{1,1}(p_1+p_2) }{\alpha_{1,1} +\alpha_{2,1}},\frac{ \alpha_{2,1} p_2 - \alpha_{1,1} p_1}{\alpha_{1,1} +\alpha_{2,1}}),  &\quad \text{along} \; r_2 \\
        (\frac{p_1}{\alpha_{1,1} + 1},\frac{p_2 + \alpha_{1,1}(p_1+p_2) }{\alpha_{1,1} + 1},\frac{ -\alpha_{1,1} p_1 }{\alpha_{1,1} + 1}),  &\quad \text{along} \; r_3 \\
        (p_1,p_2,0),  &\quad \text{along} \; r_4 
    \end{cases}
\end{equation}
So the soft limit corresponds to ray $r_4$ for which we check condition (ii) using the tropical function~\cite{Arkani-Hamed:2022cqe} -- which is simply what we obtain by tropicalizing the integrand in $\tau_1,\tau_2$ space:
\begin{equation}
    \text{Trop}_ {\I_{(1)}} = \tau_1 + \tau_2  -(3 - \D/2) \max(\tau_1+\tau_2,2\tau_1,2\tau_2) - (\D-3) \max(\tau_1,\tau_2,0).
\end{equation}
We then have that the integral is logarithmically divergent along the direction of ray $r_i$, if the tropical function evaluated for $\tau_e = (r_i)_e$ is zero, $i.e.$ $\text{Trop} = 0$. Instead, we have a power divergence  when $\text{Trop} > 0$, and no divergence otherwise. Therefore, taking the ray $r_4$ and plugging it into $ \text{Trop}_{\I_{(1)}}$, with $\D=4$, we get
\begin{equation}
    \text{Trop}(r_4) = 0 \,,  
\end{equation}
which implies an IR divergence in $\D=4$.
In the projective slice where $e^{\tau_3} \equiv \gamma_1=1$, it corresponds to the case in which $\tau_1$ and $\tau_2$ go to negative infinity (and therefore $\alpha_{1,1}/\alpha_{2,1}$ go to zero) at the same rate. We will write this scaling as follows
\begin{equation}
    r = (\alpha_{1,1},\alpha_{2,1},\gamma_1) = (-1,-1,0) \,.
\end{equation}
This means that we scale each $\alpha_e$ as $\alpha_e \to \lambda^{-r_e} \alpha_e $ and take $\lambda \to 0$ at the end of the computation.
Note that the $\text{Trop}$ function is imply equal to minus the power in $\lambda$ of the leading term of the integrand expanded in series along the ray $r$:
\begin{equation}
   \left[ 
   \frac{\d \alpha_{1,1} \d \alpha_{2,1}}{\left( s \alpha_{1,1} \alpha_{2,1} - m^2 (\alpha_{1,1}+\alpha_{2,1})^2 \right)^{3-\D/2} ( \alpha_{1,1} + \alpha_{2,1} + 1)^{\D-3}} 
   \right]_{\alpha_e \to \lambda^{-r_e} \alpha_e}= \mathcal{O}(\lambda^{-\text{Trop}(r)}) .
\end{equation} 
Due to $\GL(1)$ covariance, we can freely add any constant to all components of the vector $r$. Anticipating later examples, we will rewrite $r$ as
\begin{equation}
    r = (\alpha_{1,1},\alpha_{2,1},\gamma_1)= (1,1,2) \,,
\end{equation}
$i.e.$, taking all components to be positive, which allows for an intuitive picture of all the parameters scaling to infinity, although $\gamma_1$ scales twice as fast as $\alpha_{1,1}$ and $\alpha_{2,1}$. In loop-momentum space, using~\cref{eq:mom-interpretation}, this precisely corresponds to the limit in which photon $\gamma_1$ goes soft.
Applying this scaling to~\cref{eq:one-loop-Symanzik}, the singular configuration of the integral is simply the leading term in the expansion,
\begin{equation}
    \label{eq:one-loop-expanded}
    \I_{(1)} = \Gamma \Big(3 - \frac{\D}{2} \Big)
    \int_{\mathcal{D}_1} \frac{\d \alpha_{1,1} \d \alpha_{2,1} \d \gamma_1}{\GL (1)} \frac{1}{(\mathcal{F}_{(1)})^{3- \frac{\D}{2}} \gamma_1^{\D-3}}  + \text{(IR finite)} \,,
\end{equation}
with $\mathcal{D}_1$ is some domain in which the integrand approximation is valid.
Before we go on, it is worth commenting that the dependence of $\gamma_1$ has completely factored from the integrals in $\alpha_{1,1}$ and $\alpha_{2,1}$ and so the corresponding integral is scaleless: it vanishes in dimensional regularisation if we take the domain of integration $\mathcal{D}_1$ to be the entire $\mathbb{R}_+^3$. In particular, the IR divergence of \cref{eq:one-loop-expanded} along the ray $(1,1,2)$ is exactly cancelled by an opposite UV divergence along the opposite ray $(-1,-1,-2)$, which appeared after the integrand approximation along $r_4$ was taken.
Because in the next sections we will be mainly concerned with the IR properties of integrands, rather than integrals, from now on we assume to always be able to distinguish IR and UV divergences and systematically ignore the UV ones. We will then come back to spurious UV divergences and resolve this problem in \cref{sec:spurious_UV_and_RGE}.

Since in the coming sections we will perform our analysis in Schwinger representation, let us integrate back in the overall scale and remove the projective invariance.
We therefore define the (Schwinger) IR integrand as
\begin{equation} \label{eq:S_1}
    \d \overline{S}_{(1)}(\alpha) =  i^{3-\D/2} \frac{ \d\alpha_{1,1}\d\alpha_{2,1} \d\gamma_{1}}{\gamma_1^{\D/2}} \exp\left(\frac{i\F_{(1)}}{\gamma_1} \right).
\end{equation}
where the barred notation reminds us that in order to obtain the full integrand for a given theory we will need to add overall normalisation $\mathcal{C}$ containing the correct power of the coupling and factors of $\pi$ as well as a numerator part $\mathcal{N}$, which will be introduced in the context of parametric integrals in \cref{sec:numerators}. So in general the theory-specific one-loop soft integrand will have the form 
\begin{equation}
    \d S_{(1)}(\alpha) =  \mathcal{C} \times \d \overline{S}_{(1)}(\alpha) \times \mathcal{N}(\alpha) . 
\end{equation}
For example, in the case of QED it would be
\begin{equation}\label{eq:S_1_recall}
    \d S^{\text{QED}}_{(1)}(\beta_1,\beta_2,\gamma)  =  
    -i^{3-\D/2} \frac{\overline{\alpha}(\mu)}{4\pi}  
     \frac{(2p_1)\cdot (2p_2)}{\mu^2} \frac{\d\gamma \d\beta_{1}\d\beta_{2}}{\gamma^{\D/2}} \exp\left[ i \, \frac{s \beta_{1} \beta_{2} - m^2 (\beta_{1}+\beta_{2})^2 }{\mu^2 \, \gamma} \right],
\end{equation}
where we have restored the dim.reg.~scale $\mu$ and we used the modified coupling 
\begin{equation}
    \overline{\alpha}(\mu) \equiv  \mu^{2\epsilon}  e^2/(4\pi)^{1-\epsilon} ,
\end{equation}
with $e$ the bare electric charge.
\subsubsection*{Divergent rays of the two-loop ladder diagrams}

When detecting divergent directions more generally, one does not need to make a change of variables to the exponential variables $\tau_e$. We can directly consider the Minkowski sum $\mathbf{S} = \mathbf{U} \oplus \mathbf{F}$, where $\mathbf{U}$ and $\mathbf{F}$ are the Newton polytopes of the Symanzik polynomials. We can directly infer the divergent rays using standard algebraic geometry software such as {\tt polymake}, see attached {\tt Mathematica} notebook for an example code. 

\begin{figure}[!t]
    \centering
    \includegraphics[width=0.95\textwidth]{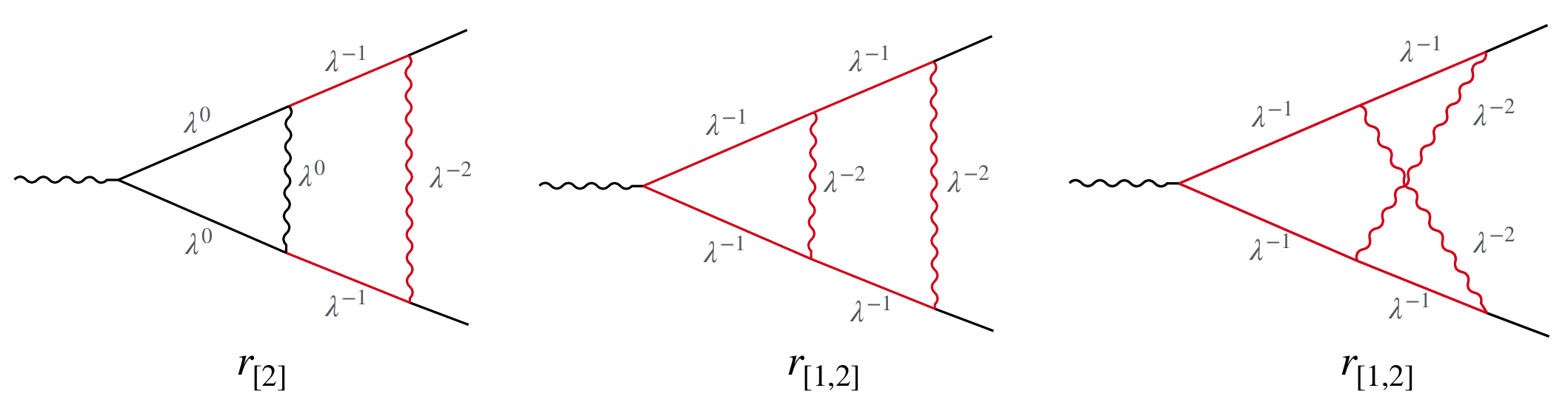}
    \caption{Explicit scalings for the rays of  the planar and non-planar two-loop ladders as given in \cref{eq:scalings2loops}. Black propagators are unscaled while red ones are scaled (soft/on-shell).}
    \label{fig:scalings2loops}
\end{figure}
At two loops, we start by focusing on the planar and non-planar ladder diagrams, respectively presented on the left and right of \cref{fig:2loop-diagrams}, for which we find the following divergent scalings:
\begin{equation}
\begin{aligned}
\text{Planar Ladder}:\quad  &r_{[2]}=(\alpha_{1,1},\alpha_{1,2},\alpha_{2,1},\alpha_{2,2},\gamma_1,\gamma_2) = (0,1,0,1,0,2);\\
&r_{[1,2]}=(\alpha_{1,1},\alpha_{1,2},\alpha_{2,1},\alpha_{2,2},\gamma_1,\gamma_2) = (1,1,1,1,2,2).\\
\text{Non-planar Ladder}:\quad  &r_{[1,2]}=(\tilde{\alpha}_{1,1},\tilde{\alpha}_{1,2},\tilde{\alpha}_{2,1},\tilde{\alpha}_{2,2},\gamma_1,\gamma_2) = (1,1,1,1,2,2).
\end{aligned}
\label{eq:scalings2loops}
\end{equation}
Note that we find two different divergent scalings for the planar ladder -- one corresponding to the case where \emph{only} the outer photon goes soft, $r_{[2]}$, and the other where both photons go soft, $r_{[1,2]}$ -- while for the non-planar ladder we only have a divergence in the second scaling. In both situations, the photons scale to infinity twice as fast as the Schwinger parameters associated with the hard jets $\alpha_{j,k}$ (see \cref{fig:scalings2loops}). As we will see in detail below, the most divergent part of the two-loop integrals will come from simultaneously scaling the planar ladder in $r_{[2]}$ and $r_{[1,2]}$, leading to the ${1}/{\epsilon^2}$ IR pole, while the non-planar ladder, only contributes a $1/\epsilon$ pole to the amplitude.

\subsection{Factorisation of the two-loop integrand in Schwinger space}
\label{sec:fact2loop}

Let us now study the soft expansion of each integrand in the respective scalings given in \cref{eq:scalings2loops}, and check how in both cases we find that the Symanzik polynomials $\F$ and $\U$ exhibit the factorisation properties from~\cref{eq:conjecture}. In doing so, we use the following recursive properties of the $\U$ and $\F$ polynomials: we can pick any edge $e$ of the graph $G$ and write 
\begin{equation}
\label{eq:recursive_U_and_F}
\begin{aligned}
    \U(G) & = \U(G / e) + \alpha_e \U(G - e), \\
    \F_0(G) & = \F_0(G / e) + \alpha_e  \F_0(G - e) \,,
\end{aligned}
\end{equation}
where $\F_0 = \F +  (\sum_e m_e^2 \alpha_e) \U$ is defined as the part of the $\F$ polynomial independent of the internal masses, $G / e$ is the diagram $G$ with edge $e$ contracted, while $G - e$ is $G$ with $e$ deleted (or cut).
These properties are exact and follow from the definitions of the Symanzik polynomials. 

\paragraph{Planar ladder, scaling $r_{[2]}$:}We begin by analysing the contributions we get from the planar ladder in the scaling where only the outer photon becomes soft, given by $r_{[2]}$. Using the recursive structure of $\mathcal{U}$, we can derive that in the $r_{[2]}$ scaling it behaves as
\begin{equation}
\U = \U
\left( 
\begin{gathered}
    \begin{tikzpicture}[line width=0.7,scale=0.35,baseline={([yshift=0.0ex]current bounding box.center)}]
        \coordinate (a) at (-0.5,0);
        \coordinate (b) at (1,0);
        \coordinate (c) at (2,0);
        \coordinate (d) at (2,1);
        \coordinate (e) at (2,-1);
        \coordinate (f) at (3,0);
        \filldraw[color=black,fill=white] (c) circle[radius=1];
        \draw[photon] (d) -- (e);
        \draw[] (f) --++ (35:1);
        \draw[] (f) --++ (-35:1);

        \filldraw[color=black,fill=gray] (b) circle[radius=0.2];
    \end{tikzpicture}
\end{gathered}
\right) + \gamma_2 \U \left( \begin{gathered}
    \begin{tikzpicture}[line width=0.7,scale=0.35,photon/.style={decorate, decoration={snake, amplitude=1pt, segment length=6pt}
	}]
        \coordinate (a) at (-0.5,0);
        \coordinate (b) at (1,0);
        \coordinate (d) at (4,1.5);
        \coordinate (e) at (4,-1.5);
        \path (b) -- (d) coordinate[midway] (midbd);
        \path (b) -- (e) coordinate[midway] (midbe);
        \path (b) -- (d) coordinate[pos=0.7] (bd_point);
        \path (b) -- (e) coordinate[pos=0.7] (be_point);
        \draw[] (b) -- (d);
        \draw[] (b) -- (e);
        \draw[photon] (midbd) -- (midbe);
        \draw[photon] (bd_point) --++ (-35:1);
        \draw[photon] (be_point) --++ (35:1);
        \filldraw[color=black,fill=gray] (b) circle[radius=0.2];
    \end{tikzpicture}
\end{gathered} \right) \xrightarrow[r_{[2]}]{} (\alpha_{1,2}+\alpha_{2,2}) \U_H + \U_H \U_S \xrightarrow[r_{[2]}]{} \U_H \U_S  + \mathcal{O}(\lambda^{-1})
\end{equation}
where $\U_H = (\alpha_{1,1}+\alpha_{2,1}+\gamma_1)$ corresponds to the inner hard loop, and  $\U_S =  \gamma_2$, which is just the leading term of the $\U$ polynomial of the outer loop in the scaling $r_{[2]}$. Similarly, using the recursive features of $\mathcal{F}_0$, we get that in the $r_{[2]}$ scaling it behaves as follows 
\begin{equation}
\begin{aligned}
\F_0 &= \F_0
\left( 
\begin{gathered}
    \begin{tikzpicture}[line width=0.7,scale=0.35,baseline={([yshift=0.0ex]current bounding box.center)}]
        \coordinate (a) at (-0.5,0);
        \coordinate (b) at (1,0);
        \coordinate (c) at (2,0);
        \coordinate (d) at (2,1);
        \coordinate (e) at (2,-1);
        \coordinate (f) at (3,0);
        \filldraw[color=black,fill=white] (c) circle[radius=1];
        \draw[photon] (d) -- (e);
        \draw[] (f) --++ (35:1);
        \draw[] (f) --++ (-35:1);
        \filldraw[color=black,fill=gray] (b) circle[radius=0.2];
    \end{tikzpicture}
\end{gathered}
\right) + \gamma_2 \F_0 \left( \begin{gathered}
    \begin{tikzpicture}[line width=0.7,scale=0.35,photon/.style={decorate, decoration={snake, amplitude=1pt, segment length=6pt}
	}]
        \coordinate (a) at (-0.5,0);
        \coordinate (b) at (1,0);
        \coordinate (d) at (4,1.5);
        \coordinate (e) at (4,-1.5);
        \path (b) -- (d) coordinate[midway] (midbd);
        \path (b) -- (e) coordinate[midway] (midbe);
        \path (b) -- (d) coordinate[pos=0.7] (bd_point);
        \path (b) -- (e) coordinate[pos=0.7] (be_point);
        \draw[] (b) -- (d);
        \draw[] (b) -- (e);
        \draw[photon] (midbd) -- (midbe);
        \draw[photon] (bd_point) --++ (-35:1);
        \draw[photon] (be_point) --++ (35:1);
        \filldraw[color=black,fill=gray] (b) circle[radius=0.2];
    \end{tikzpicture}
\end{gathered} \right) \\
&\xrightarrow[r_{[2]}]{} s \alpha_{1,2} \alpha_{2,2} \U_H  + \gamma_2\left[ \F_0 \left( \begin{gathered}
    \begin{tikzpicture}[line width=0.7,scale=0.35,photon/.style={decorate, decoration={snake, amplitude=1pt, segment length=6pt}
	}]
        \coordinate (a) at (-0.5,0);
        \coordinate (b) at (1,0);
        \coordinate (d) at (4,1.5);
        \coordinate (e) at (4,-1.5);
        \path (b) -- (d) coordinate[midway] (midbd);
        \path (b) -- (e) coordinate[midway] (midbe);
        \path (b) -- (d) coordinate[pos=0.7] (bd_point);
        \path (b) -- (e) coordinate[pos=0.7] (be_point);
        \draw[] (b) -- (d);
        \draw[] (b) -- (e);
        \draw[photon] (midbd) -- (midbe);
        \draw[photon] (midbd) --++ (-35:1);
        \draw[photon] (be_point) --++ (35:1);
        \filldraw[color=black,fill=gray] (b) circle[radius=0.2];
    \end{tikzpicture}
\end{gathered} \right) + \alpha_{1,2} \F_0 \left( \begin{gathered}
    \begin{tikzpicture}[line width=0.7,scale=0.35,photon/.style={decorate, decoration={snake, amplitude=1pt, segment length=6pt}
	}]
        \coordinate (a) at (-0.5,0);
        \coordinate (b) at (1,0);
        \coordinate (d) at (4,1.5);
        \coordinate (dp) at (3,1.2);
        \coordinate (dpp) at (5,2);
        \coordinate (e) at (4,-1.5);
        \path (b) -- (d) coordinate[midway] (midbd);
        \path (b) -- (e) coordinate[midway] (midbe);
         \path (dpp) -- (d) coordinate[midway] (middpp);
        \path (b) -- (d) coordinate[pos=0.7] (bd_point);
        \path (b) -- (e) coordinate[pos=0.7] (be_point);
        \draw[] (b) -- (dp);
        \draw[] (d) -- (dpp);
        \draw[] (b) -- (e);
        \draw[photon] (midbd) -- (midbe);
       \draw[photon] (middpp) --++ (-35:1);
        \draw[photon] (be_point) --++ (35:1);
        \filldraw[color=black,fill=gray] (b) circle[radius=0.2];
    \end{tikzpicture}
\end{gathered} \right) \right] \\
&\xrightarrow[r_{[2]}]{}  s \alpha_{1,2} \alpha_{2,2} \U_H  + \gamma_2\left[ \F_0 \left( \begin{gathered}
    \begin{tikzpicture}[line width=0.7,scale=0.35,photon/.style={decorate, decoration={snake, amplitude=1pt, segment length=6pt}
	}]
        \coordinate (a) at (-0.5,0);
        \coordinate (b) at (1,0);
        \coordinate (d) at (4,1.5);
        \coordinate (e) at (4,-1.5);
        \path (b) -- (d) coordinate[midway] (midbd);
        \path (b) -- (e) coordinate[midway] (midbe);
        \path (b) -- (d) coordinate[pos=0.7] (bd_point);
        \path (b) -- (e) coordinate[pos=0.7] (be_point);
        \draw[] (b) -- (d);
        \draw[] (b) -- (e);
        \draw[photon] (midbd) -- (midbe);
        \draw[photon] (midbd) --++ (-35:1);
        \draw[photon] (midbe) --++ (35:1);
        \filldraw[color=black,fill=gray] (b) circle[radius=0.2];
    \end{tikzpicture}
\end{gathered} \right)+ \alpha_{2,2} \F_0 \left( \begin{gathered}
    \begin{tikzpicture}[line width=0.7,scale=0.35,photon/.style={decorate, decoration={snake, amplitude=1pt, segment length=6pt}
	}]
        \coordinate (a) at (-0.5,0);
        \coordinate (b) at (1,0);
        \coordinate (d) at (4,1.5);
        \coordinate (e) at (4,-1.5);
        \coordinate (ep) at (3,-1.2);
        \coordinate (epp) at (5,-2);
        \path (b) -- (d) coordinate[midway] (midbd);
        \path (b) -- (e) coordinate[midway] (midbe);
        \path (b) -- (d) coordinate[pos=0.7] (bd_point);
        \path (b) -- (e) coordinate[pos=0.7] (be_point);
           \path (e) -- (epp) coordinate[midway] (midepp);
        \draw[] (b) -- (d);
        \draw[] (b) -- (ep);
         \draw[] (e) -- (epp);
        \draw[photon] (midbd) -- (midbe);
        \draw[photon] (bd_point) --++ (-35:1);
       \draw[photon] (midepp) --++ (35:1);
       \filldraw[color=black,fill=gray] (b) circle[radius=0.2];
    \end{tikzpicture}
\end{gathered} \right)+ \alpha_{1,2} \F_0 \left( \begin{gathered}
    \begin{tikzpicture}[line width=0.7,scale=0.35,photon/.style={decorate, decoration={snake, amplitude=1pt, segment length=6pt}
	}]
        \coordinate (a) at (-0.5,0);
        \coordinate (b) at (1,0);
        \coordinate (d) at (4,1.5);
        \coordinate (dp) at (3,1.2);
        \coordinate (dpp) at (5,2);
        \coordinate (e) at (4,-1.5);
        \path (b) -- (d) coordinate[midway] (midbd);
        \path (b) -- (e) coordinate[midway] (midbe);
         \path (dpp) -- (d) coordinate[midway] (middpp);
        \path (b) -- (d) coordinate[pos=0.7] (bd_point);
        \path (b) -- (e) coordinate[pos=0.7] (be_point);
        \draw[] (b) -- (dp);
        \draw[] (d) -- (dpp);
        \draw[] (b) -- (e);
        \draw[photon] (midbd) -- (midbe);
       \draw[photon] (middpp) --++ (-35:1);
        \draw[photon] (be_point) --++ (35:1);
        \filldraw[color=black,fill=gray] (b) circle[radius=0.2];
    \end{tikzpicture}
\end{gathered} \right) \right] \\
&\xrightarrow[r_{[2]}]{}   s \alpha_{1,2} \alpha_{2,2} \U_H + m^2 \gamma_2 \left(\alpha_{1,2}+\alpha_{2,2}\right)\U_H + \gamma_2 \F_{0,H}+ \mathcal{O}(\lambda^{-1}),
\end{aligned}
\end{equation}
where $\F_{0,H}$ stands for the one-loop $\F_0$ corresponding to the inner hard one-loop. Therefore we have that $\F$ goes as
\begin{equation}
\begin{aligned}
    \F &\xrightarrow[r_{[2]}]{}  s \alpha_{1,2} \alpha_{2,2} \U_H + m^2 \gamma_2 \left(\alpha_{1,2}+\alpha_{2,2}\right)\U_H + \gamma_2 \F_{0,H}  \\
    & \quad \quad  \quad \quad - m^2 \left(\alpha_{1,1} + \alpha_{1,2} + \alpha_{2,1} + \alpha_{2,2}\right)\left[ \left(\alpha_{1,2}+\alpha_{2,2}\right)\U_H +\U_H\U_S\right] + \mathcal{O}(\lambda^{-1})\\
    &\xrightarrow[r_{[2]}]{}  s \alpha_{1,2} \alpha_{2,2} \U_H + \gamma_2 \F_{0,H} -m^2 \left(\alpha_{1,1} + \alpha_{2,1} \right) \U_H\U_S- m^2  \left(\alpha_{1,2}  + \alpha_{2,2}\right)^2\U_H + \mathcal{O}(\lambda^{-1}) \\
   & \xrightarrow[r_{[2]}]{}   \U_H \left[ s \alpha_{1,2} \alpha_{2,2} -m^2  \left(\alpha_{1,2}+\alpha_{2,2}\right)^2 \right]+  \U_S \left[ \F_{0,H} -m^2  \left(\alpha_{1,1}  + \alpha_{2,1} \right)\U_H \right]+ \mathcal{O}(\lambda^{-1})\\
    &  \xrightarrow[r_{[2]}]{} \U_H \F_S + \U_S \F_H + \mathcal{O}(\lambda^{-1}),
\end{aligned}
\end{equation}
where $\F_H$ is now the full second Symanzik polynomial for the inner hard loop, and $\F_S$ that for the outer loop, precisely as expected.
Therefore at leading order, the two-loop planar integrand satisfies 
\begin{equation}
    \d\I_2^{\text{Pl}} \to \d \I_1^{(1)} \times  \d S_1^{(2)},
\end{equation}
where the superscripts $(i)$ stand for the photon number entering in each factor as given by \cref{fig:2loop-diagrams}. 

\paragraph{Planar and non-planar ladders, $r_{[1,2]}$:} Let's now look at scaling $r_{[1,2]}$ where both photons go soft. In this case both diagrams are divergent, but we start by analysing the planar one. For the $\U$ polynomial we find the following expansion
\begin{equation}
\begin{aligned}
\U &= \U
\left( 
\begin{gathered}
    \begin{tikzpicture}[line width=0.7,scale=0.35,baseline={([yshift=0.0ex]current bounding box.center)}]
        \coordinate (a) at (-0.5,0);
        \coordinate (b) at (1,0);
        \coordinate (c) at (2,0);
        \coordinate (d) at (2,1);
        \coordinate (e) at (2,-1);
        \coordinate (f) at (3,0);
        \draw[photon] (a) -- (b);
        \filldraw[color=black,fill=white] (c) circle[radius=1];
        \draw[] (f) --++ (35:1);
        \draw[] (f) --++ (-35:1);
        \filldraw[color=black,fill=gray] (b) circle[radius=0.2];
    \end{tikzpicture}
\end{gathered}
\right) + \gamma_2 \U \left( \begin{gathered}
    \begin{tikzpicture}[line width=0.7,scale=0.3,photon/.style={decorate, decoration={snake, amplitude=1pt, segment length=6pt}
	}]
        \coordinate (a) at (-0.5,0);
        \coordinate (b) at (1,0);
        \coordinate (d) at (4,1.5);
        \coordinate (e) at (4,-1.5);
        \path (b) -- (d) coordinate[midway] (midbd);
        \path (b) -- (e) coordinate[midway] (midbe);
        \path (b) -- (d) coordinate[pos=0.7] (bd_point);
        \path (b) -- (e) coordinate[pos=0.7] (be_point);
        \draw[] (b) -- (d);
        \draw[] (b) -- (e);
        \draw[photon] (midbd) -- (midbe);
        \draw[photon] (bd_point) --++ (-35:1);
        \draw[photon] (be_point) --++ (35:1);
        \filldraw[color=black,fill=gray] (b) circle[radius=0.2];
    \end{tikzpicture}
\end{gathered} \right) =  \U
\left( 
\begin{gathered}
    \begin{tikzpicture}[line width=0.7,scale=0.35,baseline={([yshift=0.0ex]current bounding box.center)}]
        \coordinate (a) at (-0.5,0);
        \coordinate (b) at (1,0);
        \coordinate (c) at (2,0);
        \coordinate (d) at (2,1);
        \coordinate (e) at (2,-1);
        \coordinate (f) at (3,0);
        \filldraw[color=black,fill=white] (c) circle[radius=1];
        \draw[photon] (d) -- (e);
        \draw[] (f) --++ (35:1);
        \draw[] (f) --++ (-35:1);
        \filldraw[color=black,fill=gray] (b) circle[radius=0.2];
    \end{tikzpicture}
\end{gathered}
\right) + \gamma_2 \U^{(1)} \\
&= \U
\left( 
\begin{gathered}
    \begin{tikzpicture}[line width=0.7,scale=0.35,baseline={([yshift=-0.5ex]current bounding box.center)}]
        \coordinate (a) at (-0.5,0);
        \coordinate (b) at (1,0);
        \coordinate (c) at (1,0);
        \coordinate (cp) at (2,0);
        \coordinate (d) at (2,1);
        \coordinate (e) at (2,-1);
        \coordinate (f) at (2.5,0);
        \filldraw[color=black,fill=white] (c) circle[radius=0.5];
        \filldraw[color=black,fill=white] (cp) circle[radius=0.5];
        \draw[] (f) --++ (35:1);
        \draw[] (f) --++ (-35:1);
        \filldraw[color=black,fill=gray] (0.5,0) circle[radius=0.2];
    \end{tikzpicture}
\end{gathered}
\right) + \gamma_1 \U
\left( 
\begin{gathered}
    \begin{tikzpicture}[line width=0.7,scale=0.35,baseline={([yshift=0.0ex]current bounding box.center)}]
        \coordinate (a) at (-0.5,0);
        \coordinate (b) at (1,0);
        \coordinate (c) at (2,0);
        \coordinate (d) at (2,1);
        \coordinate (dp) at (2,0.3);
        \coordinate (e) at (2,-1);
        \coordinate (ep) at (2,-0.3);
        \coordinate (f) at (3,0);
        \filldraw[color=black,fill=white] (c) circle[radius=1];
        \draw[photon] (d) -- (dp);
        \draw[photon] (e) -- (ep);
        \draw[] (f) --++ (35:1);
        \draw[] (f) --++ (-35:1);
        \filldraw[color=black,fill=gray] (b) circle[radius=0.2];
        
    \end{tikzpicture}
\end{gathered}
\right) + \gamma_2 \U^{(1)} \\
& \xrightarrow[r_{[1,2]}]{} \gamma_1 \gamma_2+ \gamma_1 \left(\alpha_{1,1}+ \alpha_{2,1}+\alpha_{1,2}+\alpha_{2,2}\right)+ \gamma_2 \left( \alpha_{1,1} + \alpha_{2,1} \right) + \mathcal{O}(\lambda^2),\\
&= \underbrace{\U_S^{(1)}\U_S^{(2)}}_{\mathcal{O}(\lambda^{-4})}+ \underbrace{\U_S^{(1)}\left(\beta_{1,2}+\beta_{2,2}\right) +\U_S^{(2)}\left(\beta_{1,1}+\beta_{2,1}\right)}_{\mathcal{O}(\lambda^{-3})}+ \mathcal{O}(\lambda^{-2}),
\end{aligned}
\label{eq:U2loopPl}
\end{equation}
where $\U^{(1)}$ stands for the $\U$ polynomial of the inner loop, involving photon $\gamma_1$, $i.e.$ $\U^{(1)} = \alpha_{1,1} + \alpha_{2,1} + \gamma_1$, and $\U_S^{(i)}$ are, respectively, the soft limits of the inner and outer loops, $\U_S^{(i)} = \gamma_i$. 
In addition, in the last line above, we see the first appearance of the worldline variables ($\beta$) associated to each jet $J$, where $\beta_{J,k} = \sum_{i=1}^k \alpha_{J,i}$, such that 
\begin{equation}
    \beta_{1,1}=\alpha_{1,1}, \quad \beta_{2,1}=\alpha_{2,1},\quad \beta_{1,2}=\alpha_{1,1}+\alpha_{1,2}, \quad \beta_{2,2}=\alpha_{2,1}+\alpha_{2,2},
    \label{eq:betas2}
\end{equation}
which can be simply interpreted as distances between the hard vertex and the vertices where the soft photons are anchored along the hard lines. 
Note that in order to prove factorisation of the $\U$ polynomial we could have kept only the $\mathcal{O}(\lambda^{-4})$ term in \cref{eq:U2loopPl}, but we also retained the next order because we will need it to show the factorisation of $\F$ below.
Indeed, for $\F_0$ we find:
\begin{equation}
\begin{aligned}
    \F_0 &= \F_0\left( 
\begin{gathered}
    \begin{tikzpicture}[line width=0.7,scale=0.35,baseline={([yshift=0.0ex]current bounding box.center)}]
        \coordinate (a) at (-0.5,0);
        \coordinate (b) at (1,0);
        \coordinate (c) at (2,0);
        \coordinate (d) at (2,1);
        \coordinate (e) at (2,-1);
        \coordinate (f) at (3,0);
        \filldraw[color=black,fill=white] (c) circle[radius=1];
        \draw[photon] (d) -- (e);
        \draw[] (f) --++ (35:1);
        \draw[] (f) --++ (-35:1);
        \filldraw[color=black,fill=gray] (b) circle[radius=0.2];
    \end{tikzpicture}
\end{gathered}
\right) + \gamma_2 \F_0 \left( \begin{gathered}
    \begin{tikzpicture}[line width=0.7,scale=0.35]
        \coordinate (a) at (-0.5,0);
        \coordinate (b) at (1,0);
        \coordinate (d) at (4,1.5);
        \coordinate (e) at (4,-1.5);
        \path (b) -- (d) coordinate[midway] (midbd);
        \path (b) -- (e) coordinate[midway] (midbe);
        \path (b) -- (d) coordinate[pos=0.7] (bd_point);
        \path (b) -- (e) coordinate[pos=0.7] (be_point);
        \draw[] (b) -- (d);
        \draw[] (b) -- (e);
        \draw[photon] (midbd) -- (midbe);
        \draw[photon] (bd_point) --++ (-35:1);
        \draw[photon] (be_point) --++ (35:1);
        \filldraw[color=black,fill=gray] (b) circle[radius=0.2];
    \end{tikzpicture}
\end{gathered} \right) \\
 &=\F_0
\left( 
\begin{gathered}
    \begin{tikzpicture}[line width=0.7,scale=0.35,baseline={([yshift=-0.5ex]current bounding box.center)}]
        \coordinate (a) at (-0.5,0);
        \coordinate (b) at (1,0);
        \coordinate (c) at (1,0);
        \coordinate (cp) at (2,0);
        \coordinate (d) at (2,1);
        \coordinate (e) at (2,-1);
        \coordinate (f) at (2.5,0);
        \filldraw[color=black,fill=white] (c) circle[radius=0.5];
        \filldraw[color=black,fill=white] (cp) circle[radius=0.5];
        \draw[] (f) --++ (35:1);
        \draw[] (f) --++ (-35:1);
        \filldraw[color=black,fill=gray] (0.5,0) circle[radius=0.2];
    \end{tikzpicture}
\end{gathered}
\right) + \gamma_1 \F_0
\left( 
\begin{gathered}
    \begin{tikzpicture}[line width=0.7,scale=0.35,baseline={([yshift=0.0ex]current bounding box.center)}]
        \coordinate (a) at (-0.5,0);
        \coordinate (b) at (1,0);
        \coordinate (c) at (2,0);
        \coordinate (d) at (2,1);
        \coordinate (dp) at (2,0.3);
        \coordinate (e) at (2,-1);
        \coordinate (ep) at (2,-0.3);
        \coordinate (f) at (3,0);
        \filldraw[color=black,fill=white] (c) circle[radius=1];
        \draw[photon] (d) -- (dp);
        \draw[photon] (e) -- (ep);
        \draw[] (f) --++ (35:1);
        \draw[] (f) --++ (-35:1);
        \filldraw[color=black,fill=gray] (b) circle[radius=0.2];
    \end{tikzpicture}
\end{gathered}
\right) + \gamma_2 \F_0 \left( \begin{gathered}
    \begin{tikzpicture}[line width=0.7,scale=0.35,photon/.style={decorate, decoration={snake, amplitude=1pt, segment length=6pt}
	}]
        \coordinate (a) at (-0.5,0);
        \coordinate (b) at (1,0);
        \coordinate (d) at (4,1.5);
        \coordinate (e) at (4,-1.5);
        \path (b) -- (d) coordinate[midway] (midbd);
        \path (b) -- (e) coordinate[midway] (midbe);
        \path (b) -- (d) coordinate[pos=0.7] (bd_point);
        \path (b) -- (e) coordinate[pos=0.7] (be_point);
        \draw[] (b) -- (d);
        \draw[] (b) -- (e);
        \draw[photon] (midbd) -- (midbe);
        \draw[photon] (bd_point) --++ (-35:1);
        \draw[photon] (be_point) --++ (35:1);
        \filldraw[color=black,fill=gray] (b) circle[radius=0.2];
    \end{tikzpicture}
\end{gathered} \right),
\end{aligned}
\end{equation}
with 
\begin{equation}
\begin{aligned}
   & \F_0
\left( 
\begin{gathered}
    \begin{tikzpicture}[line width=0.7,scale=0.35,baseline={([yshift=-0.5ex]current bounding box.center)}]
        \coordinate (a) at (-0.5,0);
        \coordinate (b) at (1,0);
        \coordinate (c) at (1,0);
        \coordinate (cp) at (2,0);
        \coordinate (d) at (2,1);
        \coordinate (e) at (2,-1);
        \coordinate (f) at (2.5,0);
        \filldraw[color=black,fill=white] (c) circle[radius=0.5];
        \filldraw[color=black,fill=white] (cp) circle[radius=0.5];
        \draw[] (f) --++ (35:1);
        \draw[] (f) --++ (-35:1);
        \filldraw[color=black,fill=gray] (0.5,0) circle[radius=0.2];
    \end{tikzpicture}
\end{gathered}
\right) =s\alpha_{1,1} \alpha_{2,2} (\alpha_{1,2}+\alpha_{2,1})+ s \alpha_{1,2} \alpha_{2,1}(\alpha_{1,1} + \alpha_{2,2}) \sim \mathcal{O}(\lambda^{-3}) ,\\
&\F_0
\left( 
\begin{gathered}
    \begin{tikzpicture}[line width=0.7,scale=0.35,baseline={([yshift=0.0ex]current bounding box.center)}]
        \coordinate (a) at (-0.5,0);
        \coordinate (b) at (1,0);
        \coordinate (c) at (2,0);
        \coordinate (d) at (2,1);
        \coordinate (dp) at (2,0.3);
        \coordinate (e) at (2,-1);
        \coordinate (ep) at (2,-0.3);
        \coordinate (f) at (3,0);
        \filldraw[color=black,fill=white] (c) circle[radius=1];
        \draw[photon] (d) -- (dp);
        \draw[photon] (e) -- (ep);
        \draw[] (f) --++ (35:1);
        \draw[] (f) --++ (-35:1);
        \filldraw[color=black,fill=gray] (b) circle[radius=0.2];
    \end{tikzpicture}
\end{gathered}
\right) = s(\alpha_{1,1}+\alpha_{1,2})(\alpha_{2,1}+\alpha_{2,2}) \sim \mathcal{O}(\lambda^{-2}),\\
&\F_0 \left( \begin{gathered}
    \begin{tikzpicture}[line width=0.7,scale=0.3,photon/.style={decorate, decoration={snake, amplitude=1pt, segment length=6pt}
	}]
        \coordinate (a) at (-0.5,0);
        \coordinate (b) at (1,0);
        \coordinate (d) at (4,1.5);
        \coordinate (e) at (4,-1.5);
        \path (b) -- (d) coordinate[midway] (midbd);
        \path (b) -- (e) coordinate[midway] (midbe);
        \path (b) -- (d) coordinate[pos=0.7] (bd_point);
        \path (b) -- (e) coordinate[pos=0.7] (be_point);
        \draw[] (b) -- (d);
        \draw[] (b) -- (e);
        \draw[photon] (midbd) -- (midbe);
        \draw[photon] (bd_point) --++ (-35:1);
        \draw[photon] (be_point) --++ (35:1);
        \filldraw[color=black,fill=gray] (b) circle[radius=0.2];
    \end{tikzpicture}
\end{gathered} \right) = m^2  \left(\alpha_{1,2}+\alpha_{2,2}\right)\U^{(i)} +  \F_0^{(i)} \sim \mathcal{O}(\lambda^{-3}),
\end{aligned}
\end{equation}
so that we get
\begin{equation}
\begin{aligned}
    \F &= \F_0 - m^2(\alpha_{1,1} +\alpha_{1,2} +\alpha_{2,1} +\alpha_{2,2} ) \left[\U
\left( 
\begin{gathered}
    \begin{tikzpicture}[line width=0.7,scale=0.35,baseline={([yshift=-0.5ex]current bounding box.center)}]
        \coordinate (a) at (-0.5,0);
        \coordinate (b) at (1,0);
        \coordinate (c) at (1,0);
        \coordinate (cp) at (2,0);
        \coordinate (d) at (2,1);
        \coordinate (e) at (2,-1);
        \coordinate (f) at (2.5,0);
        \filldraw[color=black,fill=white] (c) circle[radius=0.5];
        \filldraw[color=black,fill=white] (cp) circle[radius=0.5];
        \draw[] (f) --++ (35:1);
        \draw[] (f) --++ (-35:1);
        \filldraw[color=black,fill=gray] (0.5,0) circle[radius=0.2];
    \end{tikzpicture}
\end{gathered}
\right) + \gamma_1 \U
\left( 
\begin{gathered}
    \begin{tikzpicture}[line width=0.7,scale=0.35,baseline={([yshift=0.0ex]current bounding box.center)}]
        \coordinate (a) at (-0.5,0);
        \coordinate (b) at (1,0);
        \coordinate (c) at (2,0);
        \coordinate (d) at (2,1);
        \coordinate (dp) at (2,0.3);
        \coordinate (e) at (2,-1);
        \coordinate (ep) at (2,-0.3);
        \coordinate (f) at (3,0);
        \filldraw[color=black,fill=white] (c) circle[radius=1];
        \draw[photon] (d) -- (dp);
        \draw[photon] (e) -- (ep);
        \draw[] (f) --++ (35:1);
        \draw[] (f) --++ (-35:1);
        \filldraw[color=black,fill=gray] (b) circle[radius=0.2];
    \end{tikzpicture}
\end{gathered}
\right) + \gamma_2 \U^{(1)} \right]\\
   &\xrightarrow[r_{[1,2]}]{} \gamma_2 \left[ \F_0^{(1)} -m^2(\alpha_{1,1}+\alpha_{2,1})\U^{(1)}\right] +\gamma_1 \left[ s(\alpha_{1,1}+ \alpha_{1,2})(\alpha_{2,1}+\alpha_{2,2}) \right. \\
   & \left. \hspace{7.5cm} -m^2 (\alpha_{1,1}+\alpha_{1,2}+\alpha_{2,1}+\alpha_{2,2})^2 \right] \\
   &\xrightarrow[r_{[1,2]}]{} \U^{(2)}_S \F^{(1)}[\beta_{1,1},\beta_{2,1}] + \U^{(1)}_S \F^{(2)}[\beta_{1,2},\beta_{2,2}] + \mathcal{O}(\lambda^{-3}),
\end{aligned}
\end{equation}
where $\F^{(1)}/\F^{(2)}$ are the second Symanzik polynomials for the inner and outer loops respectively. The Schwinger parameters associated to each fermionic line are replaced by the appropriate  worldline variables in \cref{eq:betas2}.
Since the $\alpha_{J,i}$ are integrated over $\mathbb{R}^+$, we find that the $\beta$'s cover the following domain for the planar diagram (see \cref{fig:2loop_worldline} below):
\begin{equation}
    0<\beta_{1,1}< \beta_{1,2}, \quad  0<\beta_{2,1}< \beta_{2,2} \, .
\label{eq:domainBetaPlanar}
\end{equation}
Now let us look at the non-planar integrand under scaling $r_{[1,2]}$. In this case the integrand is a function of the Schwinger parameters, $\tilde{\alpha}_{J,i}$, but as we show below, when written in terms of worldline ($\beta$) variables, it precisely matches the one of the planar ladder under the soft scaling. 
Let us start with the $\U$ polynomial, which at leading and subleading order gives
\begin{equation}
\begin{aligned}
\U &= \U
\left( 
\begin{gathered}
    \begin{tikzpicture}[line width=0.7,scale=0.35,baseline={([yshift=0.0ex]current bounding box.center)}]
        \coordinate (a) at (-0.5,0);
        \coordinate (b) at (1,0);
        \coordinate (c) at (2,0);
        \coordinate (d) at (2,1);
        \coordinate (e) at (2,-1);
        \coordinate (f) at (3,0);
        \filldraw[color=black,fill=white] (c) circle[radius=1];
        \draw[] (f) --++ (35:1);
        \draw[] (f) --++ (-35:1) coordinate (g);
        \coordinate (midpoint) at ($(f)!0.5!(g)$);
        \draw[photon] (d) to[in=50, out=50] (midpoint);
        \filldraw[color=black,fill=gray] (b) circle[radius=0.2];
    \end{tikzpicture}
\end{gathered}
\right) + \gamma_2 \U \left( \begin{gathered}
    \begin{tikzpicture}[line width=0.7,scale=0.35,photon/.style={decorate, decoration={snake, amplitude=1pt, segment length=6pt}
	}]
        \coordinate (a) at (-0.5,0);
        \coordinate (b) at (1,0);
        \coordinate (d) at (4,1.5);
        \coordinate (e) at (4,-1.5);
        \path (b) -- (d) coordinate[midway] (midbd);
        \path (b) -- (e) coordinate[midway] (midbe);
        \path (b) -- (d) coordinate[pos=0.7] (bd_point);
        \path (b) -- (e) coordinate[pos=0.7] (be_point);
        \draw[] (b) -- (d);
        \draw[] (b) -- (e);
        \draw[photon] (midbd) -- (be_point);
        \draw[photon] (bd_point) --++ (-35:1);
        \draw[photon] (midbe) --++ (35:1);
        \filldraw[color=black,fill=gray] (b) circle[radius=0.2];
    \end{tikzpicture}
\end{gathered} \right) =  \U
\left( 
\begin{gathered}
    \begin{tikzpicture}[line width=0.7,scale=0.35,baseline={([yshift=0.0ex]current bounding box.center)}]
        \coordinate (a) at (-0.5,0);
        \coordinate (b) at (1,0);
        \coordinate (c) at (2,0);
        \coordinate (d) at (2,1);
        \coordinate (e) at (2,-1);
        \coordinate (f) at (3,0);
        \filldraw[color=black,fill=white] (c) circle[radius=1];
        \draw[] (f) --++ (35:1);
        \draw[] (f) --++ (-35:1) coordinate (g);
        \coordinate (midpoint) at ($(f)!0.5!(g)$);
        \draw[photon] (d) to[in=50, out=50] (midpoint);
        \filldraw[color=black,fill=gray] (b) circle[radius=0.2];
    \end{tikzpicture}
\end{gathered}
\right) + \gamma_2 \left(\tilde{\alpha}_{1,1}+\tilde{\alpha}_{2,1}+\tilde{\alpha}_{2,2} +\gamma_1 \right) \\
&\xrightarrow[r_{[1,2]}]{}  \gamma_1 \U
\left( 
\begin{gathered}
    \begin{tikzpicture}[line width=0.7,scale=0.35,baseline={([yshift=0.0ex]current bounding box.center)}]
        \coordinate (a) at (-0.5,0);
        \coordinate (b) at (1,0);
        \coordinate (c) at (2,0);
        \coordinate (d) at (2,1);
        \coordinate (dp) at (2,0.3);
        \coordinate (e) at (2,-1);
        \coordinate (ep) at (2,-0.3);
        \coordinate (f) at (3,0);
        \filldraw[color=black,fill=white] (c) circle[radius=1];
        \draw[photon] (d) -- (dp);
        \draw[] (f) --++ (35:1);
        \draw[] (f) --++ (-35:1) coordinate (g);
        \coordinate (midpoint) at ($(f)!0.5!(g)$);
        \draw[photon] (midpoint) --++ (-80:0.7);
        \filldraw[color=black,fill=gray] (b) circle[radius=0.2];
    \end{tikzpicture}
\end{gathered}
\right) + \gamma_2 \U^{(1)}(\beta_{1,1},\beta_{2,1},\gamma_1) \\
& \xrightarrow[r_{[1,2]}]{} \gamma_1 \gamma_2+ \gamma_1 \left(\tilde{\alpha}_{1,1}+ \tilde{\alpha}_{2,1}+\tilde{\alpha}_{1,2}\right)+ \gamma_2 \left( \tilde{\alpha}_{1,1} + \tilde{\alpha}_{2,1} + \tilde{\alpha}_{2,2} \right) + \mathcal{O}(\lambda^2),\\
&=  \mathcal{U}_S^{(1)} \mathcal{U}_S^{(2)}+\mathcal{U}_S^{(1)} \left(\beta_{1,2}+\beta_{2,2}\right)+ \mathcal{U}_S^{(2)} \left( \beta_{1,1} + \beta_{2,1}  \right) + \mathcal{O}(\lambda^{-2}),
\end{aligned}
\end{equation}
where now for the non-planar diagram the worldline variables are given by:
\begin{equation}
    \beta_{1,1} = \tilde{\alpha}_{1,1}, \quad \beta_{2,1} = \tilde{\alpha}_{2,1}+\tilde{\alpha}_{2,2}, \quad \beta_{1,2} = \tilde{\alpha}_{1,1}+\tilde{\alpha}_{1,2}, \quad \beta_{2,2} = \tilde{\alpha}_{2,1}.
\end{equation}
For the $\F$ polynomial we find:
\begin{equation}
\begin{aligned}
    \F_0 &= \F_0\left( 
\begin{gathered}
    \begin{tikzpicture}[line width=0.7,scale=0.35,baseline={([yshift=0.0ex]current bounding box.center)}]
        \coordinate (a) at (-0.5,0);
        \coordinate (b) at (1,0);
        \coordinate (c) at (2,0);
        \coordinate (d) at (2,1);
        \coordinate (e) at (2,-1);
        \coordinate (f) at (3,0);
        \filldraw[color=black,fill=white] (c) circle[radius=1];
        \draw[] (f) --++ (35:1);
        \draw[] (f) --++ (-35:1) coordinate (g);
        \coordinate (midpoint) at ($(f)!0.5!(g)$);
        \draw[photon] (d) to[in=50, out=50] (midpoint);
        \filldraw[color=black,fill=gray] (b) circle[radius=0.2];
    \end{tikzpicture}
\end{gathered}
\right) + \gamma_2 \F_0 \left( \begin{gathered}
    \begin{tikzpicture}[line width=0.7,scale=0.35,photon/.style={decorate, decoration={snake, amplitude=1pt, segment length=6pt}
	}]
        \coordinate (a) at (-0.5,0);
        \coordinate (b) at (1,0);
        \coordinate (d) at (4,1.5);
        \coordinate (e) at (4,-1.5);
        \path (b) -- (d) coordinate[midway] (midbd);
        \path (b) -- (e) coordinate[midway] (midbe);
        \path (b) -- (d) coordinate[pos=0.7] (bd_point);
        \path (b) -- (e) coordinate[pos=0.7] (be_point);
        \draw[] (b) -- (d);
        \draw[] (b) -- (e);
        \draw[photon] (midbd) -- (be_point);
        \draw[photon] (bd_point) --++ (-35:1);
        \draw[photon] (midbe) --++ (35:1);
        \filldraw[color=black,fill=gray] (b) circle[radius=0.2];
    \end{tikzpicture}
\end{gathered} \right) \\
 &\xrightarrow[r_{[1,2]}]{} \gamma_1 \F_0
\left( 
\begin{gathered}
    \begin{tikzpicture}[line width=0.7,scale=0.35,baseline={([yshift=0.0ex]current bounding box.center)}]
        \coordinate (a) at (-0.5,0);
        \coordinate (b) at (1,0);
        \coordinate (c) at (2,0);
        \coordinate (d) at (2,1);
        \coordinate (dp) at (2,0.3);
        \coordinate (e) at (2,-1);
        \coordinate (ep) at (2,-0.3);
        \coordinate (f) at (3,0);
        \filldraw[color=black,fill=white] (c) circle[radius=1];
        \draw[photon] (d) -- (dp);
        \draw[] (f) --++ (35:1);
        \draw[] (f) --++ (-35:1) coordinate (g);
        \coordinate (midpoint) at ($(f)!0.5!(g)$);
        \draw[photon] (midpoint) --++ (-80:0.7);
        \filldraw[color=black,fill=gray] (b) circle[radius=0.2];
    \end{tikzpicture}
\end{gathered}
\right) + \gamma_2 \F_0 \left( \begin{gathered}
    \begin{tikzpicture}[line width=0.7,scale=0.35,photon/.style={decorate, decoration={snake, amplitude=1pt, segment length=6pt}
	}]
        \coordinate (a) at (-0.5,0);
        \coordinate (b) at (1,0);
        \coordinate (d) at (4,1.5);
        \coordinate (e) at (4,-1.5);
        \path (b) -- (d) coordinate[midway] (midbd);
        \path (b) -- (e) coordinate[midway] (midbe);
        \path (b) -- (d) coordinate[pos=0.7] (bd_point);
        \path (b) -- (e) coordinate[pos=0.7] (be_point);
        \draw[] (b) -- (d);
        \draw[] (b) -- (e);
        \draw[photon] (midbd) -- (be_point);
        \draw[photon] (bd_point) --++ (-35:1);
        \draw[photon] (midbe) --++ (35:1);
        \filldraw[color=black,fill=gray] (b) circle[radius=0.2];
    \end{tikzpicture}
\end{gathered} \right),
\end{aligned}
\end{equation}
with 
\begin{equation}
\begin{aligned}
&\F_0
\left( 
\begin{gathered}
    \begin{tikzpicture}[line width=0.7,scale=0.35,baseline={([yshift=0.0ex]current bounding box.center)}]
        \coordinate (a) at (-0.5,0);
        \coordinate (b) at (1,0);
        \coordinate (c) at (2,0);
        \coordinate (d) at (2,1);
        \coordinate (dp) at (2,0.3);
        \coordinate (e) at (2,-1);
        \coordinate (ep) at (2,-0.3);
        \coordinate (f) at (3,0);
        \filldraw[color=black,fill=white] (c) circle[radius=1];
        \draw[photon] (d) -- (dp);
        \draw[] (f) --++ (35:1);
        \draw[] (f) --++ (-35:1) coordinate (g);
        \coordinate (midpoint) at ($(f)!0.5!(g)$);
        \draw[photon] (midpoint) --++ (-80:0.7);
        \filldraw[color=black,fill=gray] (b) circle[radius=0.2];
    \end{tikzpicture}
\end{gathered}
\right) = s(\tilde{\alpha}_{1,1}+\tilde{\alpha}_{1,2})\tilde{\alpha}_{2,1}+m^2\tilde{\alpha}_{2,2}(\tilde{\alpha}_{1,1}+\tilde{\alpha}_{1,2}+\tilde{\alpha}_{2,1}) \sim \mathcal{O}(\lambda^{-2}),\\
&\F_0 \left( \begin{gathered}
    \begin{tikzpicture}[line width=0.7,scale=0.35,photon/.style={decorate, decoration={snake, amplitude=1pt, segment length=6pt}
	}]
        \coordinate (a) at (-0.5,0);
        \coordinate (b) at (1,0);
        \coordinate (d) at (4,1.5);
        \coordinate (e) at (4,-1.5);
        \path (b) -- (d) coordinate[midway] (midbd);
        \path (b) -- (e) coordinate[midway] (midbe);
        \path (b) -- (d) coordinate[pos=0.7] (bd_point);
        \path (b) -- (e) coordinate[pos=0.7] (be_point);
        \draw[] (b) -- (d);
        \draw[] (b) -- (e);
        \draw[photon] (midbd) -- (midbe);
        \draw[photon] (bd_point) --++ (-35:1);
        \draw[photon] (be_point) --++ (35:1);
        \filldraw[color=black,fill=gray] (b) circle[radius=0.2];
    \end{tikzpicture}
\end{gathered} \right) = m^2  \left(\tilde{\alpha}_{1,2}\right)\U^{(1)}(\beta_{1,1},\beta_{2,1},\gamma_1) +  \F_0^{(1)}(\beta_{1,1},\beta_{2,1},\gamma_1)  \sim \mathcal{O}(\lambda^{-3}),
\end{aligned}
\end{equation}
so that we get
\begin{equation}
\begin{aligned}
    \F &= \F_0 - m^2(\tilde{\alpha}_{1,1} +\tilde{\alpha}_{1,2} +\tilde{\alpha}_{2,1} +\tilde{\alpha}_{2,2} ) \left[ \gamma_1 \U
\left( 
\begin{gathered}
    \begin{tikzpicture}[line width=0.7,scale=0.35,baseline={([yshift=0.0ex]current bounding box.center)}]
        \coordinate (a) at (-0.5,0);
        \coordinate (b) at (1,0);
        \coordinate (c) at (2,0);
        \coordinate (d) at (2,1);
        \coordinate (dp) at (2,0.3);
        \coordinate (e) at (2,-1);
        \coordinate (ep) at (2,-0.3);
        \coordinate (f) at (3,0);
        \filldraw[color=black,fill=white] (c) circle[radius=1];
        \draw[photon] (d) -- (dp);
        \draw[] (f) --++ (35:1);
        \draw[] (f) --++ (-35:1) coordinate (g);
        \coordinate (midpoint) at ($(f)!0.5!(g)$);
        \draw[photon] (midpoint) --++ (-80:0.7);
        \filldraw[color=black,fill=gray] (b) circle[radius=0.2];
    \end{tikzpicture}
\end{gathered}
\right) + \gamma_2 \U^{(1)}(\beta_{1,1},\beta_{2,1},\gamma_1) \right]\\
   &\xrightarrow[r_{[1,2]}]{} \gamma_2 \left[ \F_0^{(1)}(\beta_{1,1},\beta_{2,1},\gamma_1) -m^2(\tilde{\alpha}_{1,1}+\tilde{\alpha}_{2,1}+\tilde{\alpha}_{2,2})\U^{(1)}(\beta_{1,1},\beta_{2,1},\gamma_1)\right] \\
   &\quad \quad \quad +\gamma_1 \left[ s\beta_{1,2}\beta_{2,2} -m^2 (\beta_{1,2}+\beta_{2,1})(\beta_{1,2}+\beta_{2,2}) +m^2\tilde{\tilde{\alpha}}^b_2(\beta_{1,2}+\beta_{2,2})\right] \\
   &\xrightarrow[r_{[1,2]}]{} \U^{(2)}_S \F^{(1)} +\gamma_1 \left[ s\beta_{1,2}\beta_{2,2} -m^2 (\beta_{1,2}+\beta_{2,2})^2 \right]\\
   &\xrightarrow[r_{[1,2]}]{} \U^{(2)}_S \F^{(1)}(\beta_{1,1},\beta_{2,1}) + \U^{(1)}_S \F^{(2)}(\beta_{1,2},\beta_{2,2}) + \mathcal{O}(\lambda^{-3}),
\end{aligned}
\end{equation}
which is precisely of the same form of what we found for the planar integrand once written in terms of the $\beta$-variables!
So at leading order along $r_{[1,2]}$, both the planar and non-planar two-loop integrand satisfy 
\begin{equation}
    \d\I_2^{\pl/\npl} \to \d S_1^{(1)}(\beta_{J,1},\gamma_1) \times  \d S_1^{(2)}(\beta_{J,2},\gamma_2),
\end{equation}
where crucially each soft factor is written in terms of the respective  worldline variables. 

Note, however, that the definition of the $\beta$ variables in terms of $\tilde{\alpha}$ for the non-planar diagram is different from that of the planar diagram, which ultimately translates into a different integration region in the $\beta$'s. Concretely, since the $\tilde{\alpha} \in \mathbb{R}^+$, we have that for the non-planar diagram
\begin{equation}
    0<\beta_{1,1}< \beta_{1,2}, \quad  0<\beta_{2,2}< \beta_{2,1}.
\end{equation}
\begin{figure}[!t]
    \centering
    \includegraphics[width=\textwidth]{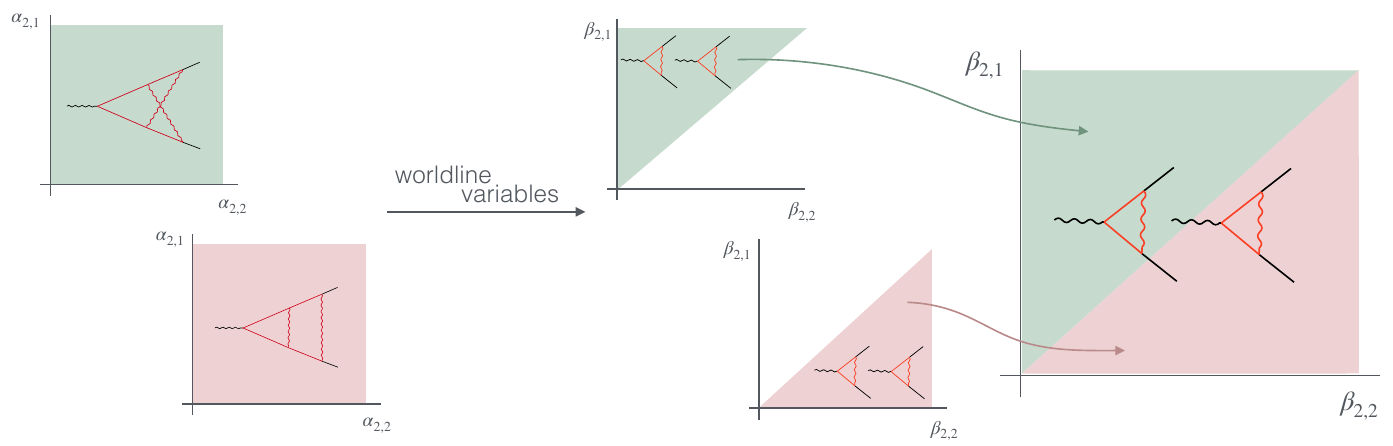}
    \caption{Illustration of how  worldline variables modify the integration region of the lower line ($p_2$) Schwinger parameters. (\textbf{left}) In Schwinger variables the planar and non-planar ladder have different (soft) integrands and both occupy the entire positive octant in $\alpha_{2,1}, \, \alpha_{2,2}$. 
    (\textbf{right}) In the worldline variables they have identical integrands and each of the diagrams fills half of the positive octant in $\beta_{2,1}, \, \beta_{2,2}$. The sum of the two diagrams fills the entire positive octant. }
    \label{fig:2loop_worldline}
\end{figure}
Therefore we see that in the scaling $r_{[1,2]}$ the contributions coming from the planar and non-planar diagrams written in the $\beta$ variables are the same, just integrated over different regions. So for the full integrand, when we sum over both diagrams, we get a single integral over the domain (see \cref{fig:2loop_worldline}):
\begin{equation}
  \mathcal{D}_\beta :  0<\beta_{1,1}< \beta_{1,2}, \quad  \beta_{2,1},\beta_{2,2} \in \mathbb{R}^+. 
  \label{eq:domainBeta2loops}
\end{equation}
Note that though the sum over diagrams has filled the entire positive octant in $\beta_{2,1}$ and $\beta_{2,2}$, the integration domain for the worldline variables of the other external leg ($\beta_{1,1}$ and $\beta_{1,2}$) remains ordered, but identically for the two diagrams. This will ultimately yield the $1/2$ (or $1/\ell!$ for $\ell$ photons) factor necessary for exponentiation.

\subsection{Matrix-tree theorem for factorisation}
\label{sec:2loopMatrix}

It is well known that the recursive structure of the Symanzik polynomials (see \cref{eq:recursive_U_and_F}), used in the previous section to derive the soft expansion of the two-loop diagrams, is a simple consequence of the matrix-tree theorem applied to the definition of $\U_G
$ and $\F_G$ via the Laplacian matrix of a graph, $G$ (see ref.~\cite{Weinzierl:2022eaz}). Concretely, given a graph $G$, the respective $\U_G$ and $\F_G$ can be defined as follows
\begin{equation}
    \U_G = \left( \prod_{e \in G} \alpha_e \right) \operatorname{det} \LL \,, \qquad \F_G =  \U_G \, \left[ p_v \cdot p_w (\mathbf{L}^{-1})_{vw} -\sum_{e \in G} m_e^2 \alpha_e  \right]  \,,
    \label{eq:matrixSymanzik}
\end{equation}
where $p_v$ is the total momentum entering in vertex $v$, and $\LL$ is the \textit{reduced Laplacian matrix}, which can be obtained by removing one row and column from the graph Laplacian matrix, $\overline{\LL}$, given by:
\begin{equation}
    \left(\overline{\LL}\right)_{vw} = \sum_{e\in G} \frac{\eta_{ve}\eta_{we}}{\alpha_e} \,,
    \label{eq:LapMatrix}
\end{equation}
where $\eta_{v,e}$ is the incidence matrix, which is a $V \times E$ matrix, with $E$ the number of internal edges in the graph, and $V$ the number of vertices in $G$, given by
\begin{equation}
      \eta_{ve} = 
    \begin{cases}
        +1,  \quad \text{$e$ directed towards $v$} \\  
        -1,  \quad \text{$e$ directed away from $v$} \\  
        \;\; \,0,  \quad \text{$e$ disconnected from $v$} 
    \end{cases}.
    \label{eq:incidenceMatrix}
\end{equation}
As one can easily check, the matrix $\overline{\LL}$ is not invertible since the sum of its rows (or columns) vanishes, but after removing a vertex from the graph (which removes a line and a column from the matrix) we are left with the invertible matrix, $\LL$, entering in \cref{eq:matrixSymanzik}. For concreteness here we choose 
    $\LL \equiv \overline{\LL}[V]$,
where $\overline{\LL}[V]$ is the matrix $\overline{\LL}$ with the $V$-th row and column deleted.
In general we can pick any vertex to drop to form the reduced Laplacian matrix, but for convenience we will always consider removing a vertex inside the hard part of the diagram.

Using this formulation, we can write the Feynman integral associated to a given graph $G$ with $E$ internal propagators and $\ell$ loops as follows
\begin{equation}
\I_G= i^{E-\ell\D/2}\int_{\alpha_e > 0}  \frac{\prod_e \mathrm{d} \alpha_e}{\left(\prod_{e} \alpha_e\right)^{\mathrm{D} / 2}\left(\operatorname{det} \mathbf{L}\right)^{\mathrm{D} / 2}}  \exp \left(i \mathcal{V} \right),
\label{eq:FeynInt2loop}
\end{equation} 
with the worldline action given by 
\begin{equation}
  \mathcal{V} =   p_v \cdot p_{w} \, (\mathbf{L}^{-1})_{vw} -\sum_{e \in G} m_e^2 \alpha_e  \equiv \frac{\F}{\U}. 
\end{equation}
From this point of view, it is then expected that the factorisation found in the previous section for $\F$ and $\U$, which ultimately ensured the correct hard-soft factorisation of the integrand in the different scalings, should translate into a simple behaviour of the matrix $\LL$ when expanded along the IR divergent rays. 
As it turns out, this is exactly the case and moreover the definition of the Feynman integral via the graph matrix $\Lb$ allows for a simple derivation of the factorisation structure of the integrand under the different IR scalings. Most importantly, this formulation allows for a completely uniform treatment of a class of different graphs -- with arbitrary number of loops and jets, as well as different web structures connecting the jets -- while also allowing for the inclusion of numerator factors, which play a role when considering different theories. This is the path we pursue in section \ref{sec:factorisation_matrix}. For now, let us stick to the two-loop example in study, and illustrate this procedure for the simplest type of the factorisation corresponding to the $r_{[1,2]}$ scaling in the planar ladder. 

For the planar ladder, if we list the vertices in the graph labelled in \cref{fig:2loop-diagrams} in the following vertex vector
\begin{equation}
    \boldsymbol{v} = (v_{h_0},v_{1,1},v_{1,2},v_{2,1},v_{2,2}),
\end{equation}
the reduced Laplacian matrix we get by deleting the first row and column, corresponding to the hard vertex, $v_{h_0}$, is simply
\begin{equation*}
 \textbf{L}=   \begin{bmatrix}
 \frac{1}{\alpha_{1,1}}+\frac{1}{\alpha_{1,2}}+\frac{1}{\gamma_1}&-\frac{1}{\alpha_{1,2}}& -\frac{1}{\gamma_1}  &0\\
-\frac{1}{\alpha_{1,2}}&\frac{1}{\alpha_{1,2}}+\frac{1}{\gamma_2}& 0 &-\frac{1}{\gamma_2} \\
 -\frac{1}{\gamma_1}&0& \frac{1}{\alpha_{2,1}}+\frac{1}{\alpha_{2,2}}+\frac{1}{\gamma_1}& -\frac{1}{\alpha_{2,2}} \\
0& -\frac{1}{\gamma_2}& -\frac{1}{\alpha_{2,2}}&\frac{1}{\alpha_{2,2}}+\frac{1}{\gamma_2}
\end{bmatrix}.
\end{equation*}
To get more insight into its structure, we can write this matrix as a sum $\textbf{L} = \Jb + \boldsymbol{\Gamma}_{J}$, where  $\Jb$ is the block diagonal part which contains the Schwinger parameters scaled as $\lambda^{-1}$ ($i.e.$ the massive ``jet'' lines), and $\boldsymbol{\Gamma}_{J}$ the part including the photon Schwinger parameters, which scale as $\lambda^{-2}$: 
\begin{equation}
   \Jb=  \begin{bmatrix}
 \frac{1}{\alpha_{1,1}}+\frac{1}{\alpha_{1,2}} & -\frac{1}{\alpha_{1,2}} & 0 &
   0 \\
 -\frac{1}{\alpha_{1,2}} & \frac{1}{\alpha_{1,2}} & 0 & 0 \\
 0 & 0 & \frac{1}{\alpha_{2,1}}+\frac{1}{\alpha_{2,2}} & -\frac{1}{\alpha_{2,2}} \\
 0 & 0 & -\frac{1}{\alpha_{2,2}} & \frac{1}{\alpha_{2,2}} \\
\end{bmatrix}, \quad 
\boldsymbol{\Gamma}_{J}=\begin{bmatrix}
 \frac{1}{\gamma_1} & 0 & -\frac{1}{\gamma_1}  & 0\\
 0 & \frac{1}{\gamma_2} & 0&-\frac{1}{\gamma_2} \\
-\frac{1}{\gamma_1} & 0 &  \frac{1}{\gamma_2} & 0 \\
 0&-\frac{1}{\gamma_2} & 0 &  \frac{1}{\gamma_2} \\
\end{bmatrix}.
\end{equation}
We see that $\Jb$ is block diagonal with blocks $\Jb_{1}$ and $\Jb_{2}$ corresponding to edges in jets 1 and 2.
Further, one can immediately see that $\LL \to \Jb$ at leading order in $\lambda$  and, since $\Jb$ is block diagonal, we can easily extract its determinant to derive the factorisation of $\U$. 
Concretely:
\begin{equation}
    \text{det}(\textbf{J}_i) = \frac{1}{\alpha_{i,1}\alpha_{i,2}} \quad \Rightarrow \quad \U \to \gamma_1 \gamma_2 \prod_{i=1}^2  \alpha_{i,1}\alpha_{i,2} \text{det}(\Jb_i) = \gamma_1 \times \gamma_2.
\end{equation}
As for the action $\mathcal{V}$, we see from the exponent of \cref{eq:FeynInt2loop} we see that we first need to invert $\LL$. Using the decomposition above, we can write the expansion of $\LL^{-1}$ under the $r_{[1,2]}$ scaling as
\begin{equation}
    \LL^{-1} = (\Jb+\boldsymbol{\Gamma}_{J})^{-1} = \Jb^{-1} -\Jb^{-1} \boldsymbol{\Gamma}_{J}\Jb^{-1} + \cdots.
\end{equation}
Plugging this result into \cref{eq:FeynInt2loop}, we find
\begin{equation}
    \mathcal{V} \to  \underbrace{p_v \cdot p_{w} \, ( \Jb^{-1})_{vw}}_{\mathcal{O}(\lambda^{-1})}-  \underbrace{p_v \cdot p_{w} \, (\Jb^{-1} \boldsymbol{\Gamma}_{J}\Jb^{-1} )_{vw}}_{\mathcal{O}(\lambda^{0})}-\underbrace{m^2 {\textstyle \sum_{e }  \alpha_e} }_{\mathcal{O}(\lambda^{-1})}.
    \label{eq:v2_1}
\end{equation}
Now, because $\Jb$ is block-diagonal we can invert it by inverting each of its blocks $\textbf{J}_i$. 
In doing so, quite nicely, we see how the worldline variables emerge automatically from the Laplacian matrix: 
\begin{equation}
    \textbf{J}_i^{-1} = \begin{bmatrix}
        \beta_{i,1} & \beta_{i,1} \\
        \beta_{i,1} & \beta_{i,2} \\
    \end{bmatrix}.
\end{equation}
Now $p_v$ is a V-dimensional vector whose entries encode the momentum entering each vertex of the graph, and therefore it takes the form $p_v = (0,p_1,0,p_2)$, where we have already dropped vertex, $v_{h_0}$. Therefore we can readily compute $\Jb^{-1} p = (\beta_{1,1}p_1,\beta_{1,2}p_1,\beta_{2,1}p_1,\beta_{2,2}p_2)^\top$, so that the first term in \cref{eq:v2_1} simply gives 
\begin{equation}
     p_v \cdot p_{w} \, ( \Jb^{-1})_{vw} = \sum_{i=1}^2 p_i^2 \beta_{i,2} = m^2 \sum_{e} \alpha_e, 
\end{equation}
which precisely cancels the last term in \cref{eq:v2_1}. 

Finally, to manifest the factorised form of $(\Jb^{-1} p)^T \boldsymbol{\Gamma}_{J} (\Jb^{-1}p)$, it is useful to swap the entries on the vertex vector such that vertices connected by a photon are adjacent to each other, $i.e.$
\begin{equation}
    \tilde{\boldsymbol{v}} = (v_{1,1},v_{2,1}|v_{1,2},v_{2,2}).
\end{equation}
In this basis, we have that $\boldsymbol{\Gamma}_{J}$ is block diagonal and each block, $\boldsymbol{\Gamma}_{i}$, is labelled by the respective photon $i$, 
\begin{equation}
   \boldsymbol{\Gamma}_{J}= \begin{bmatrix}
 \frac{1}{\gamma_1} & -\frac{1}{\gamma_1} & 0 & 0 \\
 -\frac{1}{\gamma_1} & \frac{1}{\gamma_1} & 0 & 0 \\
 0 & 0 & \frac{1}{\gamma_2} & -\frac{1}{\gamma_2} \\
 0 & 0 & -\frac{1}{\gamma_2} & \frac{1}{\gamma_2} \\
\end{bmatrix}
\equiv \begin{bmatrix}
\boldsymbol{\Gamma}_1 & 0 \\
0 & \boldsymbol{\Gamma}_2
\end{bmatrix}
,
\end{equation}
and similarly for this vertex ordering we have that 
\begin{equation}
    \Jb^{-1} p = (\beta_{1,1}p_1,\beta_{2,1}p_2\, |\, \beta_{1,2}p_1,\beta_{2,2}p_2)^\top \equiv ( \boldsymbol{P}_1 \,|\, \boldsymbol{P}_2 )^\top. 
\end{equation}
So we find
\begin{equation}
    \mathcal{V} \to \sum_{i=1}^2 \boldsymbol{P}_i^T \boldsymbol{\Gamma}_{i} \boldsymbol{P}_i = \sum_{i=1}^2 \frac{s \beta_{1,i} \beta_{2,i} - m^2 (\beta_{1,i} + \beta_{1,i})^2}{\gamma_i},
\end{equation}
which is simply the sum over $\F_i/\U_i$ for each one-loop process, written in terms of the worldline variables $\beta$,  precisely as expected.

In summary, already in this simple example we have found that the matrix formulation of the Symanzik formulation lets us derive the factorisation of the integrand on a diagram-by-diagram basis as a result of simple matrix identities coming from the fact that the graph matrices simplify -- namely become block diagonal -- under the soft scaling. 
In section \ref{sec:factorisation_matrix}, we extend this observation to a general proof at all loops as well as to more interesting soft diagram structures. 
In general, we find that the Laplacian matrix can be decomposed into a hard and a soft part -- here there was no hard part since all photons were made soft -- where now the soft part has more interesting structure coming from the presence of hard subdiagrams (or blobs) inside the photon webs. 
In addition, the Laplacian formulation will also allow us to incorporate numerators and therefore perform a uniform study of the integrand behaviour under the soft scalings in different theories. 
Nonetheless this simple two-loop case serves as a perfect example of the main features that enter in showing the factorisation of the integrands in the settings described in section \ref{sec:factorisation_matrix}. 
\subsection{Regions of integration and exponentiation}
\label{sec:regions2loop}

Having understood how the integrands for the planar and non-planar ladder behave under the different IR divergent scalings, we now proceed to discussing how putting everything together we can recover the two-loop exponentiation statement of \cref{eq:expIR}.

Note that when defining the expansions of the integrand under the different scalings we have not yet been careful in articulating in which region the expansions are valid. It is well known that if we expand a function in a certain limit, but then integrate it over a larger domain 
(namely outside the domain of applicability of the expansion) we may create new divergences that are present in the expansion but not in the original function. 
The purpose of this section is to perform the analysis of tropical rays and domains of integration carefully for the two-loop example discussed above. 
The ideas presented here are generalised to all orders and to an arbitrary number of external states in \cref{sec:regions}.
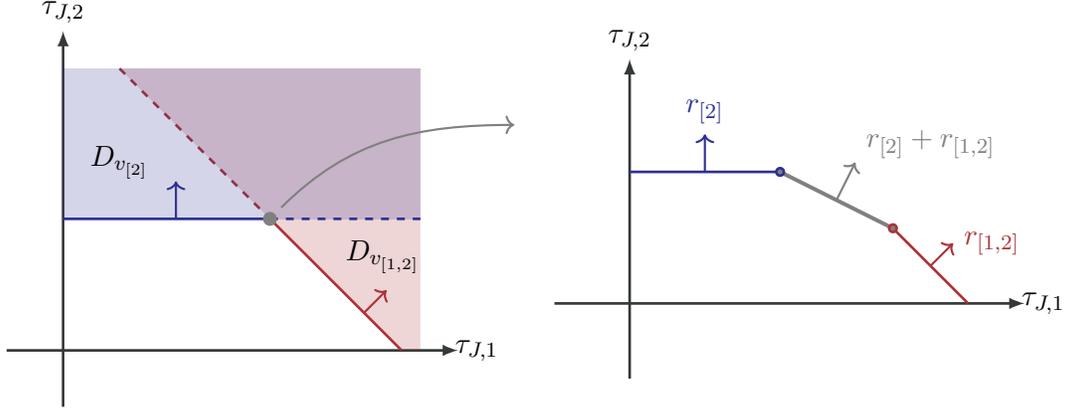
\begin{figure}[t]
\centering
 \input{Figures/2loopRegions}
\caption{(\textbf{left}) Representation of different domains in $(\tau_{J,1},\tau_{J,2})$ space, with $\alpha_{i,J}=\exp{\tau_{J,i}}$. The normals to each facet of the Newton polytope give the divergent rays (blue and red), and also determine the respective regions. 
(\textbf{right}) Resulting Newton polytope after performing a ``blow-up'' -- where the gray vertex on the left becomes a facet. It has a total of three rays/facet normals, the original $r_{[2]}$ and  $r_{[1,2]}$ and the sum of both, $r_{[2]}+r_{[1,2]}$.}
    \label{fig:2loopRegions}
\end{figure}

Let us start by looking at the planar ladder, where we have the two scalings given in \cref{eq:scalings2loops}. Calling $D_{[2]}$ and $D_{[1,2]}$ (see left of figure \ref{fig:2loopRegions}), the regions where these two scalings are valid approximations, we can approximate the full two-loop planar integral $\mathcal{I}^{\pl}_{(2)}$ as
\begin{equation}
 \int_{\mathbb{R}_+^E} \d  \mathcal{I}_{(2)}^{\pl} \simeq
   \int_{D_{[2]}} \left[\d  \mathcal{I}_{(2)}^{\pl}\right]_{r_{[2]}} +  \int_{D_{[1,2]}} \left[\d  \mathcal{I}_{(2)}^{\pl}\right]_{r_{[1,2]}}, 
   \label{eq:expRegionsD2}
\end{equation}
where here we are assuming the integrands are written in terms of the standard Schwinger parameters, $\alpha \in \mathbb{R}_+$. 
We will explicitly introduce the $\beta$ variables later on when discussing the full result after summing over the planar and non-planar ladders. In the equation above, $[\cdots]_r$ stands for the leading approximation under the ray $r$.

Now our goal is to extend the integration domain of each term above to the full integration domain.
Of course, in doing this we must be careful to avoid the over-counting of divergences coming from the overlap of the two domains -- as we now explain this can be done by introducing subtraction terms from which we define \textit{modified soft integrands}.

Starting with the integral over $D_{[2]}$, and recalling from the previous section that we have 
\begin{equation}\label{eq:v2expansion}
    \left[\d  \mathcal{I}_{(2)}^{\pl}\right]_{r_{[2]}} \simeq \d  \mathcal{I}_{(1)}[\alpha_{J,1}] \times \d  S_{(1)} [\alpha_{J,2}],
\end{equation}
we can extend it to the full integration domain as follows
\begin{equation}
     \int_{D_{[2]}} \left[\d  \mathcal{I}_{(2)}^{\pl}\right]_{r_{[2]}} =  \int_{D_{[2]}} \d  \mathcal{I}_{(1)} \times \d  S_{(1)} =  \int_{\mathbb{R}_+^E \setminus D_0} \d  \mathcal{I}_{(1)} \times \d  S_{(1)} - \int_{D_{[1,2]}} \d  \mathcal{I}_{(1)} \times \d  S_{(1)},
\end{equation}
where $D_0$ stands for the region in Schwinger parameter space for which the original integrand is IR finite, which we can be define as the complement of $(D_{[2]} \cup D_{[1,2]})$ in $\mathbb{R}_+^E$. Replacing this result back in \cref{eq:expRegionsD2} yields 
\begin{equation}
\begin{aligned}
    \int_{\mathbb{R}_+^E} \d  \mathcal{I}_{(2)}^{\pl} &\simeq \int_{\mathbb{R}_+^E \setminus D_0} \d  \mathcal{I}_{(1)} \times \d  S_{(1)} + \int_{D_{[1,2]}} \left[\d  \mathcal{I}_{(2)}^{\pl}\right]_{r_{[1,2]}} - \d  \mathcal{I}_{(1)} \times \d  S_{(1)} \\
    &\simeq \int_{\mathbb{R}_+^E \setminus D_0} \d  \mathcal{I}_{(1)} \times \d  S_{(1)} + \int_{D_{[1,2]}} \left[\d \mathcal{I}_{(2)}^{\pl}- \d  \mathcal{I}_{(1)} \times \d  S_{(1)}\right]_{r_{[1,2]}} \\
     &\simeq \int_{\mathbb{R}_+^E \setminus D_0} \d  \mathcal{I}_{(1)} \times \d  S_{(1)} + \int_{D_{[1,2]}} \left( \d  S_{(2)}[\alpha_{1,J},\alpha_{2,J}]- \d  S_{(1)}[\alpha_{1,J}] \times \d  S_{(1)}[\alpha_{2,J}]\right),
\end{aligned}
\label{eq:subtraction2loops}
\end{equation}
where in the last line we replaced the expansion of $\d \mathcal{I}_{(2)}^{\pl}$ under $r_{[1,2]}$ by $\d S_2$. As we know from the previous section, once we change variables into the $\beta$'s, this term further factors into the product of two factors of $\d S_1$. So comparing the last line of \cref{eq:subtraction2loops} to \cref{eq:expRegionsD2}, we see that the integrand approximation along ray $r_{[2]}$ is integrated on the whole IR-divergent domain, but to compensate for this in the domain $D_{[1,2]}$ we have a modified soft integrand: 
\begin{equation}
    \widetilde{ \dd S}_{(2)} = \d  S_{(2)} - \d  S_{(1)} \times \d  S_{(1)}.
\end{equation}
Now, by construction we have that $\widetilde{ \dd S}_{(2)}$ vanishes when we further expand along $v_2$, since $ [\d  S_{(2)}]_{r_{[2]}} \simeq \d  S_{(1)} \times S_{(1)}$. 
Therefore we can safely extend the integration domain of the second term in the last line of 
\cref{eq:subtraction2loops} from $D_{[1,2]}$ to $\mathbb{R}_+^E \setminus D_0$, without over-counting any divergences:
\begin{equation}
    \int_{\mathbb{R}_+^E} \d  \mathcal{I}_{(2)}^{\pl} \simeq \int_{\mathbb{R}_+^E \setminus D_0} \d  \mathcal{I}_{(1)} \times \d  S_{(1)} +  \widetilde{ \dd S}_{(2)} .
    \label{eq:final2loops}
\end{equation}
However, as we mentioned above, expanding the original integrand along the divergent rays ($r_{[2]}$ and $r_{[1,2]}$ in this case) can
produce an expression which has new IR divergent rays. Indeed, the expansion along $r_{[2]}$ given in \cref{eq:v2expansion} 
has a new ray $r_0=(1,0,1,0,2,0)$ which lies inside the previously IR-finite region $D_0$ and corresponds to the 
inner photon going soft. Still we can check that the total integrand above vanishes along this direction at leading order in $\lambda$. 
This is simply because $\d S_{(2)}$ is finite on $r_0$, but $\d  \mathcal{I}_{(1)} \times \d  S_{(1)} \xrightarrow{r_0} \d  S_{(1)} \times \d  S_{(1)}$, 
which precisely cancels with the subtraction term in $\widetilde{ \dd S}_{(2)}$. 
So we see that, quite nicely, the subtraction term added in the modified $\widetilde{ \dd S}_{(2)} $ is doing double-duty, as it both 
avoids the over-counting of the divergences when extending $D_{[1,2]} \to D_{[1,2]} \cup D_{[2]}$, as well as cancels the new divergence
of $ \left[\d  \mathcal{I}_{(2)}^{\pl}\right]_{r_{[2]}}$ once we extend the integration domain into $D_0$, recovering the full $\mathbb{R}_+^E$. 

Now writing the final result in \cref{eq:final2loops} in terms of the expansions of the integrand we find
\begin{equation}
     \mathcal{I}^{\pl}_{(2)} = \int_{\alpha \in \mathbb{R}^+}  \left[\d \I^{\pl}_{(2)}\right]_{r_{[2]}}+ \left[\d \I^{\pl}_{(2)}\right]_{r_{[1,2]}} - \left[\d \I^{\pl}_{(2)}\right]_{r_{[2]},r_{[1,2]}}  + \text{(IR finite)} .
    \label{eq:2loopPLIRPart}
\end{equation}
In summary, we found that the correct integrand capturing  the IR divergent part of the two-loop planar integrand 
is given by the sum over the expansions along the two IR divergent rays $r_{[2]}$ and $r_{[1,2]}$ of the Newton 
polytope, minus the expansion along a new ray, given by the sum of $r_{[2]}+r_{[1,2]}$.\footnote{This correspond 
respectively to the expansions $[\cdots]_{r_{[2]}}$, $[\cdots]_{r_{[1,2]}}$, and  $[\cdots]_{r_{[2]},r_{[1,2]}}$, 
where we have $[\cdots]_{r_{[2]},r_{[1,2]}} = [\cdots]_{r_{[1,2]},r_{[2]}}$.} Geometrically this corresponds to 
replacing the original Newton polytope of the Feynman integral 
with a ``blown-up'' version obtained by adding to the underlying fan this new ray, as shown on the right 
of figure \ref{fig:2loopRegions}. From this new object we can directly extract the IR divergent part by expanding 
along the different rays, with the alternating sign as explained above. 

Finally let us show how using \cref{eq:final2loops,eq:2loopPLIRPart}, together with the contribution from the non-planar ladder, we can reproduce the exponentiation result in \cref{eq:expIR}. We start by noting that for the non-planar ladder there is a single soft scaling and therefore we can trivially write the IR divergent part without the need for any subtractions
\begin{equation}
    \mathcal{I}^{\npl}_{(2)} = \int_{\tilde{\alpha} \in \mathbb{R}^+}  \left[\d \I^{\npl}_{(2)}\right]_{r_{[1,2]}} + \text{(IR finite)} ,
    \label{eq:2loopNPLIRPart}
\end{equation}
and so adding both ladders we obtain
\begin{equation}
\begin{aligned}
     \mathcal{I}_{(2)} &= \int_{\alpha \in \mathbb{R}^+} \, \left[\d \I^{\pl}_{(2)}\right]_{r_{[1,2]}} - \left[\d \I^{\pl}_{(2)}\right]_{r_{[2]},r_{[1,2]}}
     \\
     & \quad + \int_{\tilde{\alpha} \in \mathbb{R}^+}  \, \left[\d \I^{\npl}_{(2)}\right]_{r_{[1,2]}}
        + \int_{\alpha \in \mathbb{R}^+} \left[\d \I^{\pl}_{(2)}\right]_{r_{[2]},r_{[1,2]}} + \text{(IR finite)} .
\end{aligned}
\label{eq:fullPlNPl}
\end{equation}
Using \cref{eq:subtraction2loops}, we can write  the terms in the first line as
\begin{equation}
\begin{aligned}
   &\int_{\mathbb{R}_+^3}  \d \I_{(1)} (\alpha_{J,1}, \gamma_1) \times  
   \int_{\mathbb{R}_+^3} \d S_1(\alpha_{J,2}, \gamma_2) -\int_{\mathbb{R}_+^3} \d S_1(\alpha_{J,1},\gamma_1)\times 
     \int_{\mathbb{R}_+^3} \d S_{(1)}(\alpha_{J,2},  \gamma_2)\\
&= M_{(1)}S_{(1)} - S_{(1)}S_{(1)},
\end{aligned}
\label{eq:21}
\end{equation}
where we restored the dependence on the Schwinger parameters for clarity. 
On the other hand, for the two terms in the second line of \cref{eq:fullPlNPl}, changing from the $\alpha$'s to the worldline variables $\beta$, we can combine both integrals into a single integral over $\mathcal{D}_\beta$ (as defined in \cref{eq:domainBeta2loops}),
\begin{equation}
\begin{aligned}
    \int_{\mathcal{D}_\beta} \d \I_{r_{[1,2]}}  &= 
    \frac{1}{2}  \int_{\beta_{J,i} \in \mathbb{R}^+} \d \I_{r_{[1,2]}}(\beta_{J,1},\gamma_1,\beta_{J,2},\gamma_2)  \\
    &= \frac{1}{2} \int_{\beta_{1,J} \in \mathbb{R}^+} \,\d S_{(1)}(\beta_{J,1},\gamma_1)  
    \int_{\beta_{2,J} \in \mathbb{R}^+} \d S_{(1)}(\beta_{J,2},\gamma_2) 
    = \frac{1}{2} S_{(1)}^2. 
\end{aligned}
\label{eq:22}
\end{equation}
Adding \cref{eq:21,eq:22} together, we precisely reproduce the two-loop IR divergent part from~\cref{eq:expIR}, which
proves exponentiation of the IR divergences at two-loops for the form-factor amplitude.

\subsection{Diagrams beyond ladders and Renormalization}
\label{sec:Renorm2}
\begin{figure}[t]
    \centering
    \begin{tikzpicture}[line width=0.8,scale=1,line cap=round]
        \coordinate (o) at (-6,0);
        \coordinate (b1) at ($(o) + (30:1)$);
        \coordinate (b2) at ($(o) + (-30:1)$);
        \coordinate (c) at ($(o) + (30:2)$);
        \coordinate (d) at ($(o) + (-30:2)$);
        \path (b1) -- (c) coordinate[pos=0.4] (p1t); 
        \path (b2) -- (d) coordinate[pos=0.4] (p1b);
        \draw[] (o) -- (c);
        \draw[] (o) -- (d);

        \coordinate (cloop) at ($(p1t)!0.5!(p1b)$);
        \draw[color=black] (cloop) circle[radius=0.2];
        
        \draw[photon] (p1t) -- ($(p1t)!0.35!(p1b)$);
        \draw[photon] (p1b) -- ($(p1t)!0.65!(p1b)$);
     
        \filldraw[color=black,fill=gray] (o) circle[radius=0.1];
        \node[] at (-5,-1.7) {(I)} ;

        \coordinate (o2) at (-2,0);
        \coordinate (b12) at ($(o2) + (30:1)$);
        \coordinate (b22) at ($(o2) + (-30:1)$);
        \coordinate (c2) at ($(o2) + (30:2)$);
        \coordinate (d2) at ($(o2) + (-30:2)$);
        \path (b12) -- (c2) coordinate[pos=0.4] (p1t2); 
        \path (b22) -- (d2) coordinate[pos=0.4] (p1b2);
        \draw[] (o2) -- (c2);
        \draw[] (o2) -- (d2);
        
        \draw[photon] (p1t2) -- (p1b2);
        \draw[photon] ($(o2)!0.8!(p1t2)$) to[out=100,in=-200] ($(o2)!1.2!(p1t2)$);

        \filldraw[color=black,fill=gray] (o2) circle[radius=0.1];
        \node[] at (-1,-1.7) {(II)} ;

         \coordinate (o3) at (2,0);
        \coordinate (b13) at ($(o3) + (30:1)$);
        \coordinate (b23) at ($(o3) + (-30:1)$);
        \coordinate (c3) at ($(o3) + (30:2)$);
        \coordinate (d3) at ($(o3) + (-30:2)$);
        \path (b13) -- (c3) coordinate[pos=0.4] (p1t3); 
        \path (b23) -- (d3) coordinate[pos=0.4] (p1b3);
        \draw[] (o3) -- (c3);
        \draw[] (o3) -- (d3);
        
        \draw[photon] (p1t3) -- (p1b3);

        \draw[photon] ($(o3)!0.35!(p1t3)$) to[out=100,in=-200] ($(o3)!0.75!(p1t3)$);
     
        \filldraw[color=black,fill=gray] (o3) circle[radius=0.1];
         \node[] at (3,-1.7) {(III)} ;

        \coordinate (o4) at (6,0);
        \coordinate (b14) at ($(o4) + (30:1)$);
        \coordinate (b24) at ($(o4) + (-30:1)$);
        \coordinate (c4) at ($(o4) + (30:2)$);
        \coordinate (d4) at ($(o4) + (-30:2)$);
        \path (b14) -- (c4) coordinate[pos=0.05] (p1t4); 
        \path (b24) -- (d4) coordinate[pos=0.05] (p1b4);
        \draw[] (o4) -- (c4);
        \draw[] (o4) -- (d4);
        
        \draw[photon] (p1t4) -- (p1b4);
        \draw[photon] ($(o4)!1.25!(p1t4)$) to[out=100,in=100] ($(o4)!1.7!(p1t4)$);
     
        \filldraw[color=black,fill=gray] (o4) circle[radius=0.1];
        \node[] at (7,-1.7) {(IV)} ;      
    \end{tikzpicture}
    \caption{Two-loop diagrams beyond ladders.}
    \label{fig:non-ladder2loops}
\end{figure}
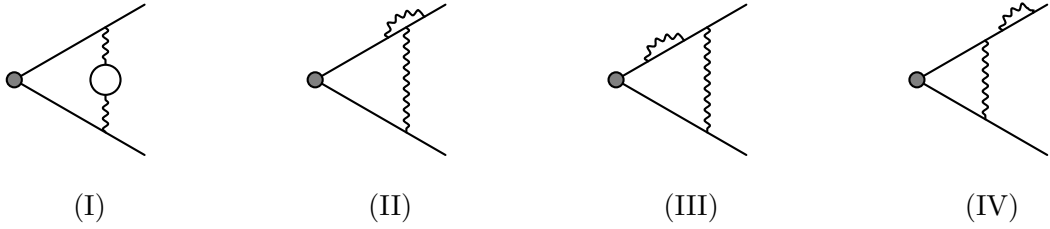

So far we have only focused on ``ladder-type" diagrams, but, already at two-loops, 
there are other types of diagrams that contribute to the amplitude, 
see figure \ref{fig:non-ladder2loops}. These diagrams are particularly important 
when considering the UV renormalization of the theory. Of course, we can eliminate the UV 
divergences in a standard way, $e.g.$ using counterterms. However, depending on which renormalization scheme we choose, these counterterms may or may not contain IR divergences as well. 

One scheme which makes our analysis particularly simple is the on-shell scheme. In this case, the counterterms for the vertex renormalization have IR divergences 
which will exponentiate, but we can fix their contribution to the cusp soft anomalous dimension by requiring that the limit as the cusp angle goes to zero is finite. We will come back to this point in~\cref{sec:spurious_UV_and_RGE}.

In more detail, diagrams (I) and (II) give the standard logarithmic UV divergences 
which allow us to derive the $\beta$-function for the coupling $\alpha_e(\mu)$. 
In addition, they are IR divergent when the photons between the jets go soft. However, these divergences 
just effectively correct $\alpha_e$ so that in the exponentiated one-loop result 
we obtain the low energy coupling, $\alpha_e(m_e)$ with $m_e$ being the mass of 
the hard lines -- in QED, the electron mass. Below the electron mass, $\alpha_e$ stops running.
Similarly, in diagram (III), we find a power-like divergence in the limit in which the photon between the two jets becomes 
soft. This, as usual, is reflecting the fact 
that the mass of internal propagator is not the physical (pole) mass.
Indeed, when we sum all the loop corrections to that internal propagator 
we shift the bare mass to the pole one and the power-like divergence disappears. 
Finally, diagram (IV), is associated to the wavefunction renormalization. Since we consider the external states to be the physical states, 
and thus only deal with amputated diagrams, we do not need to include it.

Keeping this in mind, for the rest of this paper, unless said otherwise, 
we will focus on diagrams where we have photon webs connecting different jets. 
In addition, in our IR analysis, we also assume that the UV divergences are taken 
care of by some renormalization procedure that does not affect the IR limits under consideration.

%% file: Figures/2loop.tex
\begin{equation*}
\begin{gathered}
    \begin{tikzpicture}[line width=0.8,scale=1,line cap=round]
        \coordinate (a) at (-0.5,0);
        \coordinate (b1) at (1,0.25);
        \coordinate (b2) at (1,-0.25);
        \coordinate (c) at (0.8,0);
        \coordinate (d) at (4,1.5);
        \coordinate (e) at (4,-1.5);
    
        \path (b1) -- (d) coordinate[pos=0.4] (p1t);
        \path (b2) -- (e) coordinate[pos=0.4] (p1b);
        \path (b1) -- (d) coordinate[pos=0.8] (p2t);
        \path (b2) -- (e) coordinate[pos=0.8] (p2b);
        \draw[] (b1) -- (d);
        \draw[] (b2) -- (e);

        \path (b1) -- (d) coordinate[pos=-0.18] (x);
        \path (b2) -- (e) coordinate[pos=-0.18] (y);
        \draw[] (x) -- (b1);
        \draw[] (y) -- (b2);
        \path (x) -- (y) coordinate[pos=0.5] (k);
        
        \draw[photon] (p1t) -- (p1b) node[midway,scale=0.7,xshift=-10] {$\gamma_1$};
        \draw[photon] (p2t) -- (p2b) node[midway,scale=0.7,xshift=-10] {$\gamma_2$};
        
        \node[scale=0.7,xshift=-30,yshift=3] at (p1t) {$\alpha_{1,1}$};
        \node[scale=0.7,xshift=-30,yshift=2] at (p2t) {$\alpha_{1,2}$};
    
        \node[scale=0.7,xshift=-30,yshift=-3] at (p1b) {$\alpha_{2,1}$};
        \node[scale=0.7,xshift=-30,yshift=-2] at (p2b) {$\alpha_{2,2}$};
     
        \filldraw[color=black,fill=gray] (k) circle[radius=0.1];
        \node[scale=0.65,xshift=-15,color=Maroon]  at (k) {$v_{h_0}$} ;
\node[scale=0.7,xshift=20,yshift=5] at (d) {$J=1$};
\filldraw[color=Maroon,fill=Maroon] (p1t) circle[radius=0.05] node[above,scale=0.65] {$v_{1,1}$};
\filldraw[color=Maroon,fill=Maroon] (p2t) circle[radius=0.05] node[above,scale=0.65] {$v_{1,2}$};
\filldraw[color=Maroon,fill=Maroon] (p1b) circle[radius=0.05] node[below,scale=0.65] {$v_{2,1}$};
\filldraw[color=Maroon,fill=Maroon] (p2b) circle[radius=0.05] node[below,scale=0.65] {$v_{2,2}$};
        \node[scale=0.7,xshift=20,yshift=-5] at (e) {$J=2$};
    \end{tikzpicture}
\end{gathered}
\hspace{2cm}
\begin{gathered}
    \begin{tikzpicture}[line width=0.8,scale=1,line cap=round]
        \coordinate (a) at (-0.5,0);
        \coordinate (b1) at (1,0.25);
        \coordinate (b2) at (1,-0.25);
        \coordinate (c) at (0.8,0);
        \coordinate (d) at (4,1.5);
        \coordinate (e) at (4,-1.5);
        \path (b1) -- (d) coordinate[pos=0.4] (p1t); 
        \path (b2) -- (e) coordinate[pos=0.4] (p1b);
        \path (b1) -- (d) coordinate[pos=0.8] (p2t);
        \path (b2) -- (e) coordinate[pos=0.8] (p2b);
        \draw[] (b1) -- (d);
        \draw[] (b2) -- (e);

        \path (b1) -- (d) coordinate[pos=-0.18] (x);
        \path (b2) -- (e) coordinate[pos=-0.18] (y);
        \draw[] (x) -- (b1);
        \draw[] (y) -- (b2);
        \path (x) -- (y) coordinate[pos=0.5] (k);
        
        \draw[photon] (p1t) -- (p2b) node[midway,scale=0.7,xshift=15,yshift=-5] {$\gamma_1$};
        \draw[photon] (p2t) -- (p1b) node[midway,scale=0.7,xshift=20,yshift=10] {$\gamma_2$};
        
        \node[scale=0.7,xshift=-30,yshift=3] at (p1t) {$\tilde{\alpha}_{1,1}$};
        \node[scale=0.7,xshift=-30,yshift=2] at (p2t) {$\tilde{\alpha}_{1,2}$};
    
        \node[scale=0.7,xshift=-30,yshift=-3] at (p1b) {$\tilde{\alpha}_{2,1}$};
        \node[scale=0.7,xshift=-30,yshift=-2] at (p2b) {$\tilde{\alpha}_{2,2}$};
     
        \filldraw[color=black,fill=gray] (k) circle[radius=0.1];
\node[scale=0.7,xshift=20,yshift=5] at (d) {$J=1$};
        \node[scale=0.7,xshift=20,yshift=-5] at (e) {$J=2$};
    \end{tikzpicture}
\end{gathered}
\end{equation*}

%% file: Figures/2loopRegions.tex
\begin{equation*}
    \begin{gathered}
    \begin{tikzpicture}[line width=1,draw=charcoal, scale=0.5]
    \coordinate (lx) at (-1.5,0) ;
    \coordinate (rx) at (10.5,0) ;
    \coordinate (ty) at (0,8.5) ;
    \coordinate (by) at (0,-1.5) ;
    \coordinate (c) at (0,0) ;
    \draw[-latex] (lx) -- (rx);
    \draw[-latex] (by) -- (ty);

    \draw[domain=5.5:9, smooth, samples=400, variable=\x, Maroon] plot ({\x},{-\x+9});  
    \draw[domain=1.5:5.5, smooth, samples=400, variable=\x, Maroon,dashed] plot ({\x},{-\x+9}); 
    
    \fill[fill=Maroon,,opacity=0.2] (9,0)--(1.5,7.5)--(9.5,7.5)--(9.5,0);
    
    \draw[domain=0:5.5, smooth, samples=400, variable=\y, Blue] plot ({\y}, {3.5});    
    \draw[domain=5.5:9.5, smooth, samples=400, variable=\y, Blue,dashed] plot ({\y}, {3.5}); 
    \fill[fill=Blue,,opacity=0.2] (0,3.5)--(0,7.5)--(9.5,7.5)--(9.5,3.5);   

    \node[] at (11,0) {$\tau_{J,1}$} ;
    \node[] at (0,9) {$\tau_{J,2}$} ;
    \draw[->,Blue,thick] (3.,3.5) -- (3.,4.5) ;
    \draw[->,Maroon,thick] (8,1) -- (8.6,1.6) ;
    \node[] at (1.5,5) {$D_{v_{[2]}}$} ;
    \node[] at (8.5,2.5) {$D_{v_{[1,2]}}$} ;
    \filldraw[color=gray,fill=gray] (5.5,3.5) circle[radius=0.15];
    \draw[->,gray,thick] (5.8,3.8) to[out=45,in=180] (12,6);
\end{tikzpicture}
    \end{gathered}
\hspace{0.5cm}
\begin{gathered}
\begin{tikzpicture}[line width=1,draw=charcoal, scale=0.5]
    \coordinate (lx) at (-2,0) ;
    \coordinate (rx) at (10.5,0) ;
    \coordinate (ty) at (0,6.5) ;
    \coordinate (by) at (0,-2) ;
    \coordinate (c) at (0,0) ;
    \draw[-latex] (lx) -- (rx);
    \draw[-latex] (by) -- (ty);

    \draw[domain=7:9, smooth, samples=400, variable=\x, Maroon] plot ({\x},{-\x+9});  
    \draw[domain=4:7, smooth, samples=400, variable=\x, gray,line width=1.5pt] plot ({\x},{-\x/2+5.5});
    
    
    \draw[domain=0:4, smooth, samples=400, variable=\y, Blue] plot ({\y}, {3.5});    

    \node[] at (11,0) {$\tau_{J,1}$} ;
    \node[] at (0,7) {$\tau_{J,2}$} ;
    \draw[->,Blue,thick] (2,3.5) -- (2,4.5) ;
    \draw[->,gray,thick] (5.5,2.75) -- (6,3.75) ;
    \draw[->,Maroon,thick] (8,1) -- (8.6,1.6) ;
    \node[above,Blue] at (2,4.5) {$r_{[2]}$};
    \node[right,Maroon] at (8.6,1.6) {$r_{[1,2]}$};
    \node[right,gray] at (6,4.25) {$r_{[2]}+r_{[1,2]}$};
    \filldraw[color=Blue,fill=gray] (4,3.5) circle[radius=0.1];
    \filldraw[color=Maroon,fill=gray] (7,2) circle[radius=0.1];
\end{tikzpicture}
  \end{gathered}
\end{equation*}

%% file: Sections/Schwinger_numerators.tex
\section{Diagram numerators in Schwinger representation}
\label{sec:numerators}

In this section we briefly review a general method to express any Feynman diagram, including its theory-dependent numerator, in Schwinger parameter space. We refer the reader to Appendix A of ref.~\cite{Hannesdottir:2022bmo} for a more detailed discussion. 

Starting in loop-momentum space we have the expression\footnote{Note that here we work in the conventions where the numerator is written solely in terms of internal momenta, which we can always do by momentum conservation.} 
\begin{equation} \label{eq:original_feynman_diagram}
    \I = \frac{1}{ (i \, \pi^{\D/2})^\ell}\int \prod_{e=1}^{\E} {\d^\D q_e} \, \frac{\mathcal{N}(q_e)}{\prod_{e=1}^{\E} \left(-q_e^2 + m_e^2\right)} \prod_{v=1}^{\V-1} \delta^{\D}\left( p_v + \sum \eta_{v,e} q_e \right)
\end{equation}
where $q_e$ are the momenta associated with the internal edges and $p_v$ is the external momentum entering each vertex, which is zero for all the internal vertices and non-zero for those connected to external edges. These momentum assignments are taken as independent and momentum conservation at each vertex of the graph is imposed by the delta functions. Note that we include $\V-1$ delta functions in the vertex momenta so that $\I$ uses the traditional convention of dropping an overall momentum conserving delta function. The incidence matrix $\eta_{e,v}$ takes the form given in \eqref{eq:incidenceMatrix}.
Finally, $\N$ is the numerator of the diagram arising from the Feynman rules of the corresponding theory.

In order to obtain a parametric representation for \cref{eq:original_feynman_diagram}, we start by using the identities\footnote{In \cref{eq:Schwinger_ids} we reintroduced the Feynman prescription $i\varepsilon$ to emphasise how it ensures convergence of the integral in $\alpha$ by providing a negative real part to the exponent.} 
\begin{equation}\label{eq:Schwinger_ids}
    \frac{1}{q^2 - m^2 + i\varepsilon} = -i \, \int_0^\infty \d \alpha \: e^{i (q^2 - m^2 + i\varepsilon)\alpha }, \quad\quad 
    \delta^D (P) = \int \frac{\d ^D x}{(2\pi)^D} \: e^{i x \cdot P} 
\end{equation}
to cast \cref{eq:original_feynman_diagram} into the form~\cite{Hannesdottir:2022bmo}
\begin{equation}
\begin{aligned}
     \I &= \frac{i^{E-\ell}}{\pi^{\ell\D/2}}   \int_{q_e \in \mathbb{R}^\D} \int_{\alpha_e > 0}   \prod_{e=1}^{\E} \d^{\D} q_e \, d \alpha_e \, \prod_{v=1}^{\V-1}  \int_{x_v \in \mathbb{R}^\D} \frac{\d^{\D} x_v}{(2\pi)^\D}  \, \N(q_e) \, \\
     & \quad \times \exp{\left\{ \sum_{e=1}^{\E} i\left[(q_e^2-m_e^2) \alpha_e +  \sum_{v=1}^{\V-1} x_v \cdot (\eta_{v,e} q_e)\right] +i \sum_{v=1}^{\V-1} x_v \cdot p_v  \right\}}.
\end{aligned}
\end{equation}
where $\alpha_e$ is the Schwinger parameter associated to the internal edge with momentum $q_e$.
We further express the numerator in the exponentiated form
\begin{equation}
   \mathcal{N}(q_e) = \mathcal{N}\left(  -i \frac{\partial}{\partial z_{e}}\right) \prod_{e} \exp{\left[ i q_e \cdot z_{e}\right]}\bigg\vert_{z_{e}=0}
\end{equation}
where the $z_e$ are auxiliary Schwinger parameters which are set to zero at the end and the differential 
operator $\mathcal{N}\left(  -i\partial_{z_{e}}\right)$ is obtained from $\mathcal{N}(q_e)$ 
by replacing $q_e^\mu \to (-i) \frac{\partial}{\partial {z_e}_\mu} $~\cite{Hannesdottir:2022bmo}.\\   
This leads to
\begin{equation}
\begin{aligned}
     \I &=
     \frac{i^{E-\ell}}{\pi^{\ell\D/2}}  \int_{q_e \in \mathbb{R}^\D} \int_{\alpha_e > 0} \d^{\D} q_e \,   d \alpha_e \, \prod_{v=1}^{\V-1} \int_{x_v \in \mathbb{R}^\D} \frac{\d^{\D} x_v}{(2\pi)^\D}  \, \mathcal{N}\left(-i\partial_{z_{e}}\right)  \, \\
     & \quad \times \exp{\left\{i \sum_{e=1}^{\E} \left[q_e^2 \alpha_e + q_e \cdot\left(\sum_{v=1}^{\V-1} \eta_{v,e} x_v -i z_{e}\right) -m_e^2\alpha_e \right] +i \sum_{v=1}^{\V-1} x_v \cdot p_v   \right\}}.
\end{aligned}
\end{equation}
At this point we introduce two important quantities: the \emph{Laplacian matrix} and the \emph{weighted incidence matrix}. The Laplacian matrix, $\overline{\LL}$, was already defined in \cref{sec:2loopMatrix} and takes the form given in \cref{eq:LapMatrix}. Once again, this matrix is not invertible, and so we will deal directly with the \emph{reduced Laplacian matrix}, $\LL$, which is obtained by removing a vertex from $\overline{\LL}$ -- for convenience we choose this vertex to be part of the hard diagram (not connected to any soft edge).
The {weighted incidence} matrix $\Mb$ lets us incorporate the numerators into the Schwinger representation of the integral and it is a $(\V{-}1) \times \E$ matrix defined as
\begin{equation}
    (\MM)_{ve} = \frac{\eta_{ve}}{\alpha_e} \,
\end{equation}
where the omitted vertex has to be chosen consistently with the definition of the reduced Laplacian.

Finally, performing the Gaussian integral over the $q_e$ we are left with a Gaussian integral in the $x_v$ that evaluates to the following final expression
\begin{equation}
\begin{aligned}
\I= i^{E-\ell} \int_{\alpha_e > 0} & \frac{\prod_e \mathrm{d} \alpha_e}{\left(\prod_{e} \alpha_e\right)^{\mathrm{D} / 2}\left(\operatorname{det} \mathbf{L}\right)^{\mathrm{D} / 2}} \times \mathcal{N}\left(-i\partial_{z_{e}}\right) \exp \left(i\mathcal{V}(\zz) \right),
\end{aligned}
\end{equation}
where we introduced the \emph{modified worldline action} 
\begin{equation}\label{eq:full_modified_action}
\begin{aligned}
\mathcal{V}(\zz) \equiv  \left(\pp + \frac{i}{2} \MM \zz \right)^\top \mathbf{L}^{-1}\left(\pp + \frac{i}{2} \MM \zz \right)  + \sum_{e=1}^{\E} \frac{\zz_{e}^2}{4 \alpha_e}-\sum_{e=1}^{\E} \left(m_e^2\right) \alpha_e,
\end{aligned}
\end{equation}
The standard worldline action $\mathcal{V}$ introduced in \cref{sec:2loopMatrix} can then be obtained as
\begin{equation}\label{eq:full_modified_action_atzero}
\begin{aligned}
\mathcal{V} \equiv \mathcal{V}(\zz=\mathbf{0}) =  \pp ^\top \mathbf{L}^{-1}\pp -\sum_{e=1}^{\E} m_e^2 \alpha_e  
\end{aligned}
\end{equation}
and the first and second Symanzik polynomials are given by \eqref{eq:matrixSymanzik}.
Indeed, we see that the exponential can be split in a $z$-independent part ($\mathcal{V} = \mathcal{F}/\mathcal{U}$) and a $z$-dependent one as
\begin{equation}\label{eq:I_expression_derivatives}
\begin{aligned}
\I&= i^{E-\ell}\int_{\alpha_e > 0} \frac{\prod_e \d \alpha_e}{\U^{\D / 2}} \exp \left(\frac{i\F }{\U} \right) \; \\
& \quad \times \mathcal{N}\left( -i \partial_{z_{e}}\right) \; \exp \left\{i\left[  i \pp^{\top} \LL^{-1} \MM \, \zz - \frac{1}{4} \zz^{\top} \MM^{\top} \LL^{-1} \MM \, \zz\, +  \sum_{e=1}^{\E} \frac{z_{e}^2}{4 \alpha_e} \right] \right\}  \Bigg\vert_{\zz = \mathbf{0}} 
\end{aligned}  
\end{equation}
so that an integral without any numerator corresponds to $\mathcal{N}=1$, for which we can trivially set $\zz=\mathbf{0}$.
Whenever $\mathcal{N}$ is not $1$, we find it convenient to split its terms into a $z$-independent tensor and a product of derivatives. Without loss of generality we focus on a numerator of the form 
\begin{equation}
    \mathcal{N} = T^{\mu_1\dots \mu_n} \; \frac{\partial}{\partial z_{{e_1}}^{\mu_1}} \dots \frac{\partial}{\partial z_{{e_n}}^{\mu_n}} ,
\end{equation}
where $T$ is a tensor independent of the Schwinger parameters. In general the numerator of a Feynman diagram can involve a linear combination of terms of this type.
We can then rewrite the diagram in the form 
\begin{equation}\label{eq:I_expression_Wick}
\begin{aligned}
\I= i^{E-\ell} \int_{\alpha_e > 0} & \frac{\prod_e \d \alpha_e}{\U^{\D / 2}} \exp \left(\frac{i \F}{\U} \right) T^{\mu_1\dots \mu_n} \Big\langle 
    \partial_{z_{{e_1}}^{\mu_1}} & \dots  \partial_{z_{{e_n}}^{\mu_n}} \Big\rangle ,
\end{aligned}  
\end{equation}
where we defined the quantity
\begin{equation}
\begin{aligned}
    \Big\langle 
    \partial_{z_{{e_1}}^{\mu_1}} & \dots  \partial_{z_{{e_n}}^{\mu_n}} \Big\rangle \equiv \\
    &
    \frac{\partial}{\partial z_{{e_1}}^{\mu_1}} \dots \frac{\partial}{\partial z_{{e_n}}^{\mu_n}}
    \;
    \exp \left\{i\left[  i \pp^{\top} \LL^{-1} \MM \, \zz - \frac{1}{4} \zz^{\top} \MM^{\top} \LL^{-1} \MM \, \zz\, +  \sum_{e=1}^{\E} \frac{z_{e}^2}{4 \alpha_e} \right] \right\}  \Bigg\vert_{\zz = \mathbf{0}},
\end{aligned}
\end{equation}
which can then be computed recursively as
\begin{equation}\label{eq:inductive_num}
    \Big\langle 
    \partial_{z_{{e_1}}^{\mu_1}}  \dots  \partial_{z_{{e_n}}^{\mu_n}} \Big\rangle = 
    \Delta_{e_1}^{\mu_1} 
    \Big\langle 
    \partial_{z_{{e_2}}^{\mu_2}}  \dots  \partial_{z_{{e_n}}^{\mu_n}} \Big\rangle + 
     \sum_{i=2,n} \Delta_{e_1 e_i}^{\mu_1 \mu_i}
    \Big\langle 
    \partial_{z_{{e_2}}^{\mu_2}}  \dots  
    \cancel{\partial_{z_{{e_i}}^{\mu_i}}}  \dots 
    \partial_{z_{{e_n}}^{\mu_n}} \Big\rangle,
\end{equation}
with the one- and two-point ``Wick" contractions
\begin{equation} \label{eq:num_wick_contractions}
    \Delta_{e}^{\mu} = - (\MM \LL^{-1} \pp^\mu)_e, \quad\quad 
    \Delta_{e f}^{\mu \nu} = \frac{i}{2} \left[ \left(\MM \LL^{-1} \MM  \right)_{ef}  + \frac{\delta_{ef}}{\alpha_e}\right] \, g_{\mu\nu} .
\end{equation}
Eqs.~\eqref{eq:inductive_num} and~\eqref{eq:num_wick_contractions} provide a particularly convenient way of computing the Schwinger-parameter version of Feynman diagram numerators recursively. 
Plugging this into \cref{eq:I_expression_Wick}, we find the final expression
\begin{equation}\label{eq:I_expression_Rational}
\begin{aligned}
\I= i^{E-\ell} \int_{\alpha_e > 0} & \frac{\prod_e \d \alpha_e}{\U^{\D / 2}} e^{i \mathcal{V} } \N(\alpha) ,
\end{aligned}  
\end{equation}
with the shorthand
\begin{equation}
    \N(\alpha) = T^{\mu_1\dots \mu_n} \Big\langle 
    \partial_{z_{{e_1}}^{\mu_1}}  \dots  \partial_{z_{{e_n}}^{\mu_n}} \Big\rangle .
\end{equation}
We point out that the quantity $\N(\alpha)$ is a rational function of the Schwinger parameters whose coefficients can only depend on the external momenta, polarisation vectors and space-time dimensions $\D$. The $\alpha$-dependent part of $\N$ can then be interpreted as a linear combination of scalar integrals, so that \cref{eq:I_expression_Rational} automatically yields a ``tensor-reduced'' representation of the diagram numerator. 

%% file: Sections/Factorisation.tex
\section{Integrand factorisation}
\label{sec:factorisation_matrix}

\begin{figure}[H]
    \centering
    \begin{subfigure}[t]{0.5\textwidth}
        \centering
        \includegraphics[width=\textwidth]{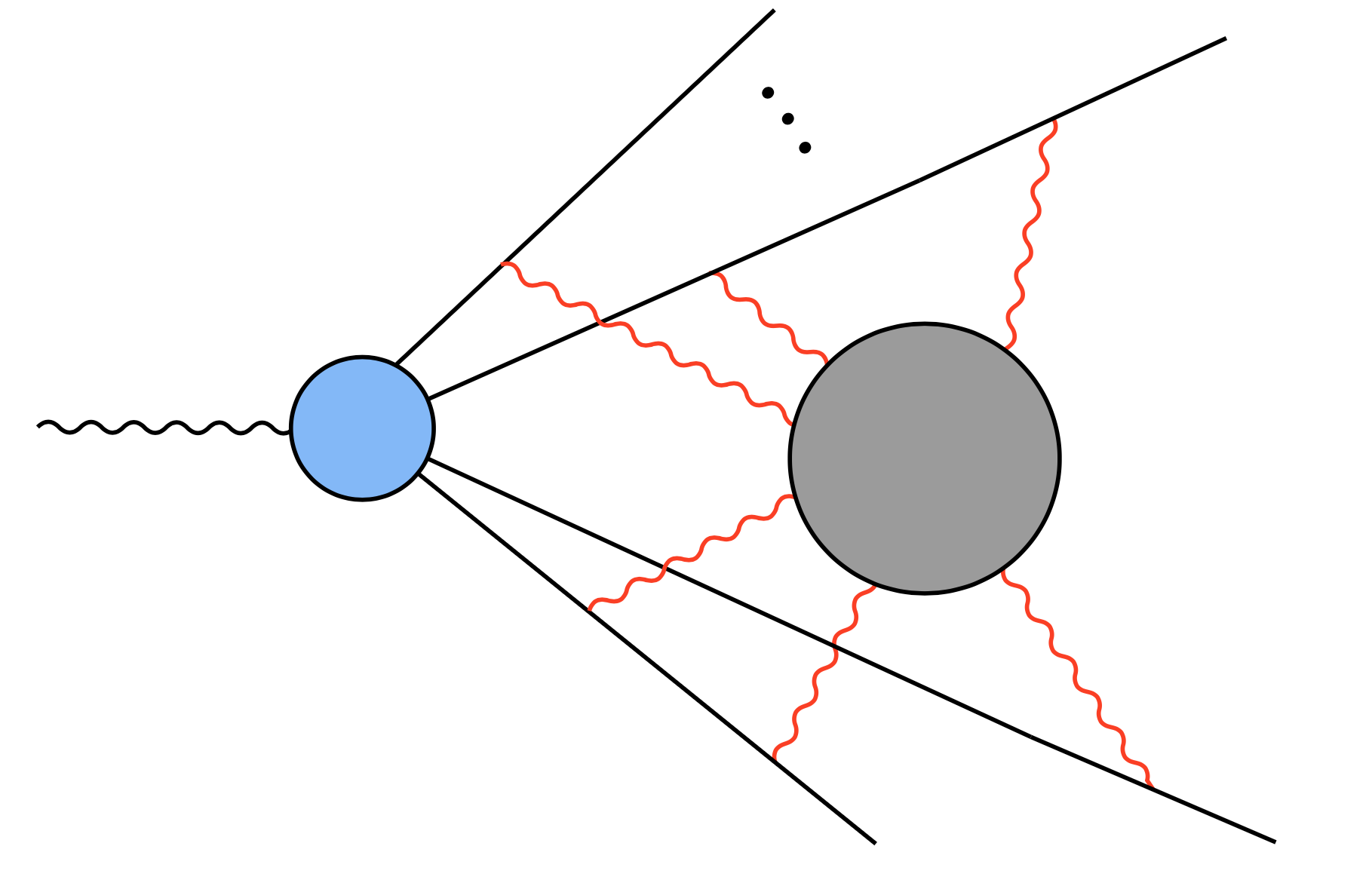}
        \caption{\label{fig:diagram_general_form}}
    \end{subfigure}%
    ~ 
    \begin{subfigure}[t]{0.5\textwidth}
        \centering
        \includegraphics[width=\textwidth]{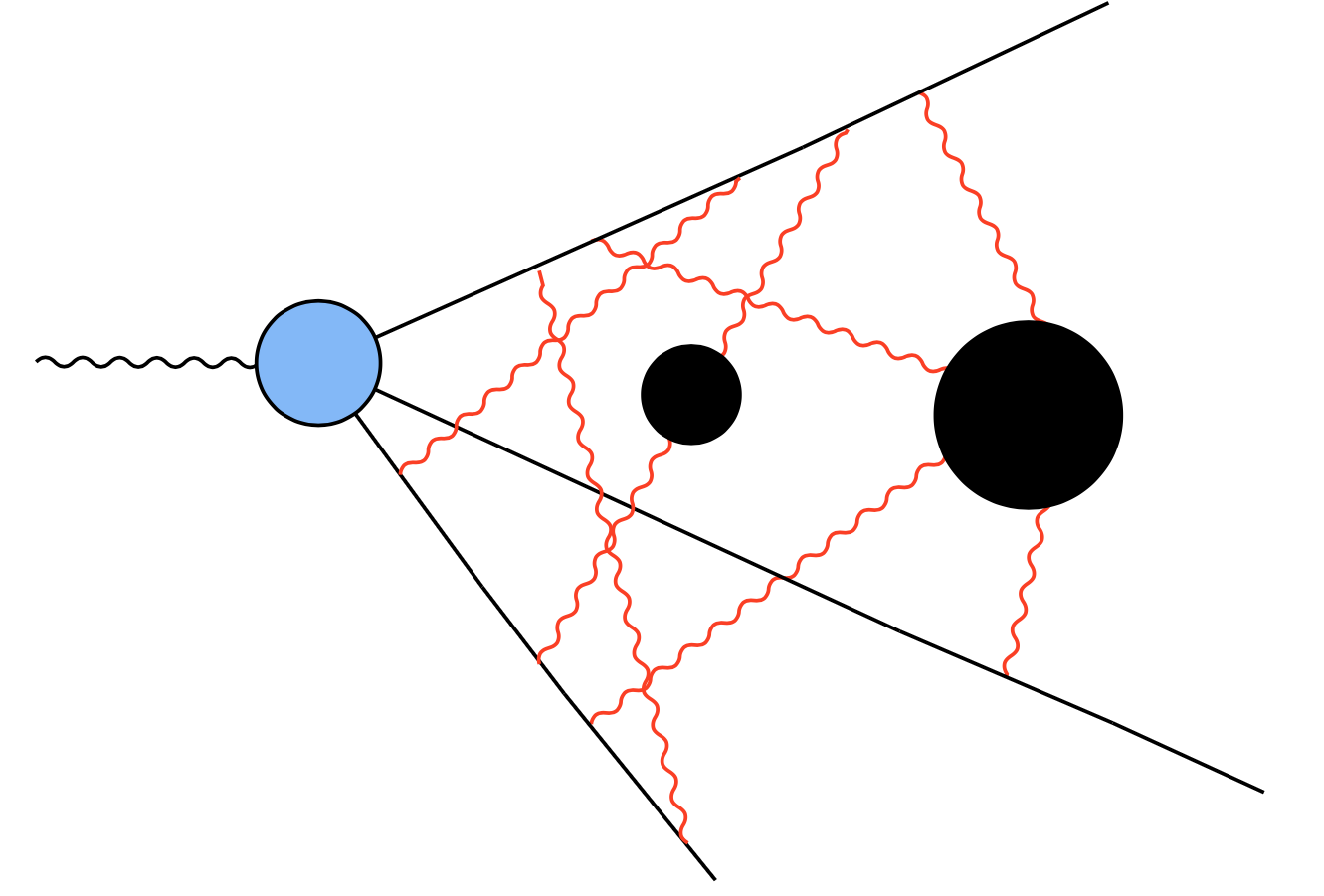}
        \caption{\label{fig:diagram_general_form_1}}
    \end{subfigure}
    \caption{(a) General form of the diagrams considered here. Red wavy lines are 
        massless and their Schwinger parameters scale as $\lambda^{-2}$, while black solid lines are massive
        and scale as $\lambda^{-1}$. The blue blob is any unscaled (hard) subdiagram. 
        The gray blob corresponds to any subdiagram of either disconnected massless lines or unscaled propagators.
        (b) Representative example of a type of diagram included in (a). The black blobs (which we will simply refer to as blobs in the rest of this 
        section) are connected subdiagrams
        of unscaled propagators.
        \label{fig:diagram_general_form_both}
        }
\end{figure}

In \cref{sec:two_loop} we showed how one- and two-loop scalar ladder diagrams 
factorise when approximated along their IR divergent rays. We 
also showed how the integrand of the full amplitude can be reorganised to 
make exponentiation of the one-loop IR divergence manifest. 
The goal of this section is to prove factorisation of the much wider class of diagrams and scalings illustrated \cref{fig:diagram_general_form} to all loops.
In the figure, the black solid lines correspond to massive lines, whose Schwinger parameters 
scale as $\lambda^{-1}$ (on-shell), while the red lines are massless and their 
Schwinger parameters scale as  $\lambda^{-2}$ (soft). 
From this point onwards we will use the shorthand \emph{jet} to refer to each collection of connected lines scaling as $\lambda^{-1}$. For instance \cref{fig:diagram_general_form_1} contains three jets.

The tropical rays corresponding to the scalings 
we consider here have entries which are either $0$, $1$ or $2$ and concretely have the form 
\begin{equation}
    r = (\underbrace{0,0,\dots,0}_{\substack{\text{all edges in}\\ \text{the hard part}}}, \underbrace{1,1,1,\dots,1,1}_{\substack{\text{massive edges} \\ \text{in jets}}}, \underbrace{2,2,\dots,2}_{\substack{\text{massless edges}\\ \text{connecting to jets}}},
    \underbrace{0,0,\dots,0}_{\substack{\text{all edges in}\\ \text{the blobs}}} ),
\end{equation}
which is defined in accordance with the (partial) edge ordering given below in \cref{eq:edge_ordering}.
The blue and gray blobs correspond to any subdiagrams whose edges are unscaled, or in the case of the gray blobs, they can also include massless edges, scaled as $\lambda^{-2}$. 
Concretely, the gray subdiagram can consist of many disconnected pieces, 
some of which are trivial (disconnected massless soft edges) while others contain a connected unscaled subgraph and 
are represented by black blobs, see \cref{fig:diagram_general_form_1} for example. 
Whenever a black connected subdiagram is encountered we refer to it simply as \emph{blob}.
We will instead use the name \emph{hard subdiagram} for the blue blob. We will
also refer to the collection of scaled edges and (black) blobs as \emph{soft subdiagram}.
An explicit example of such a diagram is given in \cref{fig:diagram_example}. 

Below will show that, on the rays defined above, the integrands of this class of 
Feynman diagrams factorises at leading order in $\lambda$ as 
\begin{equation}\label{eq:fact_1}
    \dd \mathcal{I} \simeq  \dd \mathcal{I}_H \times \dd \mathcal{I}_S ,
\end{equation}
where $\dd \mathcal{I}_H $ and $\dd \mathcal{I}_S $ are interpreted as the 
integrands of the hard subdiagram and of the soft subdiagram. We will then 
proceed to prove a further factorisation property of the soft integrand, which splits
into the product of simpler soft integrands 
related to the various connected components $\mathcal{W}$ (such as the ones in \cref{fig:diagram_general_form_1}) of the soft subgraph 
\begin{equation} \label{eq:fact_2}
    \dd \mathcal{I}_S \simeq \prod_\mathcal{W} \dd \mathcal{I}_\mathcal{W} .
\end{equation}
Further, we will prove all these properties hold at leading order in the ray approximation
not only for the scalar integrals but also for their generalised version defined in \cref{sec:numerators},
which contains information about the numerators of these diagrams.

All the information needed to prove the factorisation properties above is contained in the asymptotic structure of the $\LL^{-1}$ and $\Mb$ matrices of \cref{sec:numerators}, which in turn will let us derive the factorisation of the generalised worldline action $\mathcal{V}(\boldsymbol{z}) \simeq \mathcal{V}_H(\boldsymbol{z}_H)+\mathcal{V}_S(\boldsymbol{z}_S)$. 
Connecting the approach of this section to the diagrammatic one from \cref{sec:fact2loop} relies on the \emph{matrix-tree theorem}, see $e.g.$ ref.~\cite{Weinzierl:2022eaz}. While the two representations are mathematically equivalent for scalar integrals, we find the matrix representation to give a more transparent way to include numerators, generic hard diagrams and blobs in our analysis.

This section is organised as follows. In \cref{sec:Laplacian_structure} we organise the entries 
of the graph Laplacian matrix to highlight the structure and scalings of the diagrams.
In \cref{sec:factorisation_U} we prove the factorisation of the $\U$ polynomial. 
In \cref{sec:factorisation_Nu} we do the same for the generalised worldline action $\mathcal{V}(\boldsymbol{z})$. Finally, in \cref{sec:integrand_factorisation} we collect our results and show how they imply the factorisation properties of the integrand anticipated in  \cref{eq:fact_1,eq:fact_2}. We also comment on the form of simpler ladder diagram (where no blobs are present), which will be important for the proof of exponentiation in \cref{sec:regions}. We present multiple examples of the matrices entering in this section, as well as their respective expansion in the IR scalings, in the example {\tt Mathematica} notebook.

\paragraph{Edge/propagator naming conventions}
Before we proceed let us define a naming convention for the Schwinger parameters of the jet edges in the graphs 
we are studying. Labelling the external massive lines $J_i$ with $i=1,2,\dots$ we name the parameters of the massive propagators along $J_i$ as $\alpha_{i,j}$, with $j$ starting from $1$ at the innermost edge:
$$
\begin{tikzpicture}[line width=0.8,scale=1.2,line cap=round]
        \coordinate (A) at (-2,0);
        \coordinate (a) at (-1,0);
        \coordinate (b) at (0,0);
        \coordinate (c) at (1,0);
        \coordinate (d) at (3,0);
        \coordinate (e) at (4.3,0);
        \coordinate (B) at (5,0);
        \coordinate (a1) at (-1,-1);
        \coordinate (b1) at (0,-1);
        \coordinate (c1) at (1,-1);
        \coordinate (d1) at (3,-1);
        \coordinate (e1) at (4.3,-1);
        \draw[] (A) -- (B);
        \draw[photon,red] (a) -- (a1);
        \draw[photon,red] (b) -- (b1);
        \draw[photon,red] (c) -- (c1);
        \draw[photon,red] (d) -- (d1);
        \draw[photon,red] (e) -- (e1);
        \node[scale=0.9,xshift=-12,yshift=7] at (a) {$\alpha_{i,1}$};
        \node[scale=0.9,xshift=-12,yshift=7] at (b) {$\alpha_{i,2}$};
        \node[scale=0.9,xshift=-12,yshift=7] at (c) {$\alpha_{i,3}$};
        \node[scale=0.9,xshift=-20,yshift=7] at (e) {$\alpha_{i,l_i}$};
        \node[scale=0.9,xshift=12,yshift=0] at (B) {$J_i$};
        \fill (1.5,-0.5) circle [radius=0.03] (2,-0.5) circle [radius=0.03] (2.5,-0.5) circle [radius=0.03];
        \fill (A)  circle [radius=0.1];
        \node[scale=0.9,xshift=-12,yshift=0] at (A) {$H$};        
\end{tikzpicture}
$$
where $l_i$ is the number of edges on that line.
In the following we will not need to refer explicitly to the parameters of the soft massless edges or to those of the unscaled edges so 
we will not fix any specific convention here.

\subsection{Structure of the Laplacian matrix}
\label{sec:Laplacian_structure}

\begin{figure}[H]
    \centering
    \includegraphics[width=\linewidth]{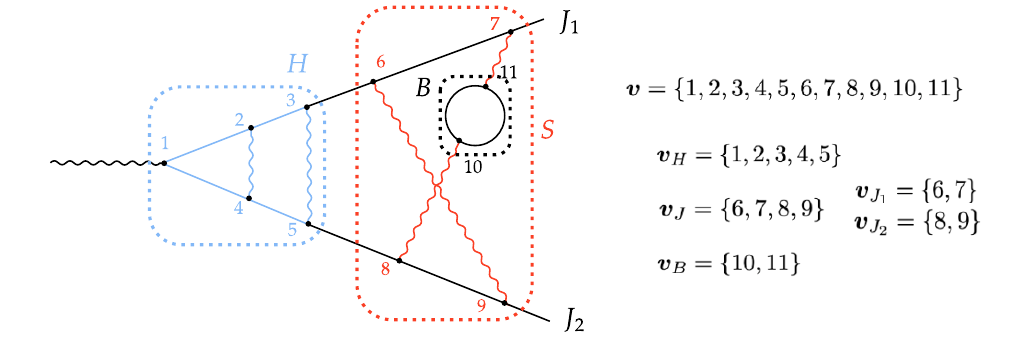}
    \caption{Simple example of a diagram of the general form  considered in this section. On the right, we present the vertex vector, which gives an ordering to the vertices and is based on which we define the graph matrices $\Hb, \, \Sb , \, \Jb , \, \Bb$, etc.}
    \label{fig:diagram_example}
\end{figure}
In order to study the structure of the Laplacian $\overline{\Lb}$ in the limit described above, we begin by defining the following ordering among vertices:
\begin{equation}
        \boldsymbol{v} =  \{ \boldsymbol{v}_H \, | \boldsymbol{v}_J \, | 
        \boldsymbol{v}_B \} \,,
    \label{eq:vertexvector}
\end{equation}
where here and in the rest of this section we use the symbol ``$|$'' to separate the columns of different blocks of a vector or matrix. Above the three lists on the r.h.s.~are respectively the vertices in the hard diagram, along the jet lines and in the blobs. 
We can then further order and split the jet and blob lists of vertices as follows:
\begin{equation}
    \begin{aligned}
        \boldsymbol{v}_J &= \{
        \underbracket[0.4pt]{v_{J_1,1},v_{J_1,2},\ldots \, }_{\substack{\text{ vertices in } \\ \text{ first jet }}}| \, \,
        \ldots \, |\,\,
        \underbracket[0.4pt]{v_{J_n,1},v_{J_n,2},\ldots \, }_{\substack{\text{ vertices in } \\ \text{ $n$-th jet }}}
        \} \,, \\
        \boldsymbol{v}_B &= \{
        \underbracket[0.4pt]{v_{B_1,1},v_{B_1,2},\ldots \, }_{\substack{\text{ vertices in } \\ \text{ first blob }}}| \, \,
        \ldots \, |\,\,
        \underbracket[0.4pt]{v_{B_n,1},v_{B_n,2},\ldots \, }_{\substack{\text{ vertices in } \\ \text{ $n_B$-th blob }}}
        \} \,, \\
    \end{aligned}
\end{equation}
with the vertices in each jet ordered from the inside out and no particular ordering among vertices in the each blob. For an example of vertex ordering see \cref{fig:diagram_example}.

With these conventions in mind we can split the reduced Laplacian matrix\footnote{To define the reduced Laplacian from the full Laplacian, we use the conventions that a vertex from $\boldsymbol{v}_H$ is removed, $e.g.$ in~\cref{fig:diagram_example} we would remove the vertex labelled with 1.} $\Lb$ into various blocks 
associated with vertices of the soft and hard subdiagrams:
\begin{equation} \label{eq:LLmatrix}
 \LL = 
 \begin{pmatrix}
 	\Hb & \Cb \\
 	\Cb^T & \Sb
 \end{pmatrix} \,.
\end{equation}
Above, $\Hb$ connects vertices of the hard subdiagram to themselves 
and $\Sb$ does the same for the soft subdiagram vertices, see 
\cref{fig:diagram_example}. 
$\Cb$ and its transpose instead are generally sparse matrices connecting hard to soft vertices. 
We will not make any assumptions about the matrix $\Hb$ ($i.e.$ on the 
shape or type of hard diagram) other than it is invertible and that its 
components are of order $\lambda^0$. 
The structure of $\Cb$ is fixed by the general form of diagram defined in \cref{fig:diagram_general_form}
and can be written in the form
\begin{equation}\label{eq:C_form}
 \Cb = 
 \left(
\begin{array}{c|c}
    \Cb_{J} & \boldsymbol{0} 
 \end{array}
 \right)
 =
\left(
\begin{array}{c|c|c|c}
    \Cb_{J_1} & \dots  & \Cb_{J_n} & \boldsymbol{0} 
 \end{array}
 \right), \quad\quad
 \text{with} \quad \quad    
 \left(\Cb_{J_i}\right)_{hj} =
 \ab_{i,1} \delta_{j1} \delta_{h h_{J_i}},
\end{equation}
where the hard vertex $h_{J_i}$ is the one where the jet $J_i$ is attached to the hard subdiagram. 
The block of zeros in this matrix corresponds to the blob(s), if present.
The structure above can be understood as follows. 
The $i$-th jet is connected to the hard diagram by a single edge 
(with associated parameter $ \alpha_{i,1}$) which, following the vertex ordering 
in \cref{fig:diagram_example} connects the first vertex of the jet to some vertex
in the hard subdiagram (two or more jets may connect to the same vertex). 
As a consequence, all entries in each $\Cb_{J_i}$ are identically zero, except for 
the ones corresponding to the vertices that directly connect the soft and hard diagrams.

\begin{figure}
    \centering
    \includegraphics[width=0.95\textwidth]{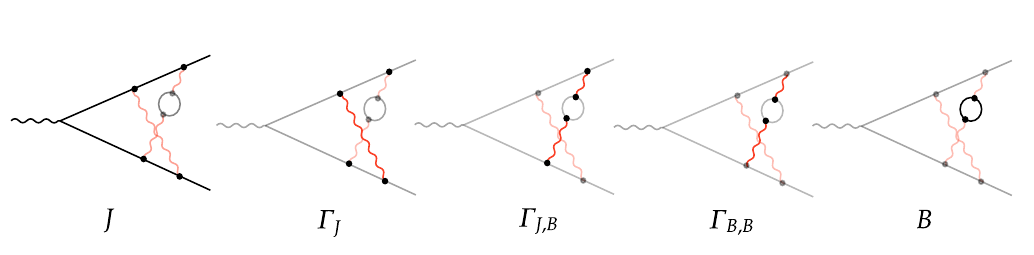}
    \caption{The soft Laplacian $\Sb$ can be split into various components corresponding to different subsets of vertices. 
    In this figure it is shown which edges and vertices of the soft diagram enter the different building blocks. 
    For each matrix all unrelated edges and vertices are shaded.
    The unshaded edges appear in the entries of the corresponding matrix, while the unshaded vertices determine its indices. The leftmost vertex is omitted as we assume it has been removed when defining the reduced Laplacian.
    \label{fig:matrix_examples} }
\end{figure}

To understand the structure of $\Sb$, we further split its entries in the following way:
\begin{equation}\label{eq:S_form}
    \Sb = 
   \begin{pmatrix}
       \Jb + \boldsymbol{\Gamma}_{J}   &  \boldsymbol{\Gamma}_{J,B} \\
       \boldsymbol{\Gamma}_{J,B}^\top &   \Bb + \boldsymbol{\Gamma}_{B}
   \end{pmatrix}
\end{equation}
As we will see shortly, this form clearly separates the components of $\Sb$ with different scalings.
\begin{figure}[!t]
    \centering
    \includegraphics[width=0.8\textwidth]{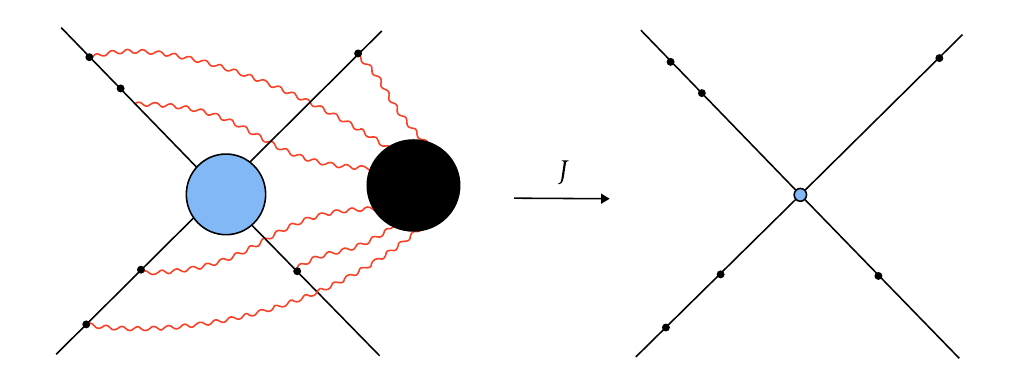}
    \caption{Graphical representation of the $\Jb$ matrix, which only contains information
    about the massive lines scaling as $\lambda^{-1}$. The diagonal blocks of $\Jb$ correspond
    to the individual lines stemming out of the hard diagram.
     \label{fig:Jmat} }
\end{figure}
The matrix $\Jb$ corresponds to the Laplacian of the soft graph where all edges except those 
that scale as $\lambda^{-1}$ have been deleted (\cref{fig:Jmat}), 
while $\Bb$ is the Laplacian of the collection of blob graphs where all edges except 
the unscaled ones have been deleted. 
The matrices $\boldsymbol{\Gamma}$ ($\boldsymbol{\Gamma}_{J}, \, \boldsymbol{\Gamma}_{J,B}$ and $\boldsymbol{\Gamma}_{B}$) 
contain the edges that scale as $\lambda^{-2}$, $e.g.$ the photons in QED and gluons in QCD. 
Figure~\ref{fig:matrix_examples} shows on which edges the various blocks of $\Sb$ depend on and which vertices they connect (for the soft subdiagram of the
example in \cref{fig:diagram_example}).
The various components therefore scale homogeneously as 
    \begin{equation}\label{eq:soft_graph_scalings}
        \Jb \xrightarrow{v} \lambda \Jb, \quad
        \Bb \xrightarrow{v}  \Bb  , \quad
        \boldsymbol{\Gamma} \xrightarrow{v} \lambda^2 \boldsymbol{\Gamma} .
    \end{equation}
Looking more in detail at the matrix $\Jb$, we find it is block-diagonal
\begin{equation} \label{eq:S_J}
    \Jb = 
   \begin{pmatrix}
       \Jb_{1}   & \cdots & 0 \\
       \vdots &  \ddots & \vdots  \\
       0 & \cdots & \Jb_{n} 
   \end{pmatrix},
\end{equation}
where the matrices on the diagonal correspond to the various branches or jets of \cref{fig:Jmat} and  
are given by tridiagonal matrices with the entries
\begin{equation} \label{eq:LJ_form}
    \Jb_{i} = 
    \begingroup
   \renewcommand*{\arraystretch}{1.6}
   \begin{pmatrix}
    \ab_{i,1} \!+\! \ab_{i,2} & -\ab_{i,2}  &  &  &   \\
    -\ab_{i,2} & \ab_{i,2} \!+\! \ab_{i,3} & -\ab_{i,3} &  & \\
        & -\ab_{i,3} & \ddots & \ddots &  \\
        &  & \ddots & \ab_{i,l_i-1} \!+\! \ab_{i,l_i-1} & -\ab_{i,l_i}\\
        &  &  &  -\ab_{i,l_i} & \ab_{i,l_i} \\
   \end{pmatrix},
   \endgroup
\end{equation}
where again $l_i$ is the number of soft photons attached to jet $J_i$.  In the case of jets with a single fermion propagator ($i.e.$ $l_i = 1$) the equation above reduces to $\Jb_i = \alpha^{-1}_{i,1}$.
Using \cref{eq:S_J,eq:LJ_form} we find
\begin{equation} \label{eq:SandJ_inv}
        \Jb^{-1} = 
       \begin{pmatrix}
           \Jb_{1}^{-1}   & \cdots & 0 \\
           \vdots &  \ddots & \vdots  \\
           0 & \cdots & \Jb_{n}^{-1} 
       \end{pmatrix},
       \quad\quad
       (\Jb_{i}^{-1})_{v,v^\prime} = 
        \begingroup
        \renewcommand*{\arraystretch}{1}
        \begin{pmatrix}
            \beta_{i,1} & \beta_{i,1}  & \beta_{i,1} & \cdots &        \\
            \beta_{i,1} & \beta_{i,2} & \beta_{i,2} & \beta_{i,2} & \cdots \\
            \beta_{i,1} & \beta_{i,2} & \beta_{i,3} & \cdots  &   \\
            \vdots  & \beta_{i,2} &  \vdots & \ddots  &  \\
                    & \vdots  &         &         & \beta_{i,l_i} 
        \end{pmatrix}
        = \beta_{i,\text{min}(v,v^\prime)} \sim \frac{1}{\lambda} \,,
        \endgroup
\end{equation}
where $\beta_{i,k} = \alpha_{i,1} + \dots + \alpha_{i,k}$.
That is, each $\beta_{i,k}$ represents the \textit{worldline distance} from the hard scattering 
to edge $k$ in jet $i$, precisely mirroring the $\beta$'s we got in the diagrammatic formalism in~\cref{sec:two_loop}. Therefore we see for a very general set of diagrams how the Laplacian formalism naturally suggests the worldline variables to describe the jet edges. 

The matrix $\Bb$ can also be written in the block diagonal form
\begin{equation} \label{eq:B_form}
    \Bb = 
   \begin{pmatrix}
       \Bb_{1}   & \cdots & 0 \\
       \vdots &  \ddots & \vdots  \\
       0 & \cdots & \Bb_{n_B} 
   \end{pmatrix},
\end{equation}
where each $\Bb_i$ is the Laplacian matrix of the $i$-th blob with all external photons deleted and $n_B$ 
the total number of blobs.

\subsubsection*{Structure of the weighted incidence matrix}
In addition to the Laplacian $\Lb$ we can also split the components of the weighted incidence matrix $\Mb$ according to the vertex ordering defined above. 
Recall that $\Mb$ is a $(V-1) \times E$ matrix connecting vertices to edges ($\Mb \to (\Mb)_{ve}$) and so, to define it, on top of the vertex vector (given in \cref{eq:vertexvector}), we also need to specify an edge vector $\boldsymbol{e}$, which we order as follows 
\begin{equation}
\begin{aligned}
    \boldsymbol{e} &= 
    \{ \boldsymbol{e}_H \, | \boldsymbol{e}_J \, |  \boldsymbol{e}_S \, | \boldsymbol{e}_B \} \,,  \\
    \boldsymbol{e}_J &= 
    \{
    \underbracket[0.4pt]{e_{J_1,1},e_{J_1,2},\ldots \, }_{\substack{\text{ edges in } \\ \text{ first jet }}}| \, \,
    \ldots \, |\,\,
    \underbracket[0.4pt]{e_{J_n,1},e_{J_n,2},\ldots \, }_{\substack{\text{ edges in } \\ \text{ $n$-th jet }}}
    \} \\
    \boldsymbol{e}_B &= 
     \{
    \underbracket[0.4pt]{e_{B_1,1},e_{B_1,2},\ldots \, }_{\substack{\text{ edges in } \\ \text{ first blob }}}| \, \,
    \ldots \, |\,\,
    \underbracket[0.4pt]{e_{B_{n_B},1},e_{B_{n_B},2},\ldots \, }_{\substack{\text{ edges in } \\ \text{ $n_B$-th blob }}}
    \} \,,
\end{aligned}
\label{eq:edge_ordering}
\end{equation}
where $\boldsymbol{e}_H$ are the unscaled edges of the hard part and 
$\boldsymbol{e}_S$ are the massless edges scaled as $\lambda^{-2}$. Just like for $\Lb$, we can begin by splitting hard and soft vertices
\begin{equation} \label{eq:M_components}
    \Mb = 
    \left(
\begin{array}{c|c}
    \Mb_H & \Mb_{S}
 \end{array}
 \right)^\top,
\end{equation}
and then further decompose the soft component into jet and blob blocks: 
\begin{equation}
    \Mb_S 
    =
    \left(
    \begin{array}{c|c}
      \Mb_{J} & \Mb_B 
    \end{array}
    \right)^\top
    =
    \left(
    \begin{array}{c|c|c|c|c|c}
      \Mb_{J_1} & \dots  & \Mb_{J_n} & \Mb_{B_1} & \dots & \Mb_{B_{n_B}} 
    \end{array}
    \right)^\top , 
\end{equation} 
where the minors $\Mb_{J_i}$ related to the jet vertices are explicitly given by
\begin{equation} \label{eq:MJ_minors}
    \Mb_{J_i}= \begingroup
\renewcommand*{\arraystretch}{1}
\begin{pmatrix}
    \alpha_{i,1}^{-1} & -\alpha_{i,2}^{-1}  &    & \\
    &\alpha_{i,2}^{-1} &-\alpha_{i,3}^{-1}  &    \\
    &     & \ddots &  \ddots \\
    &  &   & \alpha_{i,l_i}^{-1} 
\end{pmatrix} .
\endgroup
\end{equation}
Finally, the vector of external momenta $\pp$ can be decomposed  
similarly to $\Lb$ and $\Mb$ as 
\begin{align} \label{eq:p_components}
\pp &  =  (\pp_H | \pp_S)^\top \nonumber \\
    & =(\pp_H | \pp_{J_1} | \pp_{J_2} | \dots  | \pp_{J_n} |\boldsymbol{0})^\top \nonumber \\ 
    & =
    (\underbracket[0.4pt]{p_{H,1},\,p_{H,2}\,,\ldots \, }_{\substack{\text{ vertices} \\ \text{in hard diagram}}}| \,\,
    \underbracket[0.4pt]{0,0,\ldots,0,p_{J_1} \, }_{\substack{\text{ vertices in } \\ \text{ first jet }}}| \, \,
    \ldots \, |\,\,
    \underbracket[0.4pt]{0,0,\ldots,0,p_{J_n} \,}
    _{\substack{\text{ vertices in } \\ n\text{-th jet}}} |
    \underbracket[0.4pt]{0,0,\ldots\, }
    _{\substack{\text{ vertices in } \\ \text{the blobs}}} 
    )^\top \,,
\end{align}
where we have assumed some arbitrary momenta $p_{H,i}$ entering each hard vertex. 
\subsection{Factorisation of \texorpdfstring{$\mathcal{U}$}{U}}
\label{sec:factorisation_U}
Having understood the features of the underlying graph matrices from which we can extract the Symanzik polynomials and define the worldline action, we now proceed by showing the factorisation under the IR scaling of the first Symanzik polynomial. As we saw already in section \ref{sec:2loopMatrix}, since $\mathcal{U} = \prod_{e} \alpha_e \det \Lb$, the factorisation of $\mathcal{U}$
is equivalent to the factorisation of $\det \Lb$.
\subsubsection*{Factorisation into hard and soft}
Our first goal is to prove that to leading order in $\lambda$
\begin{equation}
    \mathcal{U} \simeq \mathcal{U}_H \times \mathcal{U}_S \,,
\end{equation}
where $\mathcal{U}_H$ and $\mathcal{U}_S$ are the first Symanzik polynomials of the hard and soft graphs respectively.
Using \cref{eq:LLmatrix} we rewrite the determinant of $\Lb$ as\footnote{If $M$ is an invertible block matrix,
\begin{equation}
    M = \begin{pmatrix}
        A & B \\ C & D
    \end{pmatrix}
\end{equation}
and $A$ is invertible, then $\det M = \det (D - C A^{-1} B) \det (A)$, and
\begin{equation} \label{eq:block_inversion}
    M^{-1} = \begin{pmatrix}
A^{-1} + A^{-1}B(D - CA^{-1}B)^{-1}CA^{-1} & -A^{-1}B(D - CA^{-1}B)^{-1} \\
-(D - CA^{-1}B)^{-1}CA^{-1} & (D - CA^{-1}B)^{-1}
\end{pmatrix} \,.
\end{equation}
}
\begin{equation} \label{eq:detL}
    \det \Lb = 
    \det{\widetilde{\Hb}} \det{\Sb} ,
\end{equation}
where we introduced the deformed hard matrix 
\begin{equation} \label{eq:Htilde}
    \widetilde{\Hb} = \Hb - \Cb \,\Sb^{-1} \,\Cb^\top \,.
\end{equation}
Therefore in order to show that $\U$ factors in the correct way, we simply have to argue that $ \Cb \,\Sb^{-1} \,\Cb^\top \,$ is sub-leading in the scaling as $\lambda \to 0$, so that $\widetilde{\Hb}$ can be approximated by $\Hb$ in  \cref{eq:detL}. 
To show this we use the block-inversion formula \cref{eq:block_inversion} on $\Sb$ split as in\cref{eq:S_form} and use the special form of $\Cb$ given by \cref{eq:C_form} which yields
\begin{equation} \label{eq:CdotSinvdotC}
    \Cb \Sb^{-1} \Cb^\top = 
    \Cb_J \left( \Jb + \boldsymbol{\Gamma}_{J} - \boldsymbol{\Gamma}_{J,B} (\Bb + \boldsymbol{\Gamma}_{B})^{-1} \boldsymbol{\Gamma}_{B,J} \right)^{-1} \Cb_J^\top
    \simeq
    \Cb_J \Jb^{-1} \Cb_J^\top 
    \sim 
    \lambda .
\end{equation}
To explain the scaling above, we start by looking at the matrix $(\Bb + \boldsymbol{\Gamma}_{B})$. 
Since $\Bb$ is not invertible by itself (it is the unreduced Laplacian of the blob subdiagrams), 
we must consider the full matrix $\Bb + \boldsymbol{\Gamma}_{B}$. 
This matrix is invertible, and its inverse has components starting at order $\lambda^{-2}$.\footnote{
    We will give more details on the form of the $(\Bb + \boldsymbol{\Gamma}_{B})^{-1}$ matrix in the 
    next section.
} 
Still, in the equation above it appears multiplied by two 
matrices of uniform scaling $\lambda^2$, so that the whole term is negligible when compared to $\Jb$, 
which scales as $\lambda$. 
In the last approximation we also dropped $\boldsymbol{\Gamma}_{J}$ since 
it scales as $\lambda^2$. 
Finally, we see from \cref{eq:SandJ_inv} that $\Jb^{-1}$ scales as $\lambda^{-1}$, but here it appears 
multiplied by two uniformly scaling $\mathcal{O}(\lambda)$ matrices ($\Cb_J$), so overall the result 
goes like $\lambda$.  
Eqs.~\eqref{eq:CdotSinvdotC} and~\eqref{eq:Htilde} imply 
\begin{equation}\label{eq:Htilde_approx}
    \widetilde{\Hb} = \Hb + \mathcal{O}(\lambda),
\end{equation}
which in turn gives us the desired factorisation property
\begin{equation}
    \det \Lb = 
    (\det{\Hb}+ \mathcal{O}(\lambda) ) \det{\Sb},
\end{equation}
when plugged into \cref{eq:detL}.

\subsubsection*{Factorisation into webs}
\begin{figure}
    \centering
    \includegraphics[width=0.9\textwidth]{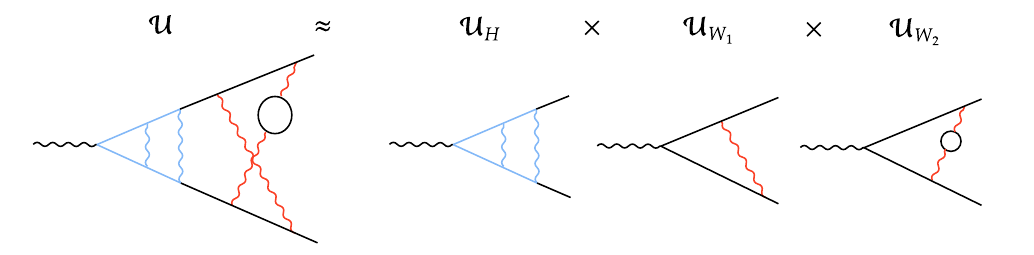}
    \caption{Example of factorisation of the $\U$ polynomial into hard and web terms.
    \label{fig:webs_first_example}}
\end{figure}

From the block diagonal structure of the leading entries of $\Sb$, it is simple to see that $\det{\Sb}$ 
further factorises as
\begin{equation}
    \det{\Sb} \simeq \prod_{i} \det{\Jb_i} \times \prod_{j} \det\left( \Bb_j+ \boldsymbol{\Gamma}_{B_j} \right)
\end{equation}
and since $\det \Jb_i = \prod_{e} \alpha_{i,e}^{-1} $ one can check that 
\begin{equation}
    \U_S \simeq \prod_{\mathcal{W}} \U_\mathcal{W},
\end{equation}
where the connected webs $\mathcal{W}$ are the graphs obtained by deleting all disconnected components of the soft 
subdiagram except for one. In the example of the graph in \cref{fig:diagram_example} we find that there are two connected 
webs and the $\U$ polynomial factorises as shown in \cref{fig:webs_first_example}.
In the next section we will show that the (modified) worldline action factorises into a sum
of terms which correspond exactly to the same types of diagrams: connected webs.

\subsection{Factorisation of \texorpdfstring{$\mathcal{V}(\zz)$}{V(z)}}
\label{sec:factorisation_Nu}
We now switch to showing how the modified action $\mathcal{V}(\zz)$
factorises in the scalings considered. All the results obtained below can be trivially extended to
the standard action $\mathcal{V}$ by setting all $z_e = 0$. 

\subsubsection*{Factorisation into hard and soft}

Let us focus on the part of the action involving the inverse Laplacian:
\begin{equation}\label{eq:Lm1_Acion}
\begin{aligned}
\mathcal{V}(\zz) =  \left(\pp + \frac{i}{2} \MM \zz \right)^\top \mathbf{L}^{-1}\left(\pp + \frac{i}{2} \MM \zz \right)  + \cdots  ,
\end{aligned}
\end{equation}
as all other terms are already trivially factorised.
To begin, let us write $\LL^{-1}$ through the block-inversion formula \cref{eq:block_inversion} as
\begin{equation} \label{eq:Linv1}
    \LL^{-1} =
    \begin{pmatrix}
       \widetilde{\Hb}^{-1} &
        {\color{white} .} \hspace{0.5cm}
        - \Sb^{-1}\, \Cb^\top\, \widetilde{\Hb}^{-1}  \\
        -\widetilde{\Hb}^{-1}\, \Cb \,\Sb^{-1} & {\color{white} .} \hspace{0.5cm}  \Sb^{-1} + \Sb^{-1} \,\Cb^\top\, \widetilde{\Hb}    ^{-1}\, \Cb\, \Sb^{-1}
    \end{pmatrix} \,,
\end{equation}
which can be further split as
\begin{equation} \label{eq:Linv2}
    \LL^{-1} =
    \Tb_S^\top
    \begin{pmatrix}
       \widetilde{\Hb}^{-1} &
        {\color{white} .} \hspace{0.5cm}
       0 \\
       0 & {\color{white} .} \hspace{0.5cm}  \Sb^{-1} 
    \end{pmatrix} 
    \Tb_S \,, \quad\text{with}\quad 
    \Tb_S = 
    \begin{pmatrix}
        \mathds{1}_H &
         {\color{white} .} \hspace{0.5cm}
         - \Cb \Sb^{-1}\, \, \\
         0 & {\color{white} .} \hspace{0.5cm}   \mathds{1}_S
     \end{pmatrix} ,
\end{equation}
where $\widetilde{\Hb}$ was defined in \cref{eq:Htilde}, and, as we will see momentarily, 
$\Tb_S$ guarantees that the external momenta feeding into each massive line propagate through 
the soft subgraph and enter into the correct vertex of the hard diagram. 

As we proceed to show, $\Cb_J \Jb^{-1}$ scales as $\lambda^0$, and since we are interested in 
retaining terms in $\mathcal{V}$ up to order $\lambda^0$, in all the terms involving $\widetilde{\Hb}$ in \eqref{eq:Linv1}, we can safely approximate $\widetilde{\Hb} \simeq \Hb$, since the corrections coming from $ \Cb \,\Sb^{-1} \,\Cb^\top$ are sub-leading.

The action $\mathcal{V}(\zz)$ then takes a factorised
form if $\Tb_S$ acting on the vector of (modified) external momenta $ \pp + i \MM \zz/2$ 
yields a vector whose hard vertex entries depend only on Schwinger parameters of the hard subgraph, 
and similarly for the soft vertex entries. More concretely, we would like 
\begin{equation}
    \Tb_S \pp_z  = \left(
    \begin{array}{c|c}
        \boldsymbol{P}_H(\boldsymbol{\alpha}_H) & \boldsymbol{P}_S(\boldsymbol{\alpha}_S) 
    \end{array}
    \right)^\top,
\end{equation}
where by $\boldsymbol{\alpha}_{H(S)}$ we mean the collection of Schwinger parameters of the hard(soft)
subgraph.  
To prove this we start by looking more in detail at the entries of the vector. 
Using \cref{eq:p_components,eq:M_components} we find 
\begin{equation} \label{eq:TdotPz}
    \begin{aligned}
        \Tb_S \pp_z   &= \Tb_S \left(\pp + \frac{i}{2} \MM \zz \right) = \\
                    &=  
                    \left(
                    \begin{array}{c|c}
                    \pp_H + \frac{i}{2} \MM_H \zz 
                    - \Cb {\Sb}^{-1} (\pp_S + \frac{i}{2} \Mb_{S}\zz  ) {\color{white} .} \hspace{0.2cm} &  {\color{white} .} \hspace{0.2cm}
                    \pp_S + \frac{i}{2} \Mb_{S}\zz 
                    \end{array}
                    \right)^\top\\
                    &\simeq  
                    \left(
                    \begin{array}{c|c}
                        \pp_H + \frac{i}{2} \MM_H \zz 
                        - \Cb {\Sb}^{-1}\pp_S   {\color{white} .} \hspace{0.2cm} &  {\color{white} .} \hspace{0.2cm}
                        \pp_S + \frac{i}{2} \Mb_{S}\zz 
                    \end{array}
                    \right)^\top\, ,
    \end{aligned}
\end{equation}
where in the first block we dropped the term proportional to $\Mb_S$ since it is sub-leading as $\lambda \to \infty$ and it will only be multiplied by $\Hb$ 
in the action, see \cref{eq:Linv1,eq:Linv2}.

We now turn our attention to the product $\Cb \Sb^{-1} \equiv (\Sb^{-1} \Cb^\top)^\top$. 
Once again, we use block inversion on $\Sb^{-1}$ to obtain 
\begin{equation} \label{eq:CdotSinv}
    \Cb \Sb^{-1} \simeq
    \left(
    \begin{array}{c|c}
        \Cb_J \Jb^{-1} {\color{white} .} \hspace{0.2cm}&
         {\color{white} .} \hspace{0.2cm}
        - \Cb_J \Jb^{-1}\, \boldsymbol{\Gamma}_{J,B} \, (\Bb + \boldsymbol{\Gamma}_{B})^{-1}
    \end{array}
    \right),
\end{equation}
where once again we neglected the $\mathcal{O}(\lambda^2)$ correction $\boldsymbol{\Gamma}_{J}$ to $\Jb$  
since the product above only appears multiplied by $\widetilde{\Hb}$ in \cref{eq:Linv1} and therefore
can be expanded to first order. It also turns out that the part of $\Cb \Sb^{-1}$ associated to the blobs, $i.e.$ 
the part to the right of the vertical bar in \cref{eq:CdotSinv}, can be ignored since it will always be 
multiplied by the zero entries of $\pp_S$. The overall scaling is deduced through arguments similar to those 
below \cref{eq:CdotSinvdotC}.
We further notice that $\Cb_{J} \Jb^{-1} = (\Cb_{J_1}  \Jb_1^{-1} \, | \, \Cb_{J_2} \Jb_2^{-1} \, | \, \dots)$ and
\begin{equation}\label{eq:CbJi}
    (\Cb_{J_i} \Jb_i^{-1})_{vw} =  
    \frac{1}{\alpha_{i,1}} 
    \sum_{v'} \delta_{v' 1 } \delta_{v h_{J_i}} \beta_{i,\min(v',w)} =
    \frac{1}{\alpha_{i,1}} 
     \delta_{v h_{J_i}} \beta_{i,\min(1,w)} =
       \delta_{v h_{J_i}},   
\end{equation}
so for instance if the vertex $h_{J_i}=1$, $i.e.$ $i$-th jet attaches to the hard subdiagram
on vertex 1, then the matrix  $\Cb_{J_i} \Jb_i^{-1}$ will be a matrix of zeros, apart from the first row which will be $(1,1,...,1)$.
Since the larger matrix $ \Cb_J \Jb^{-1}$ is made up of blocks of the type in \cref{eq:CbJi}, this proves 
the scaling stated above:
\begin{equation}
    \Cb \Sb^{-1} \sim \lambda^0 \,.
\end{equation}
Multiplying a ``jet vector'' $\pp_{J_i}$ (its entries corresponding to the vertices along a jet line) 
by the corresponding components of $ \Cb_J \Jb^{-1}$ creates a vector of momenta with the external momentum entering the jet $J_i$ is injected instead into the appropriate hard vertex. More precisely, it has the effect of injecting into the vertex $h_{J_i}$ the sum of all (modified) external momenta that entered the diagram through the jet $J_i$. 
Using this result we can rewrite \cref{eq:TdotPz} as
\begin{equation} \label{eq:TdotPz2}
        \Tb_S \Big( \pp + \frac{i}{2} \MM \zz \Big)  =
        \left(
        \begin{array}{c|c}
            \pp'_H + \frac{i}{2} \MM_H \zz 
             {\color{white} .} \hspace{0.2cm} &  {\color{white} .} \hspace{0.2cm}
            \pp_S + \frac{i}{2} \Mb_{S}\zz 
        \end{array}
        \right)^\top\, ,
\end{equation}
where $\pp'_H =  \pp_H  + \Cb \Sb^{-1} \pp_S  $ is exactly the set of external momenta of the diagram after deleting the soft sub-diagram.
Plugging \cref{eq:TdotPz2,eq:Linv2} into the modified action \cref{eq:full_modified_action} we obtain the factorised expression
\begin{equation}\label{eq:action_factorisation_proof}
    \begin{aligned}
        \mathcal{V}(\zz) &=  
        \mathcal{V}_H(\zz_H) + \mathcal{V}_S(\zz_S)
         + \mathcal{O}(\lambda) ,
    \end{aligned}
\end{equation}
with
\begin{equation} \label{eq:HandS_actions}
    \begin{aligned}
        \mathcal{V}_H(\zz_H) &=  
        \left(\pp'_H + \frac{i}{2} \MM_H \zz \right)^\top \mathbf{\Hb}^{-1}\left(\pp'_H + \frac{i}{2} \MM_H \zz\right)  
        + \sum_{e \in H} \frac{z_{e}^2}{4 \alpha_e}
        -\sum_{e \in H}m_e^2\alpha_e \, ,  \\
        \mathcal{V}_S(\zz_S) &=   \left(\pp_S+ \frac{i}{2} \MM_S \zz \right)^\top \mathbf{\Sb}^{-1}\left(\pp_S + \frac{i}{2} \MM_S \zz\right)  
        + \sum_{e \in S} \frac{z_{e}^2}{4 \alpha_e}
        -\sum_{e \in S} m_e^2 \alpha_e  ,
    \end{aligned}
\end{equation}
which confirms that the modified worldline action exactly factorises onto the sum of the worldline action of the hard and soft subdiagrams, 
as defined above. 
From \cref{eq:action_factorisation_proof,eq:HandS_actions} we also immediately see that the numerator $\mathcal{N}(\partial_{\zz})$
factorises too, since all $2$-point contractions $\Delta^{\mu \nu}_{ef}$ (defined in \cref{eq:num_wick_contractions}) mixing hard and soft edges are sub-leading.

\subsubsection*{Factorisation into connected webs}

So far we have proven that the action of the full diagram $\mathcal{V}(\zz)$ factorises into the sum
of a soft and a hard action.
We will now look deeper into $\mathcal{V}_S(\zz_S)$ and show how it further factorises into
a sum of simpler terms, similarly to what we found for the $\U$ polynomial, as depicted in \cref{fig:webs_first_example}. 
Among other things, we will find that as a function of the worldline variables
the soft action is \emph{independent} of the ordering of attachment 
of the soft ($\alpha \sim \lambda^{-2}$) photons to the jet lines ($\alpha \sim \lambda^{-1}$).\\

Let us start by observing that the matrix $\Sb^{-1}$ in $\mathcal{V}_S$ is of order $\lambda^{-1}$, 
so we will have to retain the first two orders (up to $\lambda^0$) as well as contribution coming 
from the two factors of $\Mb_S$. 
We start by block-inverting $\Sb$ using the decomposition in \cref{eq:S_form} and then further factorising it
like we did in \cref{eq:Linv2}:
\begin{equation}\label{eq:Sinv_fact}
        \Sb^{-1} 
        =
        \Tb_{S_J}^\top
        \begin{pmatrix}
            (\Jb + \boldsymbol{\Gamma}_{J})^{-1}   &  0 \\
           0 &  (\Bb + \boldsymbol{\Gamma}_{B})^{-1} + \mathcal{O}(\lambda^{-1}) 
        \end{pmatrix}
        \Tb_{S_J} \, , 
\end{equation}
with the transfer matrix
\begin{equation}
    \Tb_{S_J} = 
    \begin{pmatrix}
       \mathds{1}_J  & 0 \\
       - \boldsymbol{\Gamma}_{B,J}  {(\Jb + \boldsymbol{\Gamma}_{J})}^{-1}  &  \mathds{1}_B
    \end{pmatrix} .
\end{equation}
As before, we consider the action of the transfer matrix on the external momenta:
\begin{equation}
    \begin{aligned}
        \Tb_{S_J} &\left( \pp_S+ \frac{i}{2} \MM_S \zz  \right)= 
        \Tb_{S_J} \left( 
            \begin{array}{c|c}  \pp_J+ \frac{i}{2} \MM_J \zz  
            {\color{white} .} \hspace{0.2cm} &  {\color{white} .} \hspace{0.2cm}
             \frac{i}{2} \MM_B \zz 
            \end{array} \right) =  \\
        &=
        \left(
        \begin{array}{c|c}
            \pp_J + \frac{i}{2} \Mb_J \zz 
            {\color{white} .} \hspace{0.2cm} &  {\color{white} .} \hspace{0.2cm}
             \frac{i}{2} \Mb_B \zz -  \boldsymbol{\Gamma}_{B,J}  {(\Jb + \boldsymbol{\Gamma}_{J})}^{-1}  \left( \pp_J+ \frac{i}{2} \MM_J \zz  \right)
        \end{array}
        \right)
    \end{aligned}
\end{equation}
and plugging this into the soft action and using ${(\Jb + \boldsymbol{\Gamma}_{J})}^{-1} = \Jb^{-1} - \Jb^{-1}\boldsymbol{\Gamma}_{J}\Jb^{-1} + \cdots $, we find that the expansion of the soft action up to $\mathcal{O}(\lambda^0)$ is
\begin{equation}\label{eq:soft_action_expanded}
    \begin{aligned}
        \mathcal{V}_S(\zz) & \simeq 
        \left[ \pp_J^\top\Jb^{-1}\pp_J 
        -\sum_{e \in J} m_e^2 \alpha_e \right] \\
        &
        -
         \pp_J^\top\Jb^{-1}
         \left\{
          \boldsymbol{\Gamma}_{J}   -  \boldsymbol{\Gamma}_{J,B}(\Bb + \boldsymbol{\Gamma}_{B})^{-1}\boldsymbol{\Gamma}_{B,J} 
         \right\}
        \Jb^{-1} 
        \pp_J 
        -\sum_{e \in B} m_e^2 \alpha_e 
        \\
        &+
        i \pp_J^\top  \Jb^{-1} \Mb_J   \zz
        + 
        \left(  
            \frac{i}{2} \Mb_B \zz 
        \right)^\top
        (\Bb + \boldsymbol{\Gamma}_{B})^{-1} 
        \left(  
            \frac{i}{2} \Mb_B \zz
        \right) 
        + \sum_{e \in B} \frac{z_{e}^2}{4 \alpha_e}
        .
    \end{aligned}
\end{equation}
Here, we have already dropped some sub-leading terms using that $(\Bb + \boldsymbol{\Gamma}_{B})^{-1} \Mb_B$ is of $\mathcal{O}(\lambda^0)$, as we will explain momentarily.
The terms in square brackets are divergent, of order $\lambda^{-1}$, but exactly cancel since 
\begin{equation} \label{eq:Pw_expl}
    \begin{aligned}
        \Jb^{-1}\pp_J &= 
        \left(
        \begin{array}{c|c|c}
         \Jb^{-1}_{1}\pp_{J_1} & \dots & \Jb^{-1}_{n}\pp_{J_n}    
        \end{array}
        \right)^\top
        \\[1em]
        &=
        \left(
        \begin{array}{cccc|c|cccc}
            \overmat{\text{jet 1}}{ p_{J_1} (\beta_{1,l_1}, &  \beta_{1,l_1-1}, & \dots, &  \beta_{1,1})  }  & 
            \dots & 
            \overmat{\text{jet n}}{p_{J_n} (\beta_{n,l_n}, &   \beta_{n,l_n-1}, & \dots, &  \beta_{n,1})  }  
        \end{array}
        \right)^\top
    \end{aligned}
\end{equation}
and therefore
\begin{equation}
    \pp_J^\top\Jb^{-1}\pp_J = 
    \sum_i p_{J_i}^2 \, \beta_{i,l_i} = 
    \sum_i p_{J_i}^2 \, \sum_{e \in J_i} \alpha_{i,e} = 
    \sum_{e \in J} m_e^2 \alpha_e .
\end{equation}
The terms in the second line of \cref{eq:soft_action_expanded} are independent of $\zz$, while those on the third line carry all 
dependence on $\zz$ and affect the numerator structure.
They are all of order $\lambda^0$. 
We point out that although $(\Bb + \boldsymbol{\Gamma}_{B})^{-1} \sim \lambda^{-2}$
and $\Mb_B \zz \sim \lambda^0$, their product scales only as $\lambda^0$. 
This is because, as mentioned above, $\Bb$ is not invertible. To be more precise, the inverse of the 
Laplacian of the unscaled subgraph can be split into a direct sum ($i.e.$ in block diagonal form)
where each term corresponds to a disconnected part of the subgraph and the same is true for $\Mb_B $:
\begin{equation}
    \begin{aligned}
        (\Bb + \boldsymbol{\Gamma}_{B})^{-1}  &= 
        (\Bb_1 + \boldsymbol{\Gamma}_{B_1})^{-1}  
        \oplus 
        \dots
        \oplus 
        (\Bb_{n_W} + \boldsymbol{\Gamma}_{B_{n_W}})^{-1} \, , \\
        \Mb_{B} &= \Mb_{B_i} \oplus \dots \oplus \Mb_{B_{n_W}} \, ,
    \end{aligned}
\end{equation}
where $n_B$ is the number of disconnected subgraphs of $B$.
The matrix $\Bb_i$ of each subgraph corresponds to its \emph{unreduced} Laplacian and 
is therefore one short of full rank, with kernel $(1,1,\dots,1)^\top$. 
Because of this, to $\mathcal{O}(\lambda^{-2})$ we find
\begin{equation}\label{eq:leading_B_inv}
    (\Bb_i + \boldsymbol{\Gamma}_{B_i})^{-1}  = 
    \frac{1}{\mathrm{Tr}\,\boldsymbol{\Gamma}_{B_i} } 
    (1,1,\dots,1)^\top
    \otimes
    (1,1,\dots,1)
    + \mathcal{O}(\lambda^0).
\end{equation}
Since the vector $(1,1,\dots,1)$ is in the left kernel of $\Mb_{B_i}$ (again because 
it is the unreduced $\Mb$ matrix of $B_i$), the leading term 
is projected out in the product $ (\Bb_i + \boldsymbol{\Gamma}_{B_i}) ^{-1}\Mb_{B_i}$, which 
is therefore of $\mathcal{O}(\lambda^0)$.
The soft action is therefore finite in the scaling considered, however, as anticipated, 
we can say a lot more about its structure. Below we will show that $\mathcal{V}_S(\zz)$ can be written as a sum over a set of subdiagrams called \emph{connected webs}. We will prove this term by term. \\

\begin{figure}
    \centering
    \includegraphics[width=0.95\textwidth]{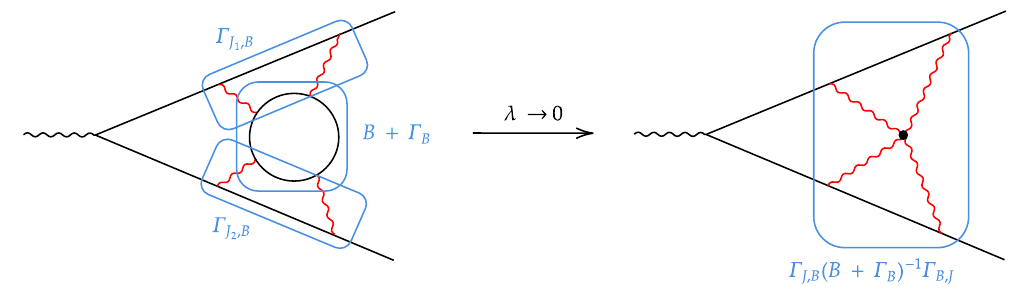}
    \caption{ 
    The matrices $(\Bb +\boldsymbol{\Gamma}_B), \boldsymbol{\Gamma}_{J_i,B}$ are associated to the sets of vertices inside the corresponding blue rectangles (\textbf{left}). 
    In particular the matrix $ (\Bb + \boldsymbol{\Gamma}_{B})$ corresponds to the Laplacian matrix of the encircled sub-graph.  
    After taking the approximation corresponding to the soft region where all four photons become soft, the graph matrices associated to the blob appear in the combination 
    $\boldsymbol{\Gamma}_{J,B}
          (\Bb + \boldsymbol{\Gamma}_{B})^{-1}
          \boldsymbol{\Gamma}_{B,J} $
          which at leading order is identical to the same quantity computed on the graph where the entire blob has been pinched to a point (\textbf{right}).
    }
    \label{fig:pinching_example}
\end{figure}

We begin from the $\zz$-independent terms of  \cref{eq:soft_action_expanded}.
Since the sum over masses is trivial we focus on the first term, 
which we recast into the form 
\begin{equation}\label{eq:term_no_z}
        \pp_J^\top\Jb^{-1}     
         \left\{
          \boldsymbol{\Gamma}_{J}   
          -  
          \boldsymbol{\Gamma}_{J,B}
          (\Bb + \boldsymbol{\Gamma}_{B})^{-1}
          \boldsymbol{\Gamma}_{B,J} 
         \right\}
        \Jb^{-1} 
        \pp_J 
        \simeq 
          \boldsymbol{P}_J^\top \,
         \boldsymbol{\gamma}_J \,
         \boldsymbol{P}_J .
\end{equation} 
Here we defined the quantities
\begin{equation} \label{eq:some_defs}
\begin{aligned}
    \boldsymbol{P}_J &=  \Jb^{-1} \pp_J , \\
    \boldsymbol{\gamma}_J 
    &=  \boldsymbol{\Gamma}_{J}  + \boldsymbol{\Gamma}'_{J} , \\
    \boldsymbol{\Gamma}'_{J}  
    &= 
    -\frac{1}{\mathrm{Tr}\boldsymbol{\Gamma_B}} 
    \boldsymbol{\Gamma}_{J,B} \left[
    (1,1,\dots,1)^\top
    \otimes
    (1,1,\dots,1) \right]
    \boldsymbol{\Gamma}_{B,J},
\end{aligned}
\end{equation}
where one can check that 
\begin{equation}
    \mathrm{Tr}\boldsymbol{\Gamma_B} = \sum_{e \in \gamma_B} \frac{1}{\alpha_e},
\end{equation}
with $\gamma_B$ the set of soft photons connecting the blob to the jets.
The approximation (up to sub-leading terms) of \cref{eq:term_no_z} is justified by the leading-order
expansion of \cref{eq:leading_B_inv}. Above, $\boldsymbol{\Gamma}'_{J}$ is independent of the unscaled 
parameters in $B$. Indeed, it is completely independent of the structure of the disconnected parts of $B$, 
and only retains information about which soft photon attaches to which unscaled subdiagram. 
It can therefore be interpreted as ``connecting'' jets through a modified version of $B$ where each 
disconnected piece has been contracted to a point. In other words, the form of $\boldsymbol{\Gamma}'_J$ encodes the action of ``pinching'' the unscaled blob to a point, as illustrated in \cref{fig:pinching_example}.

$\boldsymbol{\Gamma}_{J}$ instead connects jets through individual photon exchanges.
Finally $\boldsymbol{P}_J$ is a vector of the form given in \cref{eq:Pw_expl}, with each entry equal to some 
external four-momentum multiplied by the appropriate worldline variable $\beta$. Importantly 
the Schwinger parameters of the jet lines enter in $\mathcal{V}_S$ \emph{only} through the worldline variables, 
and never independently. A consequence of this fact is that the soft action, when expressed in terms of the worldline
variables is the same for all attachment orderings of the soft photons to the jet lines!

\begin{figure}
    \centering
    \includegraphics[width=0.95\textwidth]{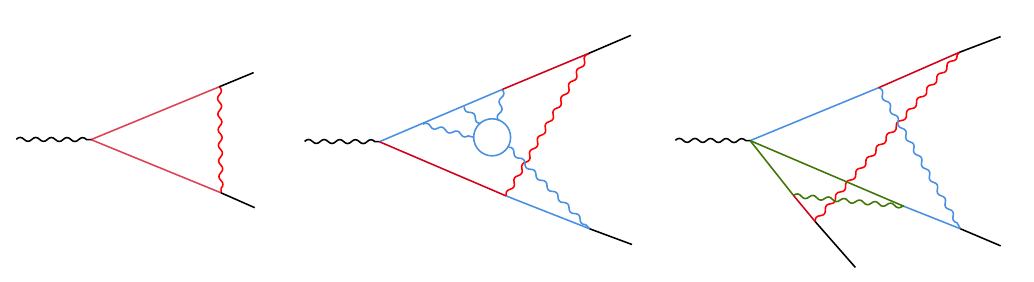}
    \caption{ Some examples of how massive edges are associated to soft connected webs.}
    \label{fig:web_jet_edges}
\end{figure}

The non-zero minors of $\boldsymbol{\gamma}_J $ identify \emph{connected webs}, sub-graphs of $S$
which cannot be disconnected by removing any number of jet edges. 
We define the jet edges associated to a web as the edges starting from a web vertex (which lies on the jet line) and going towards the hard sub-diagram, as depicted in the example of \cref{fig:web_jet_edges}.
It is therefore natural to reorganise l.h.s. of \cref{eq:term_no_z} as a sum over connected webs
\begin{equation}\label{eq:NuS_z_0}
 \boldsymbol{P}_J^\top \,
         \boldsymbol{\gamma}_J \,
         \boldsymbol{P}_J 
         = 
         \sum_{\mathcal{W}}^{\text{c-webs}} \boldsymbol{P}_\mathcal{W}^\top \,
        \boldsymbol{\gamma}_{\mathcal{W}} \,
         \boldsymbol{P}_\mathcal{W} ,
\end{equation}
where $\boldsymbol{\gamma}_{\mathcal{W}}$ are the ``connected'' minors of $\boldsymbol{\gamma}_J$ and $\boldsymbol{P}_\mathcal{W}$ 
are the components of $\boldsymbol{P}_J$ corresponding to the vertices connected by $\boldsymbol{\gamma}_{\mathcal{W}}$.

Moving to the term of $\mathcal{V}_S(z)$ linear in $\zz$, we find 
\begin{equation}\label{eq:NuS_z_1}
\begin{aligned}
     i \pp_J^\top  \Jb^{-1} \Mb_J   \zz &= 
     i \sum_{i} p_{J_i} \sum_{e \in J_i} z_e  = 
     i \sum_{w}^{\text{c-webs}} p_{J_i} \sum_{e \in J_i \cap \mathcal{W}_w} z_e  
    \, ,
            \\
    \end{aligned}
\end{equation}
where in the first equality we used the explicit form of the matrix $
\Mb_{J} = 
\left(
\begin{array}{c|c|c}
    \Mb_{J_1} & \dots & \Mb_{J_n}
\end{array}
\right)
$ 
given in \cref{eq:MJ_minors} and in the second equality we divided the sum over edges into contributions 
to different connected webs using the association of edges to webs of \cref{fig:web_jet_edges}.

Finally, the term of \cref{eq:soft_action_expanded} quadratic in $\zz$ can be simply written as a sum over webs, 
since each disconnected component of $B$ identifies a connected web:
\begin{equation} \label{eq:NuS_z_2}
    \begin{aligned}
        \left(  
        \frac{i}{2} \Mb_B \zz 
        \right)^\top&
        (\Bb + \boldsymbol{\Gamma}_{B})^{-1} 
        \left(  
        \frac{i}{2} \Mb_B \zz
        \right) + \sum_{e \in B} \frac{z_{e}^2}{4 \alpha_e} = \\
        &\sum_{\mathcal{W}}^{\text{c-webs}} \left\{
        \left(  
        \frac{i}{2} \Mb_{B_{\mathcal{W}}} \zz 
        \right)^\top
        (\Bb_{\mathcal{W}} + \boldsymbol{\Gamma}_{B_{\mathcal{W}}})^{-1} 
        \left(  
        \frac{i}{2} \Mb_{B_{\mathcal{W}}} \zz
        \right) + \sum_{e \in B_{\mathcal{W}}} \frac{z_{e}^2}{4 \alpha_e} 
        \right\} .
    \end{aligned}
\end{equation} 
Above it is implied that if c-web $\mathcal{W}$ is simply made up of a single photon, the corresponding contribution to the
sum in \cref{eq:NuS_z_2} vanishes. Otherwise $\Bb_\mathcal{W}$ is the minor of the $\Bb$ matrix associated with that web. 

To obtain an intuitive understanding of the expression in curly braces above we write $\Mb_{B_{\mathcal{W}}} \zz$ as a sum of two terms according to their scaling in $\lambda$
\begin{equation}
    \Mb_{B_{\mathcal{W}}} \zz = \Mb^{(0)}_{B_{\mathcal{W}}} \zz + \Mb^{(2)}_{B_{\mathcal{W}}} \zz,
\end{equation}
where the superscript corresponds to the scaling in $\lambda$ of the corresponding matrix. By definition $\Mb^{(0)}_{B_{\mathcal{W}}}$ depends solely on the $\alpha$-parameters of the unscaled edges in the blob, while $\Mb^{(2)}_{B_{\mathcal{W}}}$ depends only on those of the massless edges attached to it. In particular $\Mb^{(0)}_{B_{\mathcal{W}}}$ corresponds exactly to the weighted incidence matrix of the blob $B_\mathcal{W}$ and since it is the unreduced one we know that 
\begin{equation}    
(1,1,\dots) \cdot \Mb^{(0)}_{B_{\mathcal{W}}} = (0,0,\dots).
\end{equation}
Using this together with \cref{eq:leading_B_inv} for the piece $(\Bb_{\mathcal{W}} + \boldsymbol{\Gamma}_{B_{\mathcal{W}}})^{-1}$, at $\mathcal{O}(\lambda^0)$ the expression in curly brackets can be recast into the form 
\begin{equation} 
    \left\{  
        \pp_\mathcal{W}(\zz)^\top
        (\Bb_{\mathcal{W}})^{-1}_{red} \,
        \pp_\mathcal{W}(\zz) + \sum_{e \in B_{\mathcal{W}}} \frac{z_{e}^2}{4 \alpha_e} 
        \right\} , 
        \label{eq:BlobRed}
\end{equation}
where $(\Bb_{\mathcal{W}})_{red}$ is the reduced Laplacian matrix of the blob (with all scaled propagators deleted) and 
$\pp_\mathcal{W}(\zz) = (i/2)\Mb^{(0)}_{B_{\mathcal{W}}} \zz$ acts as an effective set of external momenta. So \cref{eq:BlobRed} together with the mass terms $ -\sum_{e \in B_{\mathcal{W}}} m_e^2 \alpha_e$ 
can be interpreted as an effective worldline action for the blob $B_\mathcal{W}$ in the soft limit:
\begin{equation} \label{eq:web_eff_action}
   \mathcal{V}_{B_\mathcal{W}}(\zz) = 
    \left\{  
        \pp_\mathcal{W}(\zz)^\top
        (\Bb_{\mathcal{W}})^{-1}_{red} \,
        \pp_\mathcal{W}(\zz) + \sum_{e \in B_{\mathcal{W}}} \frac{z_{e}^2}{4 \alpha_e} 
        \right\} -\sum_{e \in B_{\mathcal{W}}} m_e^2 \alpha_e  . 
\end{equation}
Collecting the results of \cref{eq:NuS_z_0,eq:NuS_z_1,eq:NuS_z_2,eq:web_eff_action} we find that the soft action can be expressed 
as a sum over connected webs:
\begin{equation}\label{eq:web_fact_1}
    \mathcal{V}_S(\zz) = \sum_{\mathcal{W}}^{\text{c-webs}} \mathcal{V}_\mathcal{W}(\zz)  + \mathcal{O}(\lambda) ,
\end{equation}
with the connected web soft actions
\begin{equation}\label{eq:web_fact_2}
        \mathcal{V}_\mathcal{W}(\zz)  = 
        - \boldsymbol{P}_\mathcal{W}^\top \,
        \boldsymbol{\gamma}_{\mathcal{W}} \,
         \boldsymbol{P}_\mathcal{W} 
        +   i \sum_i p_{J_i} \sum_{e \in J_i \cap \mathcal{W}} z_e
        + \mathcal{V}_{B_\mathcal{W}}(\zz) . 
\end{equation}
Eqs.~\eqref{eq:web_fact_1} and~\eqref{eq:web_fact_2} prove the factorisation of the soft action we promised at the beginning of this section. 
Furthermore, note that in this representation of the soft action, the Schwinger parameters of the jet line edges only enter though the 
worldline variables in $\boldsymbol{P}_\mathcal{W}$, which was defined in \cref{eq:some_defs} and whose entries are
all of the form $\beta_{J,k} p_J$ for some jet $J$ (see \cref{eq:Pw_expl} for instance). 
This implies that the soft integrand, when expressed in term of worldline variables, not only factorises neatly,
but is also independent on how the webs attach to the jets! One can shuffle the ordering of the 
soft photons emitted from any jet and the integrand remains unchanged.

\subsection{Factorisation of the integrand and ladder diagrams}
\label{sec:integrand_factorisation}

In the previous sections we showed that the first Symanzik polynomial and the modified worldline action factorise
as 
\begin{equation}
    \begin{aligned}
        \U &\simeq \U_H \times \prod_{\mathcal{W}}^{\text{c-webs}} \U_\mathcal{W} , \\
        \mathcal{V}(\zz) &\simeq \mathcal{V}_H(\zz_H) + \sum_{\mathcal{W}}^{\text{c-webs}} \mathcal{V}_\mathcal{W} (\boldsymbol{\beta};\zz_\mathcal{W}) ,
    \end{aligned}
\end{equation}
where all c-web terms are approximated along their corresponding soft rays and with the jet Schwinger 
parameter evaluated on the worldline variables, $\beta$, which measure the \emph{entire} distance to the hard subgraph.

\begin{figure}[!t]
    \centering
    \includegraphics[width=0.95\linewidth]{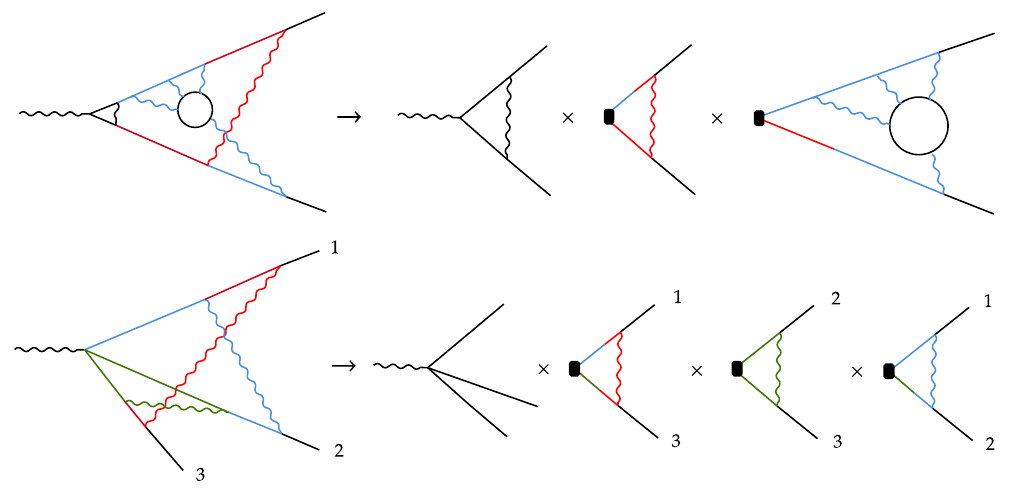}
    \caption{Examples of integrand factorisation in the soft scalings defined at the beginning of this section. Here colours are not associated to scaling, except for black lines which are either external or unscaled, but instead they help identify the different connected webs (\textbf{left}) and how the different worldline variables are defined (\textbf{right}). 
    The graph integrands factorise as the product of c-web integrands in their respective 
    soft limits and where the Schwinger parameters associated to jet lines have been replaced by the corresponding worldline variables. The black rectangles signify that an overall factor proportional to the corresponding tree-level diagram has been stripped out.}
    \label{fig:beta_variables_graph}
\end{figure}

Plugging this into the definition of the full diagram integrand we find the sought-after factorisation property
\begin{equation}
\dd \mathcal{I} \simeq  \dd \mathcal{I}_H  \times \prod_\mathcal{W}^{\text{c-webs}}
\dd \mathcal{I}_\mathcal{W}(\boldsymbol{\beta}) 
\end{equation}
The c-web integrands above take the form 
\begin{equation}
    \dd \mathcal{I}_\mathcal{W}
    \simeq
    \left.\left( \frac{\prod \d \alpha \d \beta \d \gamma}{\mathcal{U}_\mathcal{W}^{\D / 2}} 
    \mathcal{N}_{\mathcal{W}}\left(\partial_{z}\right) 
    e^{i \mathcal{V}_\mathcal{W}(\boldsymbol{\beta};\zz_\mathcal{W} ) }  \right) \right|_{\zz \to 0},
\end{equation}
which is proportional to the (scaled) integrand of the web $\mathcal{W}$ but with the jet $\alpha$ parameters swapped for the appropriate $\beta$ worldline variables and having stripped out an overall factor proportional to the tree-level diagram. 
This is graphically shown in \cref{fig:beta_variables_graph}.

\subsubsection*{Ladder diagrams}

Before proceeding let us quickly look into what happens for the simpler case where 
we only consider ladder diagrams without any blobs. This is important since in the IR 
limit these are the only diagrams in the class discussed in this paper which in four-dimensional QED 
actually induce divergences, with the exception of ladder diagrams dressed by photon propagator corrections (discussed in section \ref{sec:Renorm2}). 

In this case, where we have taken $\ell$ photons soft, which means that we have $\ell$ separate webs, 
the soft action simply becomes:
\begin{equation}
    \begin{aligned}
        \mathcal{V}_S(\zz) &= \sum_{w=1}^\ell \frac{\F_{(1),S}(\beta_{J_i,w}, \beta_{J_j,w})}{\U_{(1),S}(\beta_{J_i,w}, \beta_{J_j,w},\gamma_{w})} +i \sum_w (p_i z_{J_i,w} + p_j z_{J_j,w}) = \\
        &= \sum_{w=1}^\ell \frac{s_{ij} \beta_{J_i,w} \beta_{J_j,w} - m^2 (\beta_{J_i,w}+\beta_{J_j,w})^2}{\gamma_{w}} +i \sum_w (p_i z_{J_i,w} + p_j z_{J_j,w}),
        \end{aligned}
    \label{eq:SoftFactor-ladders}
\end{equation}
with $s_{ij} \equiv (p_i + p_j)^2$.
So, once again, we find that by writing the soft action in terms of the worldline variables 
we get the sum over the one-loop integrands, and a simple numerator part encoded by the second 
sum in the equation above. In both sums, we are summing over the webs which are labelled by the photon, 
$w$, where $w$ picks a $J_i$ and $J_j$ corresponding to the jets it is attached to, and the $\beta$ 
variables are the ones associated with the vertices where the photon is anchored to jets. 

\paragraph{Example: numerators in scalar QED}
In the case of scalar QED, we have that the numerator function is of the form
\begin{equation}
    \begin{gathered}
    \begin{tikzpicture}[line width=0.6,scale=1.2,baseline={([yshift=0.0ex]current bounding box.center)}]
        \coordinate (t1) at (-1,0.6);
        \coordinate (t2) at (1.5,0.6);
        \coordinate (b1) at (-1,-0.6);
        \coordinate (b2) at (1.5,-0.6);

        \draw[] (t1) -- (t2);
        \draw[] (b1) -- (b2);

        \coordinate (mt) at ($(t1)!0.5!(t2)$);
        \coordinate (mb) at ($(b1)!0.5!(b2)$);

        \draw[->] (-0.7,0.8)-- (-0.2,0.8);
        \draw[->] (-0.7,-0.4)-- (-0.2,-0.4); 
        \draw[->] (0.45,-0.4)-- (0.45,0.4); 
        \node[scale=0.8] at (0.75,0) {$q_{w}^\mu$};
        
        \draw[photon] (mt) -- (mb);
        
        \node[scale=0.8] at (-0.5,0.3) {$p_{J_i,w}^\mu$};
        \node[scale=0.8] at (-0.5,-0.9) {$p_{J_j,w}^\mu$};

        \node[scale=0.8,above] at (t2) {$J_i$};
        \node[scale=0.8,below] at (b2) {$J_j$};
    \end{tikzpicture}
\end{gathered} \quad \propto \quad (2p_{J_i,w} + q_w)\cdot (2p_{J_j,w} - q_w) 
\end{equation}
where we have one such factor for all the photons $w$, connecting jets $J_i$ and $J_j$. Therefore, the numerator prefactor in the Schwinger representation, $\mathcal{N}(\partial_{z_e})$, is given by
\begin{equation}
    \mathcal{N}(\partial_{z_e}) \propto \prod_{w=1}^\ell \left( 2\frac{\partial }{\partial z_{J_i,w}} + \frac{\partial }{\partial z_{w}} \right) \cdot \left( 2\frac{\partial }{\partial z_{J_j,w}} - \frac{\partial }{\partial z_{w}} \right),
\end{equation}
and so when we act with this operator on $e^{i \mathcal{V}_S(\zz)}$ with the soft action given by \eqref{eq:SoftFactor-ladders}, for each photon $w$ we find 
\begin{equation}
     \mathcal{N}(\partial_{z_e}) \; e^{i \mathcal{V}(\zz)} \propto
     (2 \, p_{J_i,w}) \cdot (2 p_{J_j,w}) \; e^{i \mathcal{V}(\zz)}
\end{equation}
which corresponds to the product of standard eikonal numerator factors for each soft photon insertions. 

\paragraph{Example: numerators in QED}
For QED we can consider a single jet line of momentum $p_J$, with soft photons $\{w_1,w_2,\dots\}$ attached to it. This yields a numerator of the form 
\begin{equation}
 \mathcal{N}(\partial_{z_e}) \propto  
 \; 
 \cdots
 \;
\left( -i\frac{\partial}{\partial \slashed{z}_{w_2}} + m \right)
\gamma^{\mu_2} 
\left( -i\frac{\partial}{\partial \slashed{z}_{w_1}}  + m \right)
\gamma^{\mu_1} u(p_J)
\end{equation}
which when acting on the exponential of the action \cref{eq:SoftFactor-ladders}, repeatedly using the Clifford algebra $\{\gamma_\mu,\gamma_\nu\} = 2 g_{\mu \nu}$ to commute the $\slashed{p}_J$ to the right and the Dirac equation $(\slashed{p}_J-m) u(p_J) = 0$, gives
\begin{equation}
\begin{aligned}
    \mathcal{N}(\partial_{z_e}) \; e^{i \mathcal{V}(\zz)} 
    &\propto  
    \left( 
    \cdots
    \;
    (\slashed{p}_{J} + m )
    \gamma^{\mu_2} 
    (\slashed{p}_{J} + m )
    \gamma^{\mu_1} u(p_J)
    \right)
    \; e^{i \mathcal{V}(\zz)} \\
    &=
    u(p_J)
    \;
    \left( 
    \cdots
    \;
    (2p_J^{\mu_2})
    (2p_J^{\mu_1})
    \right)
    \; e^{i \mathcal{V}(\zz)}
\end{aligned}
\end{equation}
which once again yields the (same) eikonal numerator factor for each photon insertion.

%% file: Sections/Exponentiation.tex
\section{Exponentiation of tropical rays}
\label{sec:regions}

In the previous section, we showed how the \emph{integrand} leading to soft divergences for ladder-type diagrams factorises into a soft and hard contribution. In this section, we show that the \emph{integrated} answer also factorises. As a by-product, we will show that this leads to the exponentiation of divergences in QED. To demonstrate the integral factorisation, we show that when summing over all diagrams, the integration regions combine in the correct way leading to exponentiation, as previously shown explicitly for the two-loop example in~\cref{sec:regions2loop}. Below, we will draw diagrams and write out explicit expressions for a two-jet process without any blobs for simplicity, but the generalisation to $N$ jets and including blobs is straightforward. Just like in~\cref{sec:regions2loop}, we need to make sure to only perform soft expansions in the corresponding integration regions, which can then be extended to the full original integration domain by defining suitably modified soft integrands.
Importantly, as we will see momentarily, we introduce new (spurious) UV divergences when we use soft approximations of the integrands, $i.e.$ those worked out in \cref{sec:factorisation_matrix}.
However, throughout this section, we will systematically
ignore them assuming that they are regulated in some way ($e.g.$ by a scale that separates the UV from the IR) and that they don't affect the IR sensitive integration regions. Thus, we will declare that an integral is finite
as long as it is free of soft (IR) divergences. We will then come back 
to the spurious UV limits in \cref{sec:spurious_UV_and_RGE}.

\subsection{Diagrams and scalings}
\label{sec:diags_and_scalings}
\begin{figure}[!t]
    \centering
    \includegraphics[width=0.8\linewidth]{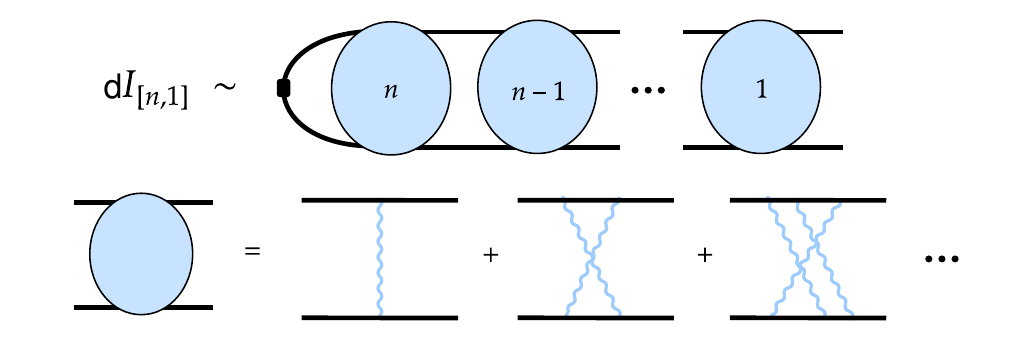}
    \caption{Organisation of the amplitude in rungs, each containing indivisible photon webs.}
    \label{fig:definition_rungs}
\end{figure}

\begin{figure}[t]
    \centering
    \includegraphics[width=0.8\textwidth]{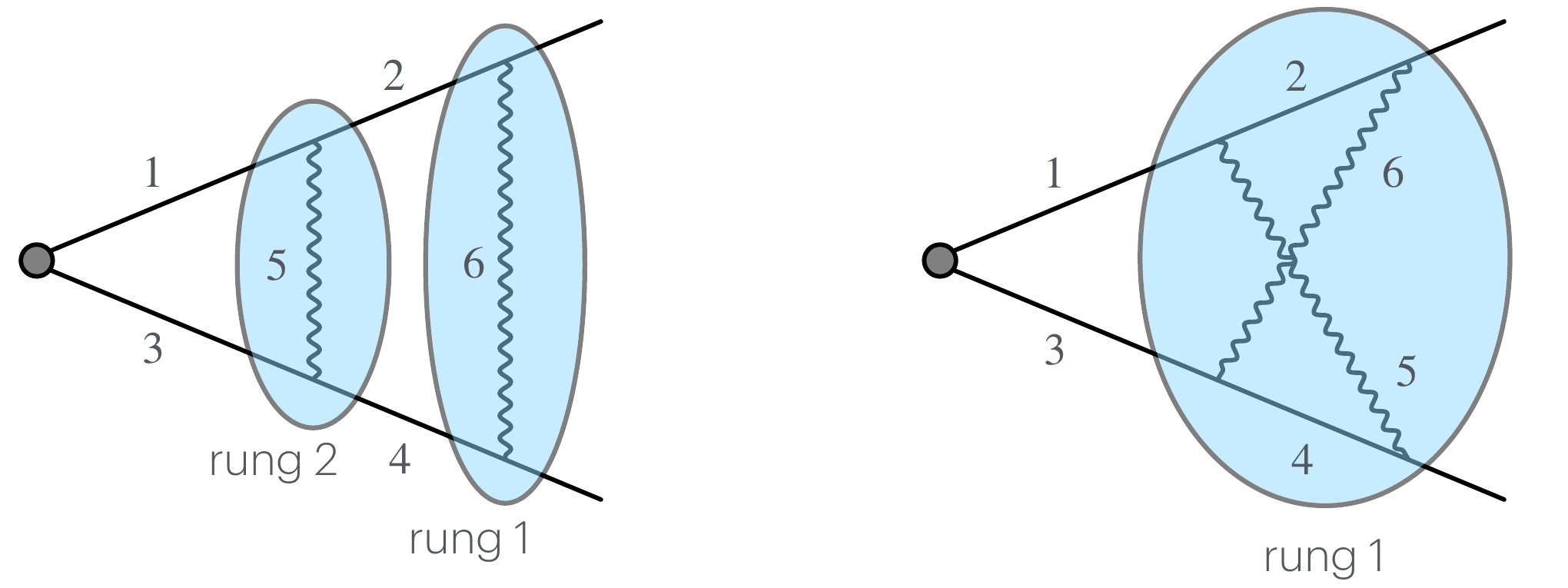}
    \caption{(\textbf{left}) The planar two-loop ladder has two rungs. Labelling the parameters by $\alpha_i$, with $i$ given by the numbering in the figure the collection of Schwinger parameters for rung 1 is given by $\rung_1 = \{ \alpha_{6}, \alpha_{4}, \alpha_{2} \}$, while that for rung 2 is $\rung_2 = \{\alpha_{5}, \alpha_{3}, \alpha_{1} \}$; (\textbf{right}) The non-planar two-loop ladder has a single rung with $\rung_1 = \{\alpha_{6},\alpha_{5},\alpha_{4},\alpha_{3},\alpha_{2},\alpha_{1} \}$. 
    So in the integrands $\d \I_{[k,h]}$, the contribution of the two-loop planar ladders comes from $\d\I_{[2,1]}$ while the non-planar one comes from  $\d\I_{[1]}$ (see fig. \ref{fig:definition_rungs}).}
    \label{fig:2loop-diagrams-omegas}
\end{figure}

In order to expose the exponentiated structure of soft divergences of ladder-type diagrams we use a different organisational principle than in~\cref{sec:regions2loop}: instead of working to fixed order in the coupling, we start by dividing each diagram into indivisible rungs,  or \textit{webs} as done in~\cref{fig:definition_rungs}. We will see how this organisation helps make the argument more transparent.
We label the webs from $1$ to $n$ starting from the \emph{outermost} rung. This labelling is different from previous sections, but will facilitate the presentation of our arguments.
Below we will refer to the Schwinger parameters of the $i$-th web of a single Feynman diagram as well as the hard (or jet) lines to its left by $\rung_i$. See~\cref{fig:2loop-diagrams-omegas},  for an example of this labelling for the two-loop ladder diagrams.  

We will refer to the sum of integrands for all diagrams with $n$ rungs as $\mathrm{d}\I_{[n,1]}$, where the notation emphasises that the integrand depends on variables $\rung_1$ through $\rung_n$. For the single-rung integrand $\d \I_{[1,1]}$ we will interchangeably use $\d \I_{[1]}$. 
According to this organisation of diagrams, the $k$-loop planar ladder enters 
in $\mathrm{d}\I_{[k,1]}$, while the fully non-planar $k$-loop ladders (where all photons are 
maximally intertwined) enters in $\mathrm{d}\I_{[1]}$.

In this section, differently from the ones above, we use the symbol $\d \I$ to refer to the \emph{full} integrand of the corresponding diagram, including all coupling and normalisation factors as well as numerators. We will however assume that the amplitude has been normalised to the tree level, which is indicated by the small black rectangle replacing the leftmost vertex in the figures of the section. In other words, the tree-level amplitude is 1 in this normalization convention.

\begin{figure}[!t]
    \centering
    \includegraphics[width=0.8\linewidth]{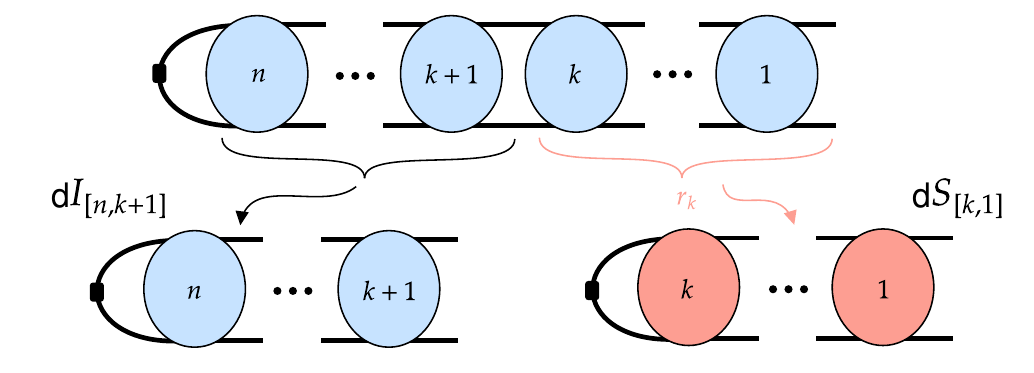}
    \caption{Depiction of integrand factorisation under the uniform scaling of the first $k$ rungs, see \cref{eq:int_factorisation}. Blue blobs represent unscaled rungs, while red ones represent scaled soft rungs.}
    \label{fig:blob_ampl_factorisation}
\end{figure}

From the tropical analysis (and by assumption),  we know that soft divergences arise when all photon webs from the outermost ($1$) 
to some inner one ($k$) become soft.  More precisely,  if we split the set of parameters $\rung_i$ into web $\web_i$ and jet $\jet_i$ components,  $\rung_i = \{\web_i,\jet_i\}$, then the divergent integration regions occur in the scalings given by the tropical rays
\begin{equation}
    r_k: \quad \web_i \sim \lambda^{-2}, \quad \quad
    \jet_i \sim \lambda^{-1}, \quad\quad
    \forall i \in \{1,\dots,k\},
\end{equation}
where $k\leq n$. For a given $k$, this prescription corresponds to the uniform soft scaling of the type illustrated in \cref{fig:blob_ampl_factorisation}: all webs from $1$ through $k$ are scaled uniformly according to $r_k$.
We call the integration region where this expansion is valid $D_{k}$. Even though there is a lot of freedom in how we exactly define the domains $D_k$, we will show below that
the their precise form is irrelevant to show integral factorisation. Our only restriction is that regions with different $k$ should not overlap and that their union 
should entirely fill the domain of integration where IR divergences arise. In other words, defining $D_0 = \mathbb{R}_+^E - \cup_k D_k $, we ask that $ \int_{D_0}\d \I  $ is IR finite.

From~\cref{sec:factorisation_matrix} we know that the integrand $\dd \I_{[n,1]}$ factorises along each ray $r_k$ as 
\begin{equation} \label{eq:int_factorisation}
 \dd \I_{[n,1]} \xrightarrow{r_k} \dd \I_{[n,k+1]} \times \dd S_{[k,1]}
\end{equation}
where
\begin{equation}
	 \dd S_{[k,1]} = \left[ \dd \I_{[k,1]} \right]_{r_k}
\end{equation}
and the symbol $[\dots]_{r}$ indicates the leading approximation with respect to the scaling induced by the ray $r$. In addition, we know that the scaled ($i.e.$~soft) diagrams also factorise under a ray $r_k$ as 
\begin{equation} \label{eq:soft_int_factorisation}
 \dd S_{[n,1]} \xrightarrow{r_k} \dd S_{[n,k+1]} \times \dd S_{[k,1]} .
\end{equation}
The expressions in~\cref{eq:int_factorisation} and~\cref{eq:soft_int_factorisation} are the ones that will be crucial for proving factorisation.

From~\cref{sec:factorisation_matrix}, we further know that, once written in terms of the worldline variables ($\beta$), each $\dd S_{[k,1]}$ further factors into products over the individual rungs in ladder-type diagrams. However, this fact will not be important until~\cref{sec:exponentiation_general}, so until then we assume that the expressions are written in terms of the original Schwinger parameters, $\rung_i \in \mathbb{R}_+$, unless explicitly stated otherwise. 
The full integral associated with the diagram in \cref{fig:definition_rungs} is then approximated by 
\begin{equation}\label{eq:int_soft_approx}
	\int_{\mathbb{R}_+^E} \dd \I_{[n,1]} \simeq \sum_{k=1}^n \int_{D_{k}} \left[ \dd \I_{[n,1]} \right]_{r_k},
\end{equation}
where the error is free of IR divergences.
In this equation, it is crucial that we are integrating each soft integrand given by the ray $r_k$ only in the corresponding subregion $D_k$. Indeed our next task is to extend the integration of each term back to the full domain $\mathbb{R}_+^E$.

\paragraph{Note} In order to avoid cluttering the notation in what follows we will omit the $\mathbb{R}_+^E$ 
(unless necessary) and use the symbol $\int$ to refer to an integral of the positive orthant.

\subsection{Modified soft integrands}
The strategy to prove factorisation will be the same as we followed in the example in \cref{sec:two_loop}.  Our first step will be to extend the integration region of the approximated integrands from the respective domains $D_{k}$ to the full positive orthant while avoiding double-counting divergent contributions.

In order to transform the r.h.s. of \cref{eq:int_soft_approx} into a sum of integrals over the full integration domain $\mathbb{R}^E_+$,  we start from region $D_1$ and write the corresponding integral as the integral over $\mathbb{R}^E_+$ minus the integral over all other regions:
\begin{equation}
	\int_{D_1} \left[ \dd \I_{[n,1]} \right]_{r_1} = \int_{D_1} \dd \I_{[n,2]} \times \widetilde{ \dd S}_1 =\int_{\mathbb{R}^E_+ - D_0} \dd \I_{[n,2]} \times \widetilde{ \dd S}_1
-\sum_{k=2}^n \int_{D_k} \dd \I_{[n,2]} \times \widetilde{ \dd S}_1 ,
\end{equation}
where we defined $ \widetilde{ \dd S}_1  =  \dd S_1$ and we remind the reader that
the region $D_0$ is the one in which the unscaled integral $\d \I_{[n,1]}$ is IR finite, 
defined as
\begin{equation}
	D_0  \equiv \mathbb{R}^E_+  - \left( \cup_{k=1}^n D_{k} \right).
\end{equation}
Plugging this into \cref{eq:int_soft_approx}, we find
\begin{equation}\label{eq:int_soft_approx_1}
\begin{aligned}
	\int \dd \I_{[n,1]} &\simeq
	\int_{\mathbb{R}^E_+ - D_0} \dd \I_{[n,2]}\times \widetilde{\dd S}_1
	+ \sum_{k=2}^n \int_{D_{k}} \left( \left[ \dd \I_{[n,1]} \right]_{r_k} 
	- \dd \I_{[n,2]}\times \widetilde{\dd S}_1 \right)  \\
	& \simeq \int_{\mathbb{R}^E_+ - D_0} \dd \I_{[n,2]}\times \widetilde{\dd S}_1
	+ \sum_{k=2}^n \int_{D_{k}} 	\left[ \dd \I_{[n,1]}  - \dd \I_{[n,2]}\times \widetilde{\dd S}_1 \right]_{r_k} 
    \\
	& \simeq \int_{\mathbb{R}^E_+ - D_0} \dd \I_{[n,2]}\times \widetilde{\dd S}_1
	+ \sum_{k=2}^n \int_{D_{k}} 	\dd \I_{[n,k+1]} \times \left( {\dd S}_{[k,1]}  - {\dd S}_{[k-1,1]} \times \widetilde{\dd S}_1 \right) ,
\end{aligned}
\end{equation}
where in the second line we used the approximation $\int_{D_k} \dd J  \simeq \int_{D_k} [\dd J]_{r_k}  $, and in the third line we used ${\dd S}_{[k,2]} \equiv {\dd S}_{[k-1,1]}$. We have also used a convention in which $\dd \I_{[n,n+1]}=1$.
Moving to the integral over region $D_2$ in the last line of the equation above, we now have the modified soft integrand  
\begin{equation}
\dd \I_{[n,3]} \times \widetilde{ \dd S}_{[2,1]}, \quad\quad \text{with \quad } \widetilde{ \dd S}_{[2,1]} =  {\dd S}_{[2,1]}  - {\dd S}_{1} \times  \widetilde{ \dd S}_1 ,
\end{equation}
which provides an approximation of the full integrand along the ray $r_2$ without double-counting the contribution from $r_1$. 
Indeed,  this integrand vanishes when the ray $r_1$ is applied to it. At the level of the regions, this integrand is such that when we extend from $D_2$ to $D_1 \cup D_2$ we do not gain any new divergences, exactly as desired. 
\begin{figure}[t]
    \centering
    \includegraphics[width=0.9\linewidth]{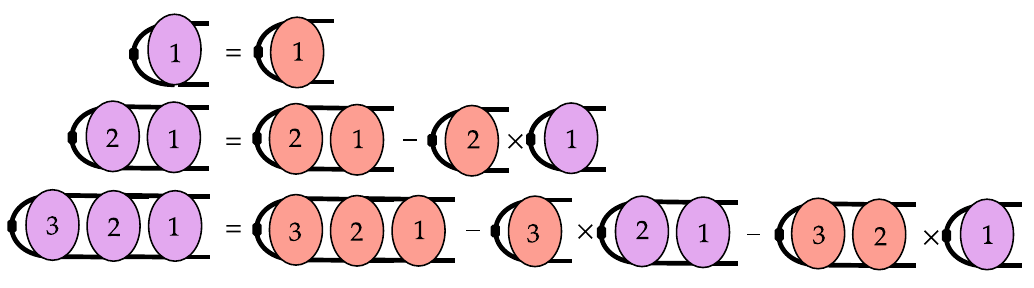}
    \caption{Example of the subtracted IR integrands $\widetilde{ \dd S}_{[k,1]}$ associated with each region vector and defined recursively by \cref{eq:subtracted_IR_integrands}. Diagrams with $k$ purple blobs correspond to the subtracted integrands $\widetilde{ \dd S}_{[k,1]}$, while diagrams with $k=1,2,3$ red blobs correspond to the soft integrands $\dd S_{[k,1]}$.}
    \label{fig:dR_definition}
\end{figure}

Recursively applying this subtraction procedure to $r_3,\, r_4,\, \text{etc.}$ we find that the full integral can be approximated by 
\begin{equation}\label{eq:int_soft_approx_3}
	\int \dd \I_{[n,1]} 
	\simeq 
     \int_{\mathbb{R}^E_+ - D_0} \sum_{k=1}^n \dd \I_{[n,k+1]} \times \widetilde{ \dd S}_{[k,1]}
\end{equation}
where the subtracted integrands are defined iteratively as
\begin{equation}\label{eq:subtracted_IR_integrands}
 	\widetilde{ \dd S}_{[k,1]} = \dd S_{[k,1]} - \sum_{j=1}^{k-1}\dd S_{[k,j+1]} \times \widetilde{ \dd S}_{[j,1]}  = \dd S_{[k,1]} - \sum_{j=1}^{k-1}\dd S_{[k-j,1]} \times \widetilde{ \dd S}_{[j,1]}.
\end{equation}

In figure \ref{fig:dR_definition} we present a graphical representation of this subtraction. Each term in the sum on the r.h.s.~of \cref{eq:int_soft_approx_3} can be interpreted as
the ``genuine'' contribution coming from the tropical ray $r_k$ since
$\dd \I_{[n,k+1]} \times \widetilde{ \dd S}_{[k,1]}$ provides an approximation of the full 
integrand $\dd \I_{[n,1]}$ along the region vector $r_k$ without double-counting any of the contributions 
coming from the previous divergent rays, $i.e.$ all rays $r_j$ with $j<k$. In other words, the modified
soft integrands have a single divergent ray: the one which scales all photons in all rungs 
at the same rate.
One can explicitly check this by verifying that $[\widetilde{ \dd S}_{[k,1]}]_{r_j} = 0 $ for all $j<k$. 

However, we have not yet succeeded in expressing the IR
approximation as an integral over the original integration domain $\mathbb{R}_+^E$, since we must still prove that the integrand in \cref{eq:int_soft_approx_3} 
yields an IR-finite result when integrated over the $D_0$ domain. This is equivalent to showing 
that the integrand 
\begin{equation}\label{eq:D0_integral}
   \left[ \sum_{k=1}^n \dd {\I}_{[n,k+1]} \times \widetilde{ \dd S}_{[k,1]} \right]_{D_0}
\end{equation}
vanishes at leading order when expanded along any of
its tropical rays associated with regions inside $D_0$. 
This can be shown by rearranging the integrand as follows:
\begin{equation}\label{eq:D0_integrand}
 \sum_{s=1}^n (-1)^{s+1} 
 \sum_{n \geq k_1\geq \dots \geq k_s \geq 1} 
 \dd {\I}_{[n,k_1+1]} \times 
 \dd S_{[k_1,k_2+1]} \times 
 \dots \times 
 \dd S_{[k_{s-1},k_s + 1]} 
 \times \dd S_{[k_{s},1]}  .
\end{equation}

From this expression it becomes clear that the new (spurious) divergent rays in $D_0$ 
correspond to the soft scalings of an internal set of rungs $[k,h]$ with $h<1$ (recall that non-soft parts of $D_0$ are already finite by assumption).
However, scaling the integrand \eqref{eq:D0_integrand} on any of these rays 
yields a vanishing result at leading order in $\lambda$ thanks to the alternating sign in the sum and to 
the factorisation properties of \cref{eq:int_factorisation,eq:soft_int_factorisation}. 
More precisely, let us focus on one particular term of the sum in \cref{eq:D0_integrand}
\begin{equation} \label{eq:D0_integrand_term}
     (-1)^{s+1} 
     \dd {\I}_{[n,k_1+1]} 
     \dots 
     \d S_{[k_{h-1},k_{h}+1]}
    \dots
      \dd S_{[k_{s},1]}  
\end{equation}
and apply the scaling given by $r_{[k_{h-1},k_{h}+1]}$, where rungs $k_{h-1}$ through $k_{h}+1$ are scaled.
The term above is invariant under this scaling and it develops a divergence due to the soft factor $\d S_{[k_{h-1},k_{h}+1]}$.
However, note that for every term of the type above, there is another in the sum in \cref{eq:D0_integrand} with one less $\d S$ factor which exactly matches \cref{eq:D0_integrand_term} when approximated along the scaling $r_{[k_{h-1},k_{h}+1]}$. 
Since this term has one less factor of $\d S$ in the product, it also has a coefficient of opposite sign, so the two terms cancel.

Using this, we can finally extend \cref{eq:int_soft_approx_3} to
\begin{equation}\label{eq:int_soft_approx_4}
	\int \dd \I_{[n,1]} 
	\simeq 
    \sum_{k=1}^n \left(\int  \dd \I_{[n,k+1]}\right) 
    \times 
    \left(\int \widetilde{ \dd S}_{[k,1]} \right)
\end{equation}
which allows us to restore the full domain of integration and for every term in the sum
uplift \emph{integrand} factorisation to \emph{integral} factorisation.  In \cref{sec:finite_remainders} we will see how \cref{eq:int_soft_approx_4} forms the basis for the (integral) factorisation of the full amplitude.

\begin{figure}
    \centering
    \includegraphics[width=0.9\linewidth]{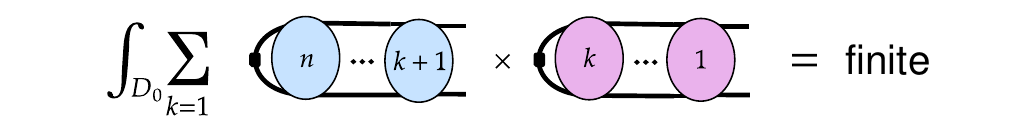}
    \caption{Graphical depiction of the finite leftover integration on the domain $D_0$, \cref{eq:D0_integral}}.
    \label{fig:D0_integral}
\end{figure}

\subsection*{Geometric interpretation} Just like we did in the two-loop example in~\cref{sec:regions2loop}, it is now interesting to 
point out that the form \cref{eq:D0_integrand} of the soft integrand can be understood
in terms of a simple geometrical picture. Starting from 
the original integrand $\d \I_{[n,1]} $ one can construct the associated 
Newton polytope $\bf N$ and identify all of its IR divergent faces. While  
the \emph{facets} (faces of codimension 1) of $\bf N$ identify the tropical rays $r_k$, 
a face of codimension $c$ is associated to a subset, $s$, of $c$ rays among the IR divergent rays where each $r_i \in s$ labels a facet that touches the face under consideration. An IR approximation of the integrand can then be obtained as
\begin{equation}
    \d \I_{[n,1]}^{\text{IR}} = \sum_{c=1} (-1)^{c+1} \sum_{s, |s| =c} \left[ \d \I_{[n,1]} \right]_{s}, 
    \label{eq:GeomFaces}
\end{equation}    
where $|s|$ is the number of rays in $s$, and $\left[ \d \I_{[n,1]} \right]_{s}$ stands for the expansion of in the integrand along the different rays in $s$, which exactly reproduces \cref{eq:int_soft_approx_4}. Note that, perhaps unsurprisingly, \cref{eq:GeomFaces} closely resembles the $R$ operation for UV divergences. In our language, the $R$ operation subtraction takes the same form of \cref{eq:GeomFaces}, but the sum runs over sets $s$ of rays which are associated to non-overlapping subdiagrams. This in our geometrical picture can be rephrased as compatibility of the corresponding rays, $i.e.$ the facets of all rays in $s$ should be adjacent, so that their scalings commute.

\subsection{Locally IR-finite remainders and factorisation}
\label{sec:finite_remainders}
Moving on, now that we have a valid IR approximation of the integrands, 
we can use it to define a IR finite integrand as the difference of the full integrand and 
its IR approximation \cref{eq:int_soft_approx_3}:
\begin{equation} \label{eq:finite_remainder_integrand}
	 \dd {H}_{[n,1]} =    \dd \I_{[n,1]} -
 \sum_{k=1}^n  \dd {\I}_{[n,k+1]} \times \widetilde{ \dd S}_{[k,1]}  .  
\end{equation}
For the case of three rungs ($n=3$) this can be represented in a similar fashion to \cref{fig:dR_definition} as
\begin{equation}
\includegraphics[width=0.9\textwidth]{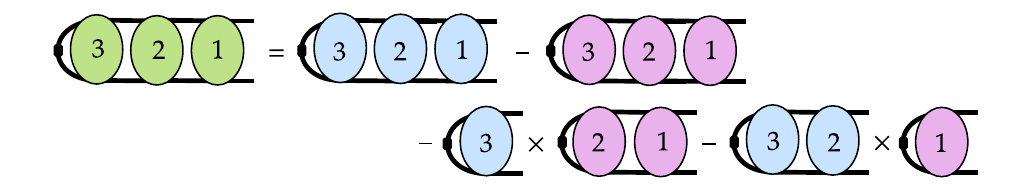}
\end{equation}
where we used the green colour to indicate the IR (locally) finite integrand.
Recursively eliminating in \cref{eq:finite_remainder_integrand} $\dd \I$ in favour of $\dd {H}$ and $\widetilde{ \dd S}$ (recall its definition \cref{eq:subtracted_IR_integrands}), one finds 
\begin{equation} \label{eq:recursive_finite_remainders}
		\int\dd \I_{[n,1]} = \sum_{k=1}^n \left( \int\dd {H}_{[n,k+1]} \right) \left( \int \dd S_{[k,1]} \right) + \int \dd H_{[n,1]}  .
\end{equation}
Note that in the equation above the IR divergent part of the integral
is captured by the original soft integrands $\d S$ rather than by 
their subtracted counterparts $\widetilde{\d S}$.
After summing over the number of webs $n$ from $n=0$ to infinity, \cref{eq:recursive_finite_remainders} implies the factorised amplitude form
\begin{equation}\label{eq:all_orders_factorisation}
 \mathcal{A}  = \mathcal{H} \times  \mathcal{S} ,
\end{equation}
with the finite remainder and the soft functions defined as 
\begin{equation}\label{eq:F_and_S_defs}
 \mathcal{H} =  \sum_{n=0}^{\infty} \int \dd 
 H_{[n,1]}, \quad\quad 
 \mathcal{S} = \sum_{n=0}^{\infty}     \int \dd S_{[n,1]},
\end{equation}
where we define $\int \dd S_{[0,1]} = 1$ and now the subscripts $[h,k]$ simply tell us how many rungs are contained in the corresponding diagram but not about which sets of parameters the integrand depends on. Let us now make some further comments.

Equations~\eqref{eq:all_orders_factorisation}~and~\eqref{eq:F_and_S_defs} provide a factorisation formula for the amplitude into an IR finite quantity ($\mathcal{H}$) and a \textit{soft} function ($\mathcal{S}$) which contains all infrared divergences. Furthermore, the finite remainder is defined explicitly and is locally IR finite in Schwinger-parameter space. 

Finally, let us highlight that in order for the amplitude to take the form of \cref{eq:all_orders_factorisation}, 
there are only two requirements:
\begin{itemize}
    \item[$(i)$] for any given divergent ray the integrand of the corresponding
        diagram(s) should factorise as in \cref{eq:int_factorisation}, 
        up to IR finite corrections. In general, if the ray $r$ is captured by the scaling $\alpha_i \sim \lambda^{-r_i}$ with $\lambda \gg 1$, the sufficient condition is
        \begin{equation}
        [\dd \I(\lambda)]_r = \dd \I' \times \dd S(\lambda) + \mathcal{O}(\lambda^0) ,
        \end{equation}
        with $\dd \I'$ the integrand of the Feynman diagram obtained by deleting the scaled subgraph and $\dd S$ any integrand of the scaled edges. In the case of logarithmic divergences, the equation above simply requires the factorisation of the leading term on $r$;
    \item[$(ii)$] the expansions of the integrand along different divergent rays 
        should commute
        \begin{equation}
        \left[ [\d \I_{[n,1]}]_{r_k} \right]_{r_h} = \left[ [\d \I_{[n,1]}]_{r_h} \right]_{r_k} .
        \end{equation}
\end{itemize}

\subsection{Exponentiation of the soft function}
\label{sec:exponentiation_general}

As a final step, we rearrange the terms in $\mathcal{S}$ in powers of the coupling, so that at each perturbative order $\ell$ it becomes a sum over planar and non-planar ladder-like diagrams approximated in their uniform soft scaling and integrated on the whole integration domain.
In particular, we can split these ladder diagrams first by loop order (number of photons) $\ell$ and then by number of photons attached to each couple of external legs: $\ell_{1,2}$ photons attached to jets 1 and 2, $\ell_{1,3}$ attached to jets 1 and 3, and so on. 
Using $\sigma$ to identify the various permutations of attachment order of the photons to the jets and collecting the number of photons connecting each pair of legs in a vector 
\begin{equation}
    \Vec{\ell} =(\ell_{1,2},\ell_{1,3},\dots,\ell_{1,N},\ell_{2,3},\ell_{2,4},\dots,\ell_{2,N},\dots,\ell_{N-1,N}),
\end{equation} 
with $N$ being the number of jets, we can write 
\begin{equation} \label{eq:sum_of_soft_ladders}
    \mathcal{S} = 1 + \sum_{\ell=1} \sum_{\substack{\Vec{\ell} \text{ s.t.}\\|\Vec{\ell}|=\ell}}  \sum_{\sigma} \int \d \mathcal{L}_{\Vec{\ell}}^{\sigma} ,
\end{equation}
where\footnote{Here and below we use the notation $(i,j)$ to refer to unordered couples with $i \neq j$.} $|\Vec{\ell}| = \sum_{(i,j)} \ell_{i,j}$, $i.e.$ the sum over all the components of $\Vec{\ell}$, and  $\d \mathcal{L}_{\Vec{\ell}}^{\sigma} $ stands for the integrand of an individual diagram, containing the photon attachments given by $\Vec{\ell}$ and ordered according to $\sigma$. 
Invoking the result from \cref{sec:factorisation_matrix}, we know that the integrand of 
each uniformly soft ladder-like diagram is given by a product of one-loop single-photon (dipole) exchanges, where the fermion Schwinger
parameters are replaced by the corresponding worldline variables:
\begin{equation} \label{eq:general_dipole_product}
    \d \mathcal{L}_{\Vec{\ell}}^{\sigma}  = 
    \prod_{w=1}^\ell  \d S_{(1)} (\boldsymbol{\beta_w}; p_{w},p^\prime_{w}).
\end{equation}
Here, just like in \cref{eq:SoftFactor-ladders}, we defined $\boldsymbol{\beta_w} \equiv (\beta_{i,w},\beta_{j,w},\gamma_w)$ where $i$ and $j$ are the two jets to which photon $w$ is attached, and $\beta_{i,w}/\beta_{j,w}$ are the corresponding worldline variables. 
In addition $p_{w} \equiv p_{i}$ and $p^\prime_{w} \equiv p_{j}$ are the external momenta entering the two jet lines. 
Remarkably we have that the $r.h.s.$ of \cref{eq:general_dipole_product} is \textit{independent} of $\sigma$, as emphasised in \cref{sec:factorisation_matrix}. Of course, for different $\sigma$'s the worldline variables are integrated over different domains.

In particular, for the case in which there are multiple jets, at a loop order specified by $\Vec{\ell}$, we can express the worldline variables in terms of the original $\alpha$-parameters for a given diagram with the attachments given by permutation $\sigma$ as 
\begin{equation}
    \beta_{j,k} = \sum_{h=1}^{\ell_j} \Theta^{\sigma_j}_{k, h} \, \alpha_{j,h},
\end{equation}
with $\ell_j$ being the number of soft photons attached to jet $j$, $i.e.$ $\ell_j = \sum_{k\neq j}^{N} \ell_{j,k}$. Above, $\Theta^{\sigma_j}_{k,h}$ is $1$ if the edge $h$ is between the hard part and the vertex where the $k$-th photon is inserted on jet $j$, and 0 otherwise.
\begin{figure}[t]
    \centering
    \includegraphics[width=\linewidth]{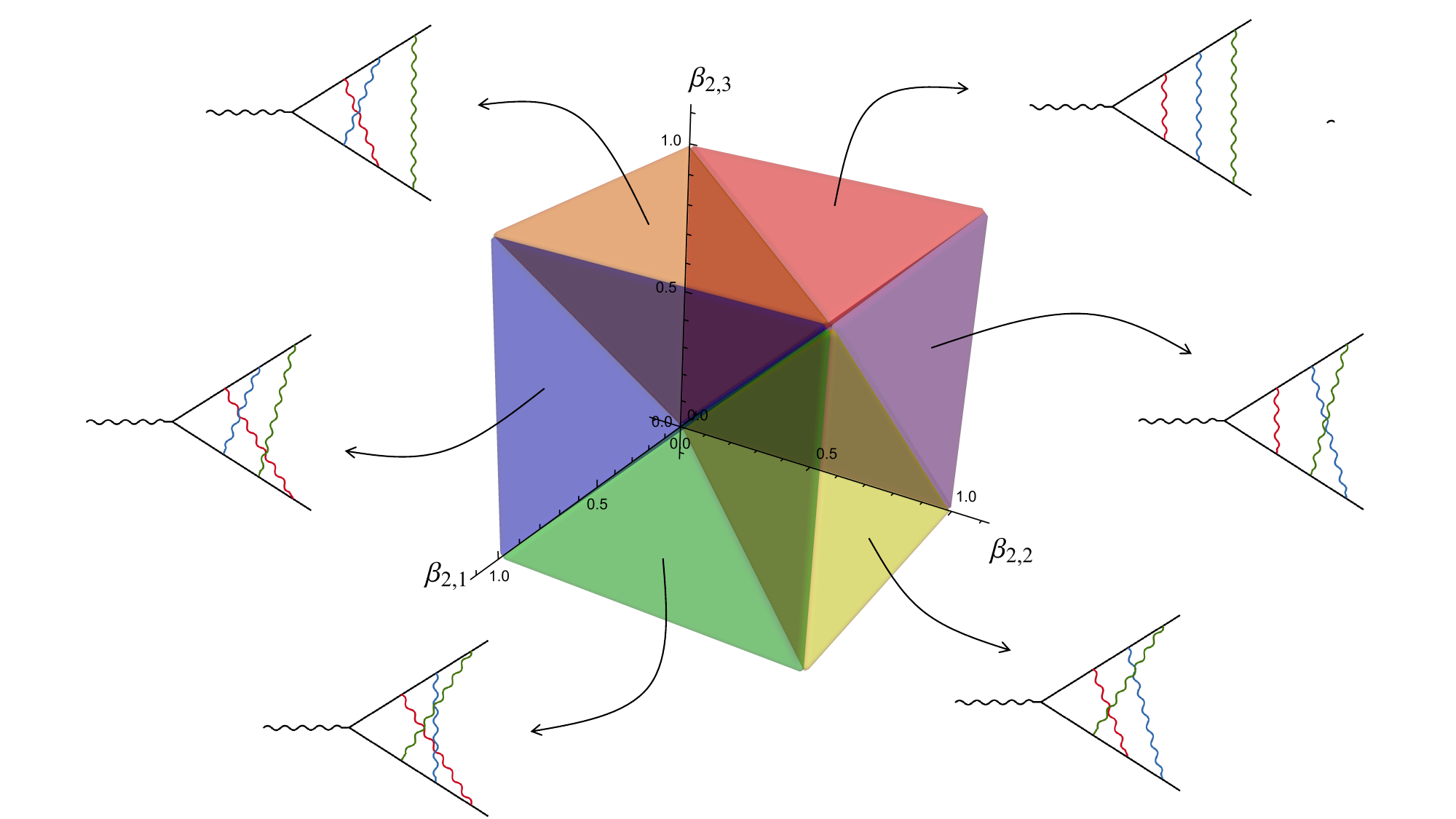}
    \caption{We fix the photon attachments along the top jet line $(J=1)$ and use it to define the labelling of the photons, $i.e.$ photon $1$ is represented in red, photon $2$ in blue and photon $3$ in green. Then we consider all the possible permutations of the attachments along bottom jet line $(J=2)$, $\sigma_2(w)$. The resulting $\beta$ variables for each ordering range inside a different domain, and the union of these domain tiles the full positive orthant.}
    \label{fig:3loopBetas}
\end{figure}

Therefore, given an order $\sigma_j$ for the order in which the photons are attached along jet $j$, $i.e.$ with $\sigma_j(1)$ the number of the photon attached to the first soft vertex, closer to the hard part, and $\sigma_j(\ell_j)$ the number of the photon attached to the last soft vertex in jet $j$, the worldline variables are defined in the domain 
\begin{equation} \label{eq:worldline_orderings}
   C^{\Vec{\ell}}_{\sigma_j}: \quad 0< \beta_{j,\sigma_j(1)} < \beta_{j,\sigma_j(2)} < \cdots < \beta_{j,\sigma_j(\ell_j)}.
\end{equation}
It is then easy to see that if we sum over all possible orderings $\sigma_j$, the final domain for half of the worldline variables is precisely the positive orthant. The integration over the remaining ones is ordered and yields a factor of $1/\prod_{(i,j)} \ell_{i,j}!$, which is crucial for exponentiation. 
In \cref{fig:2loop_worldline} we showed how the two-loop planar and non-planar ladders in $\beta$ space covered the positive quadrant. In \cref{fig:3loopBetas}, we show another example of how the domains over the $\beta$ variables correctly cover the positive octant for the case of diagrams with two jet lines at three-loops. 

Proceeding in the same way for each jet line, we can then define the domain of integration for diagram $\d \mathcal{L}_{\Vec{\ell}}^{\sigma}$ as  $C^{\Vec{\ell}}_\sigma = \prod_{j=1}^N C^{\Vec{\ell}}_{\sigma_j}$, and therefore for fixed $\Vec{\ell}$ (at loop-order $\ell=|\Vec{\ell}|$) we have
\begin{equation}
\begin{aligned}
    \sum_\sigma \int \d \mathcal{L}_{\Vec{\ell}}^{\sigma}  &=     
    \int_{\cup_\sigma C_{\sigma}^{\Vec{\ell}}} \prod_{w=1}^\ell  \d S_{(1)} (\boldsymbol{\beta_w}; p_{w},p'_{w}) \\
    &=     
    \frac{1}{\ell_{1,2}!\ell_{1,3}!\dots \ell_{n-1,n}!}
    \left(\int   \d S_{(1)} (\boldsymbol{\beta}; p_{1},p_{2}) \right)^{\ell_{1,2}} 
    \cdots
    \left(\int   \d S_{(1)} (\boldsymbol{\beta}; p_{n-1},p_{n}) \right)^{\ell_{n-1,n}}
 \end{aligned}
\end{equation}
where in the first equality we used \cref{eq:general_dipole_product}, in the second we used
the fact that the sum over $\sigma$ above corresponds to a sum over the orderings \eqref{eq:worldline_orderings}, and so the integral over the union of diagram domains
 $\cup_\sigma C_{\sigma}^{\Vec{\ell}}$ can be written as a product of one-loop
 integrals multiplied by an overall factor to compensate for over-counting. 
 Next, summing over all numbers of photon insertions on each line and using the multinomial formula we find 
\begin{equation}
     \sum_{\Vec{\ell}} \sum_\sigma \int\d \mathcal{L}_{\Vec{\ell}}^{\sigma}  = 
     \frac{1}{\ell!} \left( \sum_{(i,j)}  \int \d S_{(1)} (\boldsymbol{\beta}; p_i,p_j )  \right)^\ell.
\end{equation} 
Finally, summing over the loop order $\ell$ we get
\begin{equation}
    \mathcal{S} = 1+ \sum_{\ell=1} \frac{1}{\ell !} \left( \sum_{(i,j)} \int \d S_{(1)} (\boldsymbol{\beta}; p_i,p_j ) \right)^\ell  = \exp \left( \sum_{(i,j)}  \int \d S_{(1)} (\boldsymbol{\beta}; p_i,p_j )  \right),
 \end{equation} 
which is the sought-after exponentiated form of abelian soft divergences!

\subsection{Spurious UV divergences and soft RGE}
\label{sec:spurious_UV_and_RGE}
With the analysis above we proved that the sum of all ladder (planar and non-planar) diagrams can be cast in the form 
\begin{equation} \label{eq:recap_result}
    \mathcal{A} = \left( \int \d \mathcal{H} \right) \times \exp\left( \int \d S_{(1)} \right)
\end{equation}
where the integrand $\d \mathcal{H}$ has no IR divergences. However, looking more closely at the form of the one-loop soft QED integrand (recall \cref{eq:S_1_recall} and definitions below)
\begin{equation}
    \d S_{(1)}(\beta_1,\beta_2,\gamma)  =  
    -i^{3-\D/2} \frac{\overline{\alpha}(\mu)}{4\pi}  
     \frac{(2p_1)\cdot (2p_2)}{\mu^2} \frac{\d\gamma \d\beta_{1}\d\beta_{2}}{\gamma^{\D/2}} \exp\left[ i \, \frac{s \beta_{1} \beta_{2} - m^2 (\beta_{1}+\beta_{2})^2 }{\mu^2 \, \gamma} \right],
\end{equation}
we see that compared to its unscaled counterpart, it has developed a new UV divergence corresponding to the tropical ray
$r_{\text{cusp}} = (-1,-1,-2)$
which points exactly in the opposite direction of the soft tropical ray $r_{\text{IR}} = (1,1,2)$. More generally, this holds for any number of photons, both in the planar and non-planar case: each of the soft integrands $\d S_{[n,1]}$ defined in \cref{sec:diags_and_scalings}, which is the approximation along the soft ray $r_n$ of the corresponding unscaled integrand $\d\I_{[n,1]}$, picks up a corresponding UV divergence on the tropical ray $r_{UV,n} = -r_n$. Further, one can check that all these UV divergences are logarithmic. 

As a result, even after UV renormalisation, $\d \mathcal{H}$ is plagued by spurious UV divergences coming from the IR subtraction terms, which we should regulate if we want $\mathcal{H}$ to be a convergent integral in four space-time dimensions. Still, we would like to remove these UV poles while preserving the geometrical interpretation of exponentiation via worldline variables. Therefore whichever regularisation we choose should not involve the $\beta$ variables. 
One effective way to achieve this is to modify the soft integrands as 
\begin{equation} \label{eq:reg_dS}
    \d S_{[n,1]} \to  \d S_{[n,1]}^\Theta \equiv \d S_{[n,1]} \prod_i \Phi(\gamma_i) , 
\end{equation}
where the product runs over all soft photon (whose parameters are $\gamma_i$) and $\Theta$ is a function with the following properties:
\begin{equation}
    \Phi(\gamma) \xrightarrow{\gamma \to + \infty} 1 \, 
     \quad\quad \text{and} \quad \quad
     \Phi(\gamma) \xrightarrow{\gamma \to 0} 0 ,
\end{equation}
so that the soft integrand modification in \cref{eq:reg_dS} leaves all soft limits unchanged but removes the spurious UV divergence appearing as the photon Schwinger parameters vanish.
Because this does not interfere with the domains of integration over the worldline variables, and it preserves the factorisation and infrared approximation of the amplitude integrand, we can simply replace  $\d S_{[n,1]}$ with $\d S_{[n,1]}^\Phi$ everywhere in the proof above. As a consequence \cref{eq:recap_result} is modified to 
\begin{equation} \label{eq:recap_result_reg}
    \mathcal{A} = \left( \int \d \mathcal{H}^{\Phi} \right) \times \exp\left( \int \d S_{(1)}^{\Phi} \right),
\end{equation}
where now $\mathcal{H}^{\Phi} = \int \d \mathcal{H}^{\Phi}$ is a convergent integral in four dimensions.

\subsubsection*{Toy example of RGE of the finite remainder}
Though in theory one can freely choose the functional regulator $\Phi$ we will now focus on the simple choice
\begin{equation}
    \Phi(\gamma) = \theta(\gamma-\eta)   
    ,
\end{equation}
where $\eta$ is an $\mathcal{O}(1)$ parameter and $\theta$ the Heaviside function. This choice of $\Phi$ has the effect of simply truncating the integration in each photon Schwinger parameter $\gamma_k$.

In general it is well known that regulating the spurious UV divergences of the soft function induces an RGE equation of the finite remainder, see $e.g.$ \cite{Gardi:2009qi,Gardi:2009zv,Becher:2009cu,Becher:2009qa}, that this RGE can be written in terms of an IR regularisation scale $\mu_\text{IR}$. Further, by setting $\mu_\text{IR}$ to the renormalisation scale, the RGE can be further expressed into an evolution with respect to the running coupling. It is not our goal to show this here. Instead we simply want to demonstrate how, regardless of the regulator chosen, the RGE is controlled by the same anomalous dimension.

The RGE is obtained using the fact that $\mathcal{A}$ is independent of $\eta$, and therefore we can take a derivative of \cref{eq:recap_result_reg} with respect to $\eta$ to find 
\begin{equation}
\begin{aligned}
    \frac{1}{\mathcal{H}^{\Theta}} \partial_\eta \mathcal{H}^{\Theta} &= -  \partial_\eta \left( \int \d S_{(1)}^{\Theta} \right) \\
    &= 
    \frac{\overline{\alpha}(\mu)}{4\pi} 
    \frac{1}{{\eta^{1-\epsilon}} } (s-2m^2) \int \frac{\d b_{1}\d b_{2}}{\GL(1)} \frac{1}{s b_{1} b_{2} - m^2 (b_{1}+b_{2})^2}   \\
    &=  
    {-}\frac{\overline{\alpha}(\mu)}{4\pi} 
    \frac{1}{{\eta^{1-\epsilon}} } 
    \frac{2(s-2m^2) }{\sqrt{-s}\sqrt{4m^2 - s}}\log\left[ \frac{\sqrt{-s} - \sqrt{4m^2 - s}}{\sqrt{-s} + \sqrt{4m^2 - s}}\right] \\
    &= - \frac{1}{{\eta^{1-\epsilon}} } \gamma^b_{\text{cusp}} (p_1,p_2,m^2) . 
\end{aligned}    
\end{equation}
Rewriting it in terms of a derivative with respect to $\log \eta$ we find the familiar\footnote{The $\eta^{\epsilon}$ is an effect of the ``naive'' regularisation scheme we are using here.} RGE
\begin{equation}
    \frac{\partial \mathcal{H}^{\Theta}}{\partial \log \eta} =
    -  {\eta^{\epsilon}}  \; \gamma^b_{\text{cusp}} (p_1,p_2,m^2) \;  {\mathcal{H}^{\Theta}} , 
\end{equation}
where $\gamma^b_{\text{cusp}}$ is the QED bare cusp anomalous dimension
\begin{equation}
     \gamma^b_{\text{cusp}} = \frac{\overline{\alpha}(\mu)}{4\pi} \left(\beta - i \pi \theta(\beta) \right) \coth(\beta) \,,
     \label{eq:Cusp}
\end{equation}
and the cusp angle\footnote{Here we use $\beta$ as it is the a common symbol for the cusp angle in the literature 
but it should not be confused with the worldline variables.} 
$\beta$ between two momenta $p_1$ and $p_2$ is defined by
\begin{equation}
    \cosh \beta_{ij} = \frac{p_1 \cdot p_2}{|p_1| |p_2|}  = \frac{s-2m^2}{2m^2} \,.
\end{equation}
In the case of on-shell (OS) renormalisation the coupling in \cref{eq:Cusp} is replaced by the renormalised coupling $\overline{\alpha}_R$ and the electron-positron-photon vertex counterterms introduce a new class of diagrams whose soft divergences also exponentiate and have the overall effect of subtracting the $\beta \to 0$ (or $s \to 0$) limit  
from \cref{eq:Cusp}. This yields the familiar on-shell renormalised QED cusp anomalous dimension
\begin{equation}
         \gamma^{\text{OS}}_{\text{cusp}} = \frac{\overline{\alpha}_R}{4\pi} \left[ \left(\beta - i \pi \theta(\beta) \right) \coth(\beta)  - 1  \right]\,.
     \label{eq:Cusp_ren}
\end{equation}

Though we were successful in deriving the expected RGE equation in the regulator $\eta$ for the finite remainder $\mathcal{H}$, the result above is clearly not enough to prove that the soft singularities of QED amplitudes are governed by the UV singularities of Wilson line cusps. Indeed, the regulator $\eta$ we introduced above is useful to illustrate how in order to regulate the spurious UV divergences one is forced to introduce a new parameter or scale, but it does not have a direct physical interpretation. Other regulators might be better suited for this purpose, but we postpone the study of this connection to future investigation.

%% file: Sections/Conclusions.tex
\section{Conclusions}

In this paper, we have demonstrated how the Schwinger parametrization of Feynman integrals provides a powerful and systematic framework for deriving the factorisation and exponentiation of infrared divergences. A key advantage of our approach is the use of tropical geometry to systematically identify the potential divergences of general Feynman integrals, which can be translated into specific scalings of Schwinger parameters. 
By working directly at the level of graph Laplacians, we translated the complex combinatorics of factorisation and exponentiation into more tractable matrix manipulations, which also allow for the treatment of the numerators in a more unified way. While we applied it directly to QED exponentiation, our method offers a novel perspective that generalises to other gauge theories.

The graph Laplacian formalism not only facilitates these derivations but also naturally reveals variables interpretable as \emph{worldline distances}. Specifically, starting from the Schwinger parameters $\alpha_e$, our factorisation procedure automatically yields effective parameters, such as $\beta_j = \sum_{e \in \text{jet}} \alpha_{j,e}$, which represent the integrated proper time (or ``distance'') along a jet from the hard interaction. In terms of these worldline variables, the integrands for all ladder-type diagrams become identical in the soft limit. The distinct IR behaviour of different topologies arises from their corresponding integration regions in Schwinger parameter space. In particular, in this formalism it becomes manifest that planar ladder diagrams exhibit the most divergent behaviour, corresponding to a hierarchical softening of photons as they are further from the hard vertex. Further, combining the integration regions from all different diagrams disentangles the hierarchies among soft photons, yielding exponentiation of the one-loop soft anomalous dimension in QED.

Our proof of factorisation and exponentiation is diagrammatic and at the integrand level. As a by-product, it provides a local (in Schwinger parameter space) subtraction procedure which yields IR finite integrands at all loop orders. The subtraction terms, which we give in \cref{eq:D0_integrand,eq:GeomFaces,eq:finite_remainder_integrand}, can be understood geometrically as a recursive subtraction (and addition) of limits of the original integrand along tropical rays associated with faces of increasing codimension of the Symanzik Newton polytope. Thus, as we suggest in \cref{sec:two_loop}, a natural object to consider is not the standard Newton polytope, but rather the one we obtain by ``blowing-up'' all the IR faces -- associated to collections of the IR tropical rays -- into full facets. This is reminiscent of the ideas recently proposed in the context of Cosmohedra for the wavefunction in cosmology~\cite{Arkani-Hamed:2024jbp}.
The same procedure can be further extended from the IR case to that of UV divergences, for which it yields exactly the $R$ operation of BPHZ.
Coupled with local UV renormalisation of the diagrams our procedure can provide an integrand for the finite remainder which is convergent in four space-time dimensions.\footnote{Up to a proper contour deformation which satisfied the Feynman prescription.}
This could be interesting in view of recent progress in tackling direct numerical evaluation of QCD scattering amplitudes via local subtractions in momentum space \cite{Anastasiou:2018rib,Anastasiou:2020sdt,Anastasiou:2022eym,Anastasiou:2024xvk,Kermanschah:2024utt}.

\begin{figure}[!t]
    \centering
    \includegraphics[width=0.8\textwidth]{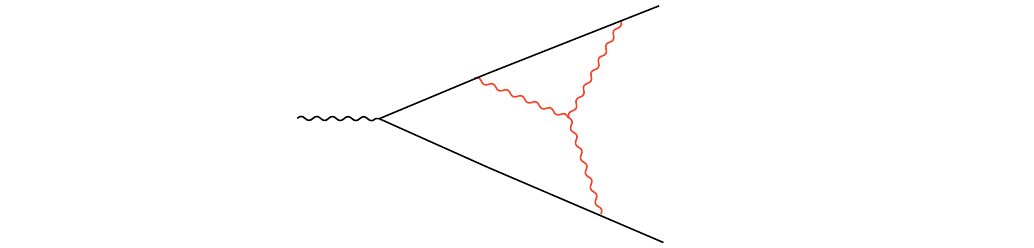}
    \caption{Simplest example of a soft purely non-abelian diagram.}
    \label{fig:non-abelian}
\end{figure}

An important future direction is to properly work out the factorisation of non-Abelian gauge theories like QCD using our methods. 
The formalism presented in this paper allows us to systematically identify all potentially divergent scalings using tools like the Landau equations and tropical geometry, and then accurately keep track of their interplay at the amplitude level, which is a key advantage as one ventures to the more complicated multipole structure of non-Abelian gauge theories. We hope this can contribute to developing a more general, and perhaps more algorithmic, way to derive factorisation theorems for a wider range of observables and in different quantum field theories. Indeed, the methods discussed can already be extended to a large class of diagrams in QCD by suitably (anti-)symmetrising over colour operators. We will explore this in a future work, where we will also extend our factorisation analysis in terms of worldline variables to purely non-abelian diagrams of the type in \cref{fig:non-abelian}.

Currently, we have used computational tools like {\tt polymake}~\cite{Gawrilow2000, polymake2} to help identify divergent scalings on a case-by-case basis. In generic kinematics, ref.~\cite{Arkani-Hamed:2022cqe} gives a complete list of the divergent rays for a class of Feynman integrals, stemming from the facet inequalities of the Newton polytope. However, since we consider diagrams for which the external particles are on shell with the same mass as the internal particles, we are no longer in the case of generic kinematics, and we get new rays compared to those found in ref.~\cite{Arkani-Hamed:2022cqe}. It would be natural to extend the rigorous methods of $e.g.$ ref.~\cite{Ma:2023hrt} to the case of strictly on-shell amplitudes in order to prove that the scalings we consider here are the only ones that lead to IR divergences.

So far, we have combined Feynman diagrams at the level of fixed orders in perturbation theory. Nevertheless, there is a simple RG equation relating different orders in perturbation theory. For example, imagining regulating our integrals by cutting off our Schwinger-parametrized worldline distances using a large wavelength $\Lambda$ for the photons, we can differentiate in this scale to get an equation of the form
\begin{equation}
    \frac{\partial \mathcal{A} (\Lambda)}{\partial \log \Lambda} =  \Gamma   \,\, \mathcal{A} (\Lambda) \, ,
\end{equation}
with $\Gamma$ proportional to the cusp anomalous dimension.
Diagrammatically, proving this equation directly involves comparing the amplitudes obtained by including particles cut off by the wavelength $\Lambda$ versus using the cutoff $\Lambda + \dd \Lambda$, order by order. Since $\Gamma$ contains powers of the coupling constant, this equation relates different orders in perturbation theory on the left and right-hand sides: adding photons propagating with an extra wavelength of $\dd \Lambda$ ``factor out'' of the Feynman diagrams and leaves the amplitude $\mathcal{A} (\Lambda)$, which includes only photons with a maximum wavelength of $\Lambda$, multiplied by $\Gamma \, \dd \log \Lambda$. It would be fascinating to investigate beyond the toy example in \cref{sec:spurious_UV_and_RGE} how this property translates to properties of the Newton polytopes for the IR divergences representing different orders in perturbation theory.

Looking further ahead, it would be interesting to understand how exponentiation relates directly to the geometry of Newton polytopes for Feynman integrals. One natural step in this direction is to further study the blown-up Newton polytope that is relevant for the IR structure, and ask whether it can be combined with the recent techniques which allow us to put large sets of Feynman diagrams into single geometrical objects, analogous to refs.~\cite{Arkani-Hamed:2023lbd,Arkani-Hamed:2023mvg,Salvatori:2024nva}. Establishing this connection could provide a more unified geometric picture, potentially allowing us to tackle both the identification of divergent scalings and the derivation of properties like factorisation and exponentiation.

%% file: Sections/App_Denominators_Diagram.tex
\section{Diagrammatic proof of scalar-integrand factorisation}
\label{app:factorisation_diagram}

In this appendix we prove the IR factorisation of the $\mathcal{F}$ and $\mathcal{U}$ polynomials as described in \cref{eq:conjecture}, for the class of diagrams corresponding to two-jet ladders, but following a diagrammatic approach analogous to what was done in \cref{sec:fact2loop}. We will also restrict to the scalar case, where we don't have any numerators, for simplicity.

\subsection{Proof at all planar loops}

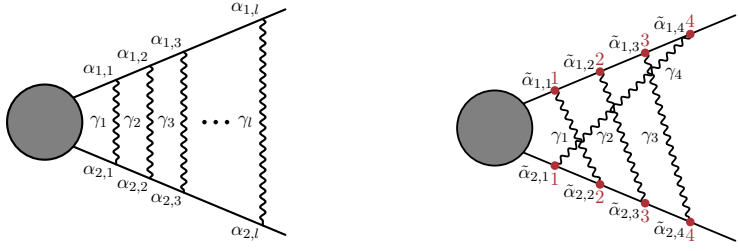
\begin{figure}[t]
    \centering
    \input{Figures/big_outerloop}
    \caption{The labelling of a $\ell$-loop planar diagram (\textbf{left}) and the labelling of an example four-loop non-planar diagram (\textbf{right}). The gray blob represents any Feynman diagram, which does not need to be a ladder. For the non-planar diagram, we define the permutations $\sigma(1)=2$, $\sigma(2)=3$, $\sigma(3)=4$ and $\sigma(4)=1$.}
    \label{fig:Pl-NPldiagrams}
\end{figure}

Assume that we have a Feynman diagram $G$, with $\ell$ outer loops connecting photons and the hard lines. Let us label the outer photon loops, $\gamma_i$, with indices from $1$ to $\ell$, and the Schwinger parameters of each outer loop $i$ as $\alpha_{1,i}$, $\alpha_{2,i}$ for jets $1$ and $2$, respectively, see \cref{fig:Pl-NPldiagrams}. The edge vector is then,
\begin{equation}
    \textbf{e} = (\underbracket[0.4pt]{x_1,x_2,\ldots,x_N}_{N  \text{ hard edges}}|\alpha_{1,1},\alpha_{1,2},\ldots,\alpha_{1,\ell}|\alpha_{2,1},\alpha_{2,2},\ldots,\alpha_{2,\ell}|\gamma_1,\gamma_2, \ldots, \gamma_\ell)
\end{equation}
and we consider the limit given by the scaling vector $r$ that splits it into inner and outer loops,
\begin{equation}
    r = (\underbracket[0.4pt]{0,0,\ldots,0}_{\substack{N \text{ hard edges}}}|\underbracket[0.4pt]{1,1,\ldots,1|1,1,\ldots,1}_{2\ell\text{ soft jet edges }}|\underbracket[0.4pt]{2,2,\ldots,2}_{\ell\text{ soft photons }} ) \,,
\end{equation}
so that we have
\begin{equation}
    x_i \xrightarrow[r]{} x_i \,, \qquad \alpha_{J,i} \xrightarrow[r]{} \lambda^{-1} \, \alpha_{J,i}\,, \qquad \gamma_i \xrightarrow[r]{} \lambda^{-2} \, \gamma_i \,.
\end{equation}
with $\lambda \to 0$. We now want to show that at leading order in the scaling parameter $\lambda$, the Symanzik polynomials behave as
\begin{align}
    \F_G & \xrightarrow[r]{}  \U_H \F_S + \U_S \F_H\,, \\
    \U_G & \xrightarrow[r]{} \U_H \U_S \,,
\end{align}
where, $\U_H$ and $\F_H$ are the first and second Symanzik polynomials of the inner diagram, and, for the soft parts we further have
\begin{equation}
    \F_S \xrightarrow[r]{} \medmath{\sum_{k=1}^\ell \prod_{i \neq k}^\ell} \gamma_i \F_1 (\beta_{J,k},\gamma_k) \,, \qquad
    \U_S \xrightarrow[r]{} \medmath{\prod_{i=1}^\ell} \gamma_i \,,
\end{equation}
where $\F_1 (\beta_{J,k},\gamma_k)$ is $\F$ polynomial for the one-loop ladder, written in terms of the worldline variables $\beta$. This ultimately leads to factorisation 
\begin{equation}
     \frac{\F_G}{\U_G} \to \frac{\F_H}{\U_H} + \sum_{k=1}^\ell \frac{\F_1 (\beta_{J,k},\gamma_k)}{\gamma_k}
\end{equation}
Let us start by looking at the $\U$ polynomial. Using the graphical identities obeyed by the Schwinger parameters summarized in \cref{eq:recursive_U_and_F}, for the planar ladder, we get
\begin{equation}
    \U \left(
    \input{Figures/outerloop}
    \right) =
    \Big(\medmath{\prod_{i=1}^\ell} \gamma_i \Big) \,
    \U \left(
    \input{Figures/cut-outerloop}
    \right) + \medmath{\sum_{j=1}^\ell} \Big( \medmath{\prod_{i \neq j}}  \gamma_i \Big) \U \left(
    \input{Figures/contr-outerloop}
    \right)  + \mathcal{O}(\lambda^{-(2\ell-2)}) \,,
\end{equation}
for the $\U$-polynomial. For the $\F_0$-polynomial, we get
\begin{equation}
    \F_0 \left( \input{Figures/outerloop} \right) 
    = 
    \Big(\medmath{\prod_{i=1}^\ell} \gamma_i \Big) \F_0 \left(
    \input{Figures/cut-outerloop}
    \right)
    +
    \medmath{\sum_{j=1}^\ell} \Big(\medmath{\prod_{i \neq j}} \gamma_i \Big) \F_0 \left(
    \input{Figures/contr-outerloop}
    \right) 
    + \mathcal{O}(\lambda^{-(2\ell-2)})
\end{equation}
Note that for both $\U_G$ and $\F_{0,G}$, not all the terms have the same scaling in $\lambda$. In particular, as discussed in the main text, the leading terms that scale like $\sim \lambda^{-2\ell}$ will cancel out in the computation of $\F_G$.

To break up the terms on the right-hand side, we sum over all different ways of forming the $\F_0$ polynomials. Recall that we have to split the diagrams into two trees. For the term with all the outer photon edges deleted, we get
\begin{equation}
    \F_0 \left(
    \input{Figures/cut-outerloop}
    \right) = \F_{0,H} + m^2 \, \medmath{\sum_{i=1}^\ell} \big( \alpha_{1,i} + \alpha_{2,i}\big) \, \U_H \,,
    \label{eq:F0-1}
\end{equation}
while for the $\F_0$ of the diagrams with a contracted edge, we get
\begin{equation}
    \F_0 \left(
    \input{Figures/contr-outerloop}
    \right) = s \beta_{1,j} \beta_{2,j} \U_H
    +
    m^2 \left(\beta_{1,j}+\beta_{2,j}\right) \medmath{\sum_{k=j+1}^\ell} (\alpha_{1,k} + \alpha_{2,k}) + \mathcal{O}(\lambda^{-(2\ell-2)})\,.
    \label{eq:F0-2}
\end{equation}
Now, $\F_G$ can be written as
\begin{equation}
    \F_G = \F_0 - \U \times \left( \medmath{\sum_{i=1}^\ell} m^2 (\alpha_{1,i} + \alpha_{2,i}) + \medmath{\sum_{e=1}^{N}} m_e^2 x_e \right),
\end{equation}
where we recall that $x_e$ is the Schwinger parameter for edge $e$ in the hard part. Entering the expansions of $\F_0$ and $\U$ derived in \cref{eq:F0-1,eq:F0-2}, we obtain at leading order
\begin{equation}
\begin{aligned}
    \F_G \xrightarrow[r]{} & \medmath{\sum_{j=1}^\ell} \Big(\medmath{\prod_{i\neq j}} \gamma_i\Big) \Big[s \beta_{1,j} \beta_{2,j} + m^2(\beta_{1,j} + \beta_{2,j}) \medmath{\sum_{k=j+1}^\ell} (\alpha_{1,k} + \alpha_{2,k}) \Big] \U_H \\
    &+ \Big(\medmath{\prod_{i=1}^\ell} \gamma_i \Big) \left[ \F_{0,H} + \medmath{\sum_{j=1}^\ell} (\alpha_{1,j}+\alpha_{2,j})m^2 \U_H\right]-\Big( \medmath{\sum_{e=1}^{N}} m_e^2 x_e\Big) \Big(\medmath{\prod_{i=1}^\ell} \gamma_i \Big) \U_H \\
    &-m^2\left(\medmath{\sum_{j=1}^{\ell}}(\alpha_{1,j}+\alpha_{2,j})\right)\left[\Big(\medmath{\prod_{i=1}^\ell} \gamma_i \Big) \U_H + \medmath{\sum_{k=1}^\ell} (\beta_{1,k}+\beta_{2,k}) \Big(\medmath{\prod_{i\neq k}} \gamma_i\Big)\U_H\right] \\
\end{aligned}
\end{equation}
from which we can directly identify the first term in the conjecture $\U_S\F_H$:
\begin{equation}
    \left(\medmath{\prod_{i=1}^\ell} \gamma_i\right) \left[\F_{0,H} - \Big( \medmath{\sum_{e=1}^{N}} m_e^2 x_e \Big) \right] \equiv \left(\medmath{\prod_{j=1}^\ell} \gamma_j \right) \F_H \equiv \U_S \F_H,
\end{equation}
in addition note that the most leading term, $-m^2\left(\medmath{\sum_{j=1}^{\ell}}(\alpha_{1,j}+\alpha_{2,j})\right)\left(\medmath{\prod_{i=1}^\ell} \gamma_i\right) \U_H$, which is of order $\sim \mathcal{O}(\lambda^{-(2n+1)})$, cancels so that the answer goes at leading order as $\lambda^{-2n}$, and simplifies to 
\begin{equation}
\begin{aligned}
    \F \to & \, \U_S\F_H + \medmath{\sum_{j=1}^\ell} \left(\medmath{\prod_{i\neq j}} \gamma_i\right) \left[s \beta_{1,j} \beta_{2,j} + m^2(\beta_{1,j} + \beta_{2,j})\left(\medmath{\sum_{k=j+1}^\ell} (\alpha_{1,k}+\alpha_{2,k})\right) \right] \U_H \\
    &-m^2\left(\medmath{\sum_{j=1}^{\ell}}(\alpha_{1,j}+\alpha_{2,j})\right)\left( \medmath{\sum_{k=1}^\ell} (\beta_{1,k}+\beta_{2,k})\right) \left(\medmath{\prod_{i\neq k}^\ell} \gamma_i\right)\U_H, \\
\end{aligned}
\end{equation}
we can finally rewrite the last term in the following way
\begin{equation}
\begin{aligned}
    &\medmath{\sum_{k=1}^\ell} \, \U_H \left(\medmath{\prod_{i\neq k}^\ell} \gamma_i\right)\left[ s\beta_{1,k} \beta_{2,k} - m^2\left(\medmath{\sum_{j=1}^k}(\alpha_{1,j}+\alpha_{2,j})\right)(\beta_{1,k}+\beta_{2,k}) \right] =\\
    &=\medmath{\sum_{k=1}^\ell} \, \U_H \left(\medmath{\prod_{i\neq k}^\ell} \gamma_i\right)\left[ s\beta_{1,k} \beta_{2,k} - m^2(\beta_{1,k}+\beta_{2,k})^2 \right] = \U_H \left(\medmath{\sum_{k=1}^\ell}  \left(\medmath{\prod_{j\neq k}^\ell} \gamma_j \right)\F_1(\beta_k,\gamma_k)\right)\\
    &= \U_H \F_S,
\end{aligned}
\end{equation}
making manifest the form conjectured.

\subsection{Incorporating non-planar diagrams: Region of integration}

Let us now consider a non-planar diagram with the same hard part, but where now the photons going soft can be anchored to the hard lines in a non-planar way. One way of labelling all such contributions is by considering all possible permutations, $\sigma(i)$, corresponding to the places in the bottom hard line where photon $i$ is anchored. So we always label photon $i$ by the vertex it is anchored to on the top hard line and consider all the possible vertices it can be anchored in the bottom line, $\sigma(i)$ (see \cref{fig:Pl-NPldiagrams}).

As described in the main text, for a given non-planar diagram, we can still define the worldline variables from the hard part to the vertex photon $i$ is anchored:
\begin{equation}
    \beta_{1,i} = \sum_{j=1}^i \tilde{\alpha}_{1,j}, \quad \quad \beta_{2,i}= \sum_{k=1}^{\sigma(i)} \tilde{\alpha}_{2,k}.
\end{equation}
Now, while for the planar diagram both $\beta_{1,i}$ and $\beta_{2,i}$ are defined in the following domain
\begin{equation}
   0< \beta_{J,1} < \beta_{J,2} < \cdots < \beta_{J,\ell},
\end{equation}
for a non-planar diagram, this domain still holds for the $\beta_{1,i}$ (since all photons are anchored in the standard ordering on the top line), but for the bottom ones, we have different domains depending on the permutation, $\sigma(i)$. For example, let us consider the case in which $\sigma(1) = 2, \, \sigma(2) = 1$ and $\sigma(i)=i$ for all $i \neq 1,2$, this is the case in which photons 1 and 2 are crossed. For this diagram the domain for the $\beta_{2,i}$ becomes
\begin{equation}
   0< \beta_{2,2} < \beta_{2,1}< \beta_{2,3} < \cdots < \beta_{2,\ell}.
\end{equation}
This makes it clear that if we consider the union of all the domains corresponding to all the different non-planar diagrams/permutations, we simply get $\beta_{2,i}>0$ for all $i\in\{1,2,\cdots,\ell\}$.

To conclude this appendix, let us show that in the limit where the $\ell$ outer photons go soft, the non-planar diagrams written in terms of the $\beta$ variables precisely agree with what we obtained for the planar one -- the only difference being the domains of the $\beta$'s in each case. This means that, since the soft part of the integrand factors from the hard one, when adding all diagrams together we can write the soft part as a single integral over a larger domain in $\beta$ space given by $ \beta_{2,i}>0$. 

So let's consider a given non-planar diagram, $G$, associated with permutation $\sigma(i)$. In terms of the $\beta$  variables we still have that in the soft limit:
\begin{equation}
    \U_G \to 
    \Big(\medmath{\prod_{i=1}^\ell} \gamma_i \Big) \,
    \U_H + \medmath{\sum_{j=1}^\ell} \Big( \medmath{\prod_{i \neq j}}  \gamma_i \Big) (\beta_{1,j}+\beta_{2,j}) \U_H  + \mathcal{O}(\lambda^{-(2\ell-2)}) \,,
\end{equation}
and similarly \cref{eq:F0-1} also holds with the replacement $\alpha \to \tilde{\alpha}$, now \cref{eq:F0-2} becomes
\begin{equation}
    \F_0  = s \beta_{1,j} \beta_{2,j} \U_H
    +
    m^2 \left(\beta_{1,j}+\beta_{2,j}\right) \left(\medmath{\sum_{k=j+1}^\ell} \tilde{\alpha}_{1,k} +\medmath{\sum_{m=\sigma(j)+1}^\ell} \tilde{\alpha}_{2,m}\right)\U_H + \mathcal{O}(\lambda^{-(2n-2)})\,.
\end{equation}
so that at leading order we obtain:
\begin{equation}
\begin{aligned}
    \F_G \xrightarrow[r]{} & \medmath{\sum_{j=1}^{\ell}} \Big(\medmath{\prod_{i\neq j}} \gamma_i\Big) \Big[s \beta_{1,j} \beta_{2,j}
    +
    m^2 \left(\beta_{1,j}+\beta_{2,j}\right) \left(\medmath{\sum_{k=j+1}^\ell} \tilde{\alpha}_{1,k} +\medmath{\sum_{m=\sigma(j)+1}^\ell} \tilde{\alpha}_{2,m}\right)\Big] \U_H \\
    &+ \Big(\medmath{\prod_{i=1}^\ell} \gamma_i \Big) \left[ \F_{0,H} + \medmath{\sum_{j=1}^\ell} (\tilde{\alpha}_{1,j}+\tilde{\alpha}_{2,j})m^2 \U_H\right]-  \left(\medmath{\sum_{e=1}^{N}} m_e^2 x_e \right)\Big(\medmath{\prod_{i=1}^\ell} \gamma_i \Big) \U_H \\
    &-m^2\left(\medmath{\sum_{j=1}^{\ell}}(\tilde{\alpha}_{1,j}+\tilde{\alpha}_{2,j})\right)\left[\Big(\medmath{\prod_{i=1}^\ell} \gamma_i \Big) \U_H + \medmath{\sum_{k=1}^\ell} (\beta_{1,k}+\beta_{2,k}) \Big(\medmath{\prod_{i\neq k}} \gamma_i\Big)\U_H\right] \\
\end{aligned}
\end{equation}
where similarly the leading order term cancels, we get precisely what we expected
\begin{equation}
\begin{aligned}
    \F_G \to & \, \U_S\F_H + \medmath{\sum_{j=1}^\ell} \left(\medmath{\prod_{i\neq j}} \gamma_i\right) \left[s \beta_{1,j} \beta_{2,j} + m^2(\beta_{1,j} + \beta_{2,j})\left(\medmath{\sum_{k=j+1}^\ell} \tilde{\alpha}_{1,k} +\medmath{\sum_{m=\sigma(j)+1}^\ell} \tilde{\alpha}_{2,m}\right) \right] \U_H \\
    &-m^2\left(\medmath{\sum_{j=1}^{\ell}}(\tilde{\alpha}_{1,j}+\tilde{\alpha}_{2,j})\right)\left( \medmath{\sum_{k=1}^\ell} (\beta_{1,k}+\beta_{2,k})\right) \left(\medmath{\prod_{i\neq k}} \gamma_i\right)\U_H \\
    =\,& \U_S\F_H + \medmath{\sum_{k=1}^\ell} \, \U_H \left(\medmath{\prod_{i\neq k}} \gamma_i\right)\left[ s\beta_{1,k} \beta_{2,k} - m^2\left(\medmath{\sum_{j=1}^k}\tilde{\alpha}_{1,j}+\medmath{\sum_{m=1}^{\sigma(k)}}\tilde{\alpha}_{2,m}\right)(\beta_{1,k}+\beta_{2,k}) \right]\\
     =\,& \U_S\F_H + \medmath{\sum_{k=1}^\ell} \, \U_H \left(\medmath{\prod_{i\neq k}} \gamma_i\right)\left[ s\beta_{1,k} \beta_{2,k} - m^2(\beta_{1,k}+\beta_{2,k})^2 \right] = \U_S\F_H +  \U_H\F_S.  \\
\end{aligned}
\end{equation}

%% file: Figures/big_outerloop.tex
\begin{tikzpicture}[line width=0.8,scale=1,line cap=round]
    \coordinate (a) at (-0.5,0);
    \coordinate (b1) at (1,0.25);
    \coordinate (b2) at (1,-0.25);
    \coordinate (c) at (0.8,0);
    \coordinate (d) at (4,1.5);
    \coordinate (e) at (4,-1.5);
    \path (b1) -- (d) coordinate[pos=0.25] (p1t); 
    \path (b2) -- (e) coordinate[pos=0.25] (p1b);
    \path (b1) -- (d) coordinate[pos=0.4] (p2t);
    \path (b2) -- (e) coordinate[pos=0.4] (p2b);
    \path (b1) -- (d) coordinate[pos=0.55] (p3t);
    \path (b2) -- (e) coordinate[pos=0.55] (p3b);
    \path (b1) -- (d) coordinate[pos=0.9] (p4t);
    \path (b2) -- (e) coordinate[pos=0.9] (p4b);
    \draw[] (b1) -- (d);
    \draw[] (b2) -- (e);
    \draw[photon] (p1t) -- (p1b) node[midway,scale=0.65,xshift=-10] {$\gamma_1$};
    \draw[photon] (p2t) -- (p2b) node[midway,scale=0.65,xshift=-10] {$\gamma_2$};
    \draw[photon] (p3t) -- (p3b) node[midway,scale=0.65,xshift=-10] {$\gamma_3$};
    \draw[photon] (p4t) -- (p4b) node[midway,scale=0.65,xshift=-10] {$\gamma_l$};
    \node[scale=0.65,xshift=-10,yshift=5] at (p1t) {$\alpha_{1,1}$};
    \node[scale=0.65,xshift=-10,yshift=5] at (p2t) {$\alpha_{1,2}$};
    \node[scale=0.65,xshift=-10,yshift=5] at (p3t) {$\alpha_{1,3}$};
    \node[scale=0.65,xshift=-10,yshift=5] at (p4t) {$\alpha_{1,l}$};
    \node[scale=0.65,xshift=-10,yshift=-5] at (p1b) {$\alpha_{2,1}$};
    \node[scale=0.65,xshift=-10,yshift=-5] at (p2b) {$\alpha_{2,2}$};
    \node[scale=0.65,xshift=-10,yshift=-5] at (p3b) {$\alpha_{2,3}$};
    \node[scale=0.65,xshift=-10,yshift=-5] at (p4b) {$\alpha_{2,l}$};
    \filldraw[color=black,fill=gray] (c) circle[radius=0.5];
    \fill[xshift=-5] (3.1,0) circle [radius=0.03] (3.25,0) circle [radius=0.03] (3.4,0) circle [radius=0.03];
\end{tikzpicture}
\hspace{2cm}
\begin{tikzpicture}[line width=0.7,scale=1,line cap=round]
    \coordinate (a) at (-0.5,0);
    \coordinate (b1) at (1,0.25);
    \coordinate (b2) at (1,-0.25);
    \coordinate (c) at (0.8,0);
    \coordinate (d) at (4,1.5);
    \coordinate (e) at (4,-1.5);
    \path (b1) -- (d) coordinate[pos=0.2] (p1t); 
    \path (b2) -- (e) coordinate[pos=0.4] (p1b);
    \path (b1) -- (d) coordinate[pos=0.4] (p2t);
    \path (b2) -- (e) coordinate[pos=0.6] (p2b);
    \path (b1) -- (d) coordinate[pos=0.6] (p3t);
    \path (b2) -- (e) coordinate[pos=0.8] (p3b);
    \path (b1) -- (d) coordinate[pos=0.8] (p4t);
    \path (b2) -- (e) coordinate[pos=0.2] (p4b);
    \draw[] (b1) -- (d);
    \draw[] (b2) -- (e);
    \draw[photon] (p1t) -- (p1b) node[midway,scale=0.65,xshift=-10] {$\gamma_1$};
    \draw[photon] (p2t) -- (p2b) node[midway,scale=0.65,xshift=-10] {$\gamma_2$};
    \draw[photon] (p3t) -- (p3b) node[midway,scale=0.65,xshift=-10] {$\gamma_3$};
    \draw[photon] (p4t) -- (p4b) node[midway,scale=0.65,xshift=30,yshift=15] {$\gamma_4$};
    \node[scale=0.65,xshift=-10,yshift=5] at (p1t) {$\tilde{\alpha}_{1,1}$};
    \node[scale=0.65,xshift=-12,yshift=5] at (p2t) {$\tilde{\alpha}_{1,2}$};
    \node[scale=0.65,xshift=-12,yshift=5] at (p3t) {$\tilde{\alpha}_{1,3}$};
    \node[scale=0.65,xshift=-12,yshift=5] at (p4t) {$\tilde{\alpha}_{1,4}$};
    \node[scale=0.65,xshift=-12,yshift=-5] at (p1b) {$\tilde{\alpha}_{2,2}$};
    \node[scale=0.65,xshift=-12,yshift=-5] at (p2b) {$\tilde{\alpha}_{2,3}$};
    \node[scale=0.65,xshift=-12,yshift=-5] at (p3b) {$\tilde{\alpha}_{2,4}$};
    \node[scale=0.65,xshift=-12,yshift=-5] at (p4b) {$\tilde{\alpha}_{2,1}$};
    \filldraw[color=black,fill=gray] (c) circle[radius=0.5];
    \filldraw[color=Maroon,fill=Maroon] (p1t) circle[radius=0.045] node[above,scale=0.65] {$1$}  (p1b) circle[radius=0.045] node[below,scale=0.65] {$2$} (p2t) node[above,scale=0.65] {$2$} circle[radius=0.045] (p2b) node[below,scale=0.65] {$3$} circle[radius=0.045] (p3t) node[above,scale=0.65] {$3$} circle[radius=0.045] (p3b) node[below,scale=0.65] {$4$} circle[radius=0.045] (p4t) node[above,scale=0.65] {$4$} circle[radius=0.045] (p4b) node[below,scale=0.65] {$1$}circle[radius=0.045];
\end{tikzpicture}

%% file: Figures/outerloop.tex
\begin{gathered}
    \begin{tikzpicture}[line width=0.7,scale=0.35,line cap=round]
        \coordinate (a) at (-0.5,0);
        \coordinate (b) at (1,0);
        \coordinate (d) at (4,1.5);
        \coordinate (e) at (4,-1.5);
        \path (b) -- (d) coordinate[pos=0.4] (p1t);
        \path (b) -- (e) coordinate[pos=0.4] (p1b);
        \path (b) -- (d) coordinate[pos=0.6] (p2t);
        \path (b) -- (e) coordinate[pos=0.6] (p2b);
        \path (b) -- (d) coordinate[pos=0.9] (p3t);
        \path (b) -- (e) coordinate[pos=0.9] (p3b);
        \draw[photon] (a) -- (b);
        \draw[] (b) -- (d);
        \draw[] (b) -- (e);
        \draw[photon] (p1t) -- (p1b);
        \draw[photon] (p2t) -- (p2b);
        \draw[photon] (p3t) -- (p3b);
        \filldraw[color=black,fill=gray] (b) circle[radius=0.5];
        \fill (3.1,0) circle [radius=0.03] (3.25,0) circle [radius=0.03] (3.4,0) circle [radius=0.03];
    \end{tikzpicture}
\end{gathered}

%% file: Figures/cut-outerloop.tex
    \begin{gathered}
    \begin{tikzpicture}[line width=0.7,scale=0.35,line cap=round]
        \coordinate (a) at (-0.5,0);
        \coordinate (b) at (1,0);
        \coordinate (d) at (4,1.5);
        \coordinate (e) at (4,-1.5);
        \path (b) -- (d) coordinate[pos=0.4] (p1t);
        \path (b) -- (e) coordinate[pos=0.4] (p1b);
        \path (b) -- (d) coordinate[pos=0.6] (p2t);
        \path (b) -- (e) coordinate[pos=0.6] (p2b);
        \path (b) -- (d) coordinate[pos=0.9] (p3t);
        \path (b) -- (e) coordinate[pos=0.9] (p3b);
        \draw[photon] (a) -- (b);
        \draw[] (b) -- (d);
        \draw[] (b) -- (e);
        \draw[photon] (p1t) -- ++ (-90:0.35) (p1b) -- ++ (90:0.35)
        (p2t) -- ++ (-90:0.55) (p2b) -- ++ (90:0.55)
        (p3t) -- ++ (-90:0.85) (p3b) -- ++ (90:0.85) (1,0)--(1.5,0);
        \filldraw[color=black,fill=gray] (b) circle[radius=0.5];
        \fill (3.1,0) circle [radius=0.03] (3.25,0) circle [radius=0.03] (3.4,0) circle [radius=0.03];
    \end{tikzpicture}
    \end{gathered}

%% file: Figures/contr-outerloop.tex
\begin{gathered}
    \begin{tikzpicture}[line width=0.7,scale=0.35,line cap=round]
        \coordinate (a) at (-0.5,0);
        \coordinate (b) at (1,0);
        \coordinate (c) at (4,0);
        \draw[] (b) to[out=90, in=90] (c);
        \draw[] (b) to[out=-90, in=-90] (c);
        \draw[] (c) --++ (30:2);
        \draw[] (c) --++ (-30:2);
        \node[scale=0.7,yshift=-1em] at (c) {$j$};
        \draw[photon] (a) -- (b);
        \coordinate (p1t) at (2,0.75);
        \coordinate (p1b) at (2,-0.75);
        \coordinate (p2t) at (2.5,0.8);
        \coordinate (p2b) at (2.5,-0.8);
        \coordinate (p3t) at (3.5,0.68);
        \coordinate (p3b) at (3.5,-0.68);
        \draw[photon] (p1t) -- ++ (-90:0.5) (p1b) -- ++ (90:0.5);
        \draw[photon] (p2t) -- ++ (-90:0.5) (p2b) -- ++ (90:0.5);
        \draw[photon] (p3t) -- ++ (-90:0.5) (p3b) -- ++ (90:0.5);
        \filldraw[color=black,fill=gray] (b) circle[radius=0.5];
        \fill (2.8,0) circle [radius=0.02] (2.95,0) circle [radius=0.02] (3.1,0) circle [radius=0.02];
    \end{tikzpicture}
    \end{gathered}

%% file: Sections/App_Loop_momentum_space.tex
\section{Exponentiation in loop-momentum space}
\label{app:exp-loopmom}
In this Appendix, we briefly discuss the proof of factorisation and exponentiation of IR divergences in loop-momentum space. The proof of Abelian exponentiation is often attributed to Yennie, Frautschi and Suura~\cite{Yennie:1961ad} as well as Weinberg's seminal paper from ref.~\cite{Weinberg:1965nx}. The former includes a discussion on the role of the numerators and renormalization. When deriving exponentiation of infrared divergences in loop-momentum space, previous work (see $e.g.$~\cite[Ch.~13]{Weinberg:1995mt}) takes a slightly different route from this work: they symmetrise over the loop momenta to simplify the discussion of nested divergences. The purpose of this Appendix is to review previous work and connect it to our derivation of exponentiation in Schwinger-parameter space.

\paragraph{Two loops}

We use the following labeling for the photons,
\begin{equation}
\begin{gathered}
    \begin{tikzpicture}[line width=0.8,scale=0.65,line cap=round]
        \coordinate (a) at (-0.5,0);
        \coordinate (b1) at (1,0.25);
        \coordinate (b2) at (1,-0.25);
        \coordinate (c) at (0.8,0);
        \coordinate (d) at (4,1.5);
        \coordinate (e) at (4,-1.5);
        \path (b1) -- (d) coordinate[pos=0.4] (p1t); 
        \path (b2) -- (e) coordinate[pos=0.4] (p1b);
        \path (b1) -- (d) coordinate[pos=0.8] (p2t);
        \path (b2) -- (e) coordinate[pos=0.8] (p2b);
        \draw[] (b1) -- (d);
        \draw[] (b2) -- (e);

        \path (b1) -- (d) coordinate[pos=-0.18] (x);
        \path (b2) -- (e) coordinate[pos=-0.18] (y);
        \draw[] (x) -- (b1);
        \draw[] (y) -- (b2);
        \path (x) -- (y) coordinate[pos=0.5] (k);
        
        \draw[photon] (p1t) -- (p1b) node[midway,scale=1,xshift=-10] {$k_1$};
        \draw[photon] (p2t) -- (p2b) node[midway,scale=1,xshift=-10] {$k_2$};
    
        \filldraw[color=black,fill=gray] (k) circle[radius=0.1];
\node[scale=1,xshift=20,yshift=5] at (d) {$J=1$};
        \node[scale=1,xshift=20,yshift=-5] at (e) {$J=2$};
    \end{tikzpicture}
\end{gathered}
\hspace{2cm}
\begin{gathered}
    \begin{tikzpicture}[line width=0.8,scale=0.65,line cap=round]
        \coordinate (a) at (-0.5,0);
        \coordinate (b1) at (1,0.25);
        \coordinate (b2) at (1,-0.25);
        \coordinate (c) at (0.8,0);
        \coordinate (d) at (4,1.5);
        \coordinate (e) at (4,-1.5);
        \path (b1) -- (d) coordinate[pos=0.4] (p1t); 
        \path (b2) -- (e) coordinate[pos=0.4] (p1b);
        \path (b1) -- (d) coordinate[pos=0.8] (p2t);
        \path (b2) -- (e) coordinate[pos=0.8] (p2b);
        \draw[] (b1) -- (d);
        \draw[] (b2) -- (e);

        \path (b1) -- (d) coordinate[pos=-0.18] (x);
        \path (b2) -- (e) coordinate[pos=-0.18] (y);
        \draw[] (x) -- (b1);
        \draw[] (y) -- (b2);
        \path (x) -- (y) coordinate[pos=0.5] (k);
        
        \draw[photon] (p1t) -- (p2b) node[midway,scale=1,xshift=15,yshift=-5] {$k_1$};
        \draw[photon] (p2t) -- (p1b) node[midway,scale=1,xshift=20,yshift=10] {$k_2$};
     
        \filldraw[color=black,fill=gray] (k) circle[radius=0.1];
\node[scale=1,xshift=20,yshift=5] at (d) {$J=1$};
        \node[scale=1,xshift=20,yshift=-5] at (e) {$J=2$};
    \end{tikzpicture}
\end{gathered}
\end{equation}
The most divergent region will come from the configuration where $k_2 \to 0$ and $k_1 \to 0$, but where $k_2$ tends to zero faster than $k_1$ does. This configuration, which is only relevant for the planar diagram, leads to a ${1}/{\epsilon_{\text{IR}}^2}$ pole of the amplitude at this order in perturbation theory. The scaling where $k_1$ and $k_2$ are of roughly equal magnitude and both tend to zero is relevant for both diagrams. If we exclude the integration region where $k_2$ is much smaller than $k_1$, this uniform scaling leads to a ${1}/{\epsilon_{\text{IR}}}$ divergence.

\paragraph{Soft scalings of loop momenta}
Two soft scalings will be relevant for the two-loop ladder, which in Schwinger-parameter space (in the notation of~\cref{sec:div2loop}) are
\begin{equation}
    r_{[1,2]} = (1,1,1,1,2,2)\,, \qquad r_{[2]} = (0,1,0,1,0,2) \,.
\end{equation}
The scaling $r_{[1,2]}$ is the uniform scaling applied to both loops, $i.e.$ the case in which both loop momenta $k_1$ and $k_2$ become soft, while $r_{[2]}$ corresponds to only the outermost photon in the loop labelled 2 becoming soft. This corresponds to the following scalings of the loop momenta,
\begin{align}
    k_1 & \mathrel{\underset{r_{[1,2]}}{\sim}} \; \lambda \,, \qquad k_2 \mathrel{\underset{r_{[1,2]}}{\sim}} \lambda\,,
    \\
    k_1 & \mathrel{\underset{r_{[2]}}{\sim}} 1\,, \qquad k_2 \mathrel{\underset{r_{[2]}}{\sim}} \lambda\,.
\end{align}
The most divergent part of the integral comes from applying both scalings simultaneously, $i.e.$ from when the outer loop momentum scales faster to zero than the inner one,
\begin{equation}
    k_1  \mathrel{\underset{r_{[1,2]}, r_{[2]}}{\sim}} \lambda\,, \qquad k_2 \mathrel{\underset{r_{[1,2]}, r_{[2]}}{\sim}} \lambda^2\,.
\end{equation}
While this discussion parallels the one in Schwinger-parameter space, we will soon see that if we first symmetrize over $k_1$ and $k_2$, then it turns out that the double scaling given by simultaneously applying $r_{[1,2]}$ and $r_{[2]}$ is automatically included in the $r_{[1,2]}$ scaling.

\paragraph{Factorisation of the integrand}
Recall that throughout the derivation in Schwinger-parameter space (see $e.g.$~\cref{sec:fact2loop}), the uniform scaling $r_{[1,2]}$ played an important role. In that scaling, we found  that the integrand factorises in terms of worldline variables  $\beta$, representing the worldline distance from the last hard scattering point to the photon emission and absorptions. 
In loop-momentum space, when adding the planar and non-planar integrands, an analogous expansion along the direction $r_{[1,2]}$ gives, 
\begin{equation}
    \d \I_{(2)}^{\text{pl}} + \d \I_{(2)}^{\text{npl}} 
    \xrightarrow[r_{[1,2]}]{}
    \frac{1}{k_1^2} \frac{1}{k_2^2} \frac{1}{2 p_1 \cdot k_1} \frac{1}{2 p_1 \cdot (k_1 + k_2)}
    \frac{1}{2 p_2 \cdot k_1}  
    \frac{1}{2 p_2 \cdot k_2}
    \d^Dk_1\d^Dk_2
    \,.
\end{equation}
The trick to simplify the integrand is to symmetrise over $k_1$ and $k_2$, which according to the derivation from~\cite[Ch.~13]{Weinberg:1995mt} gives:
\begin{align}
    \d \I_{(2)}^{\text{pl}} + \d \I_{(2)}^{\text{npl}} & \xrightarrow[r_{[1,2]}, \text{ symm in } k_1 \leftrightarrow k_2]{}
    \frac{1}{2} \d S_{(1)} \d S_{(1)} \,,
    \label{eq:I2pl-loopspace}
\end{align}
where $\d S_{(1)}$ is the eikonal factor for a one-loop diagram,
\begin{equation}
    \d S_{(1)}(k_i) \propto \frac{1}{k_i^2} \frac{1}{p_1 \cdot k_i} \frac{1}{p_2 \cdot k_i} \, \d^Dk_i\,,
\end{equation}
where the index $i$ labels the loop.
Here we have used the notation $\d S_{(1)}$ to indicate that the loop-momentum integrand above, 
assuming that the corresponding domain of integration is left untouched, can be translated in Schwinger representation
and yields exactly \cref{eq:S_1} (up to normalisation factors).

The expansion along $r_{[2]}$ is
\begin{align}
    \d \I_{(2)}^{\text{pl}} & \xrightarrow[r_{[2]}]{}
    \d \I_{(1)} \d S_{(1)} \,,
\end{align}
Furthermore, in the double scaling $r_{[1,2]}, r_{[2]}$, the integrand behaves as
\begin{equation}
\label{eq:I2-doublescaling}
    \d \I_{(2)}^{\text{pl}} + d \I_{(2)}^{\text{npl}} \xrightarrow[r_{[1,2]}, r_{[2]}]{}
    \d S_{(1)}(k_1) \, \d S_{(1)}(k_2) \,.
\end{equation}

\paragraph{The integration region}

\begin{figure}[t]
\centering
\begin{tikzpicture}[line width=1,draw=charcoal, scale=0.5]
    \coordinate (lx) at (-2,0) ;
    \coordinate (rx) at (8.5,0) ;
    \coordinate (a) at (0,3.5) ;
    \coordinate (mid) at (3.5, 3.5) ;
    \coordinate (b) at (7.5,3.5) ;
    \coordinate (c) at (3.5,0) ;
    \coordinate (d) at (3.5,7.5) ;
    \coordinate (o) at (0,0) ;
    \coordinate (x) at (7.5,0) ;
    \coordinate (y) at (0,7.5) ;
    \coordinate (ty) at (0,8.5);
    \draw[-latex] (lx) -- (rx);
    \draw[-latex] (by) -- (ty);
    \draw[Maroon, thick] (c) -- (mid) -- (b) ;
    \draw[Blue, thick] (a) -- (mid) -- (d) ;

    \fill[fill=Maroon, opacity=0.2] (c) -- (mid) -- (b) -- (x);
    \fill[fill=Blue, opacity=0.2] (a)-- (mid) -- (d) -- (y);
    \fill[fill=Gray, opacity=0.2] (o) -- (a) -- (mid) -- (c);

    \node[] at (9.5,0) {$|\vec{k}_1|$} ;
    \node[] at (0,9) {$|\vec{k}_2|$} ;
    \node[] at (5.5,1.3) {$D_{[2]}$} ;
    \node[scale = 0.5] at (5.5,0.35) {$1/\epsilon_{\text{IR}}$ divergence} ;
    \node[] at (1.5,5) {$D_{[1]}$} ;
    \node[scale = 0.5] at (1.6,4) {not divergent} ;
    \node[] at (1.7,1.9) {$D_{[1,2]}$} ;
    \node[scale = 0.75] at (-2.5,1.8) {$\substack{1/\epsilon_{\text{IR}}^2 \text{ divergence,} \\ \text{from region} \\ \text{where } k_2 \ll k_1}$} ;
    \draw[domain=0:3.5, dashed, samples=100, variable=\x, Gray, opacity=0.6] plot ({\x}, {1/3.5 * \x*\x});
    \draw[->, thick, bend left=20] (-1,1.6) to (2,0.5);
    \node[scale=0.75] at (4.5,9) {Non-symmetric integrand:};
\end{tikzpicture}
\hspace{0.5cm}
\begin{tikzpicture}[line width=1,draw=charcoal, scale=0.5]
    \coordinate (lx) at (-2,0) ;
    \coordinate (rx) at (8.5,0) ;
    \coordinate (a) at (0,3.5) ;
    \coordinate (mid) at (3.5, 3.5) ;
    \coordinate (b) at (7.5,3.5) ;
    \coordinate (c) at (3.5,0) ;
    \coordinate (d) at (3.5,7.5) ;
    \coordinate (o) at (0,0) ;
    \coordinate (x) at (7.5,0) ;
    \coordinate (y) at (0,7.5) ;
    \draw[-latex] (lx) -- (rx);
    \draw[-latex] (by) -- (ty);
    \draw[Maroon, thick] (c) -- (mid) -- (b) ;
    \draw[Maroon, thick] (a) -- (mid) -- (d) ;

    \fill[fill=Maroon, opacity=0.2] (c) -- (mid) -- (b) -- (x);
    \fill[fill=Maroon, opacity=0.2] (a)-- (mid) -- (d) -- (y);
    \fill[fill=Gray, opacity=0.2] (o) -- (a) -- (mid) -- (c);

    \node[] at (9.5,0) {$|\vec{k}_1|$} ;
    \node[] at (0,9) {$|\vec{k}_2|$} ;
    \node[] at (5.5,1.3) {$D_{[2]}$} ;
    \node[scale = 0.5] at (5.5,0.35) {$1/\epsilon_{\text{IR}}$ divergence} ;
    \node[] at (1.5,5) {$D_{[1]}$} ;
    \node[scale = 0.5] at (1.6,4) {$1/\epsilon_{\text{IR}}$ divergence} ;
    \node[] at (1.7,1.9) {$D_{[1,2]}$} ;
    \node[scale = 0.75] at (-2.5,1.8) {$\substack{1/\epsilon_{\text{IR}}^2 \text{ divergence,} \\ \text{from region} \\ \text{where } k_1,k_2 \to 0}$} ;
    \draw[->, thick, bend left=20] (-0.5,1.6) to (1,1);
    \node[scale=0.75] at (4,9) {Symmetric integrand:};
\end{tikzpicture}
\caption{2-loop integration regions in loop-momentum space. Before symmetrizing over $k_1$ and $k_2$ (\textbf{left panel}), the most divergent region contributing to a ${1}/{\epsilon_{\text{IR}}^2}$ divergence is obtained when $k_2 \ll k_1$ while $k_1,k_2 \to 0$. Since the region where $k_1 > k_2$ is not divergent, we end up picking up a factor of $1/2!$ from ignoring that part of the integration region. After symmetrizing over $k_1$ and $k_2$ (\textbf{right panel}), we manually add a factor of $1/2!$, but now the ${1}/{\epsilon_{\text{IR}}^2}$ comes from the region where $k_1,k_2 \to 0$, regardless of the hierarchy between $k_1$ and $k_2$.}
    \label{fig:2loopRegions_lspace}
\end{figure}
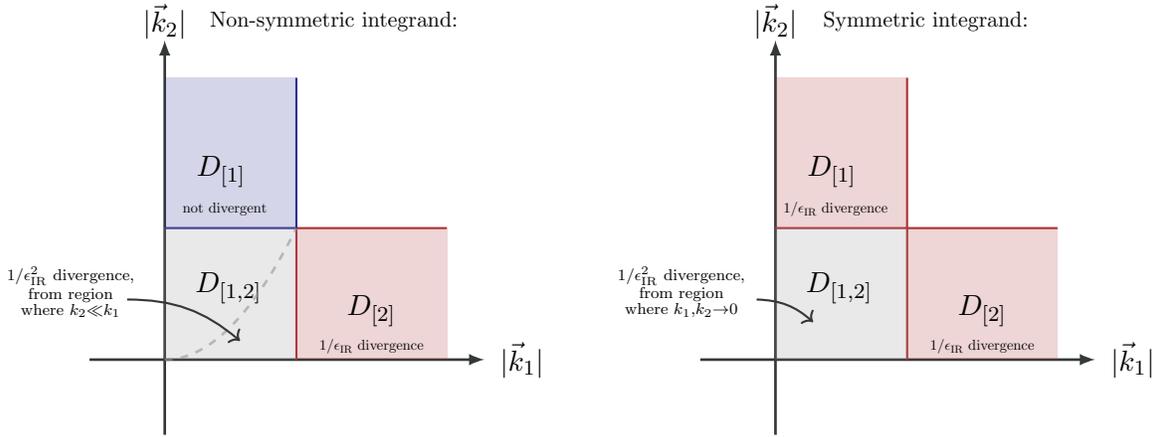
Analogously to what we did previously in Schwinger-parameter space, we split the integration region according to which momenta are going soft, see~\cref{fig:2loopRegions_lspace}: We define $D_{[1]}$ as the region where all components of $k_1$ are small but at least one component of $k_2$ is not small, $D_{[2]}$ as the one where all components of $k_2$ are small but $k_1$ is not, and $D_{[1,2]}$ as the one where both $k_1$ and $k_2$ are small. We use a slight abuse of notation and write $k_2 \ll k_1$ when $|\vec{k}_2| \ll |\vec{k}_1|$ and $k_1$, $k_2$ are approximately on shell. Similarly, we write $k_i \to 0$ when all components of $k_i$ tend to zero. Before symmetrizing over $k_1$ and $k_2$, the most divergent part of the integral, contributing to the ${1}/{\epsilon_{\text{IR}}^2}$ divergence, comes from the $k_2 \ll k_1$ part of the integration region. The trick often employed in loop-momentum space is to instead leverage the symmetric form of the integrand from~\cref{eq:I2pl-loopspace}. When doing so, we have to add a factor of $1/2!$, but the ${1}/{\epsilon_{\text{IR}}^2}$ divergence is now present regardless of the hierarchy of $k_2$ and $k_1$. Then, we can write 
\begin{equation}
    \int \dd \I_{(2)} = \int \dd \I_{(2)}^{\pl} + \int \dd \I_{(2)}^{\npl} \simeq
    \int_{D_{[2]}} \dd \I_{(1)} \dd S_{(1)} + \frac{1}{2} \int_{D_{[1,2]}} \dd S_{(1)} \dd S_{(1)} \,,
\end{equation}
analogously to what we had in \cref{sec:integrand_factorisation}.
Next, extend the domains of integration and rearrange the terms to obtain
\begin{equation}
\label{eq:dI2}
\begin{aligned}
    \int \dd \I_{(2)} & \simeq
    \int_{D_{[2]} \cup D_{[1,2]} \cup D_{[1]}} \dd \I_{(1)} \dd S_{(1)} - \frac{1}{2} \int_{D_{[1,2]} \cup D_{[1]} \cup D_{[2]}} \dd S_{(1)} \dd S_{(1)} \\
    & \hspace{1cm} - \int_{D_{[1]}} \dd S_{(1)} \dd S_{(1)} + \frac{1}{2} \int_{D_{[1]} \cup D_{[2]}} \dd S_{(1)} \dd S_{(1)}
    \,.
\end{aligned}
\end{equation}
Due to the symmetry in $D_{[1]}$ and $D_{[2]}$ for the integrand $\dd S_{(1)} \dd S_{(1)}$, the second line in~\cref{eq:dI2} adds up to zero. Furthermore, by assumption, we can extend the first line to include the domain where $|\vec{k}_1|$ and $|\vec{k}_2|$ are large and write
\begin{equation}
    \int \dd \I_{(2)} \simeq
    \int \dd \I_{(1)} \dd S_{(1)} - \frac{1}{2} \int \dd S_{(1)} \dd S_{(1)} = \I_{(1)} \mathcal{S}_{(1)} - \frac{1}{2} \mathcal{S}_{(1)}^2 \,,
\end{equation}
in agreement with the exponentiation statement of \cref{eq:expIR}.

\paragraph{Generalisation to $\ell$ loops}
To generalise the argument above to $\ell$ loops, we have to follow a procedure similar to that in~\cref{sec:integrand_factorisation}: recursively add and subtract the divergent integration regions. We can either leverage a symmetrisation in the uniform scaling analogous to~\cref{eq:I2pl-loopspace}, or, we can simply use the analogues of~\cref{eq:I2-doublescaling}. In either case, we have to be careful to include the relevant subtraction terms whenever we extend the integration regions to be over the full space.